\documentclass[preprint,
showpacs,preprintnumbers,amsmath,amssymb]{revtex4}

\usepackage{graphicx}
\usepackage{dcolumn}
\usepackage{bm,morefloats,color}

\usepackage{amssymb,bm}

\def\Eqref#1{Eq.~(\ref{#1})}

\def\Eq#1{\begin{equation} #1 \end{equation}}
\def\Eqr#1{\begin{eqnarray} #1 \end{eqnarray}}
\def\Eqrsubl#1#2{\begin{subequations}\label{#1}\Eqr{#2}\end{subequations}}

\newcommand{\nn}{\nonumber}
\newcommand{\pd}{\partial}

\newcommand{\bea}{\begin{eqnarray}}
\newcommand{\eea}{\end{eqnarray}}

\def\Xsp{{\rm X}}

\def\Ysp{{\rm Y}}
\def\Zsp{{\rm Z}}
\def\X5sp{{\rm X}_5}
\def\Y3sp{{\rm Y}_3}
\def\Z3sp{{\rm Z}_3}

\def\Msp{{\rm M}}
\def\Nsp{{\rm N}}
\def\lap{{\triangle}}
\def\e{{\rm e}}

\begin{document}

\preprint{YITP-12-60}

\title{Cosmological brane systems in warped spacetime}

\author{Masato Minamitsuji}
\affiliation{
Yukawa Institute for Theoretical Physics, 
Kyoto University, Kyoto 606-8502, Japan
}%

\author{Kunihito Uzawa}
\affiliation{
Department of Physics,
School of Science and Technology,
Kwansei Gakuin University, Sanda, Hyogo 669-1337, Japan
}%

\date{\today}

\begin{abstract}
In this paper, we discuss 
the time-dependent brane solutions
in higher-dimensional supergravity theories.
We particularly focus on the dynamical extensions of 
the intersecting brane solutions involving three branes.
We also show that
in the near-horizon limits, where the time dependence is negligible,
these branes describe warped anti-de Sitter spacetimes 
as in the corresponding static solutions. 
We finally examine
the lower-dimensional cosmological dynamics
obtained after compactifications of the higher-dimensional
solutions 
and show
the solutions we have found give the four-dimensional 
universe with power-law expansion.
\end{abstract}

\pacs{11.25.-w, 11.27.+d, 98.80.Cq}
\maketitle


\section{Introduction}
\label{sec:introduction}

This paper is devoted to a study of dynamical solutions of the D-brane 
in the higher-dimensional supergravity model.  
We discuss the cosmological models constructed from 
time-dependent D-branes in which intersect 
with NS-branes, a pp-wave, and a Kaluza-Klein monopole, that is, 
a Taub-NUT with charge $M_{\rm K}$. 
The exact dynamical solutions in the 
eleven- and ten-dimensional supergravity theory 
have been solved and studied recently \cite{Lu:1996jk, Behrndt:2003cx, 
Gibbons:2005rt, Chen:2005jp, Kodama:2005fz, Kodama:2005cz, Kodama:2006ay, 
Binetruy:2007tu, Binetruy:2008ev, Maeda:2009tq, Maeda:2009zi, 
Gibbons:2009dr, Maeda:2009ds, Maeda:2010yk, Maeda:2010ja, 
Minamitsuji:2010fp, Maeda:2010aj, 
Minamitsuji:2010kb, Minamitsuji:2010uz, 
Maeda:2011sh, Minamitsuji:2011jt, Maeda:2012xb, Blaback:2012mu}. 
However, we will see that there are still things which are not solved about 
delocalized or partially localized intersecting brane systems yet.
To state our result in ten-dimensional string theory, 
we will explain a simple T-duality mapping 
from IIA theory to IIB theory or vice versa in ten dimensions,
and dimensional reduction from eleven-dimensional theory to 
ten-dimensional type II theory.
The dynamical solutions in string theory will 
coincide with intersecting brane solutions, in the form that 
these have been presented in \cite{Cvetic:2000cj} and 
further studied in \cite{Youm:1999ti}. 
Mathematically,  there are time-dependent warp factors 
with field strengths and scalar field in higher-dimensional spacetime, 
and the coupling of scalar field takes the same value as the static case.
It has been known that 
the time dependence of the warp factor becomes negligible 
in the near-horizon limit. 
This is a consequence of 
works \cite{Binetruy:2007tu, Binetruy:2008ev, Maeda:2009zi, 
Minamitsuji:2010kb} and was used in \cite{Minamitsuji:2011jt} 
to obtain precise formulas for partially localized brane systems 
involving two branes. However, studies about the dynamical brane system 
have suggested that the relation 
to the supergravity models is not limited to the case of pair intersection 
for two branes. These are the main properties 
that we would like to understand.
Therefore,
the purpose of the present paper is 
to show the time-dependent generalizations of
the intersecting brane systems involving more than two branes,
and to understand the dynamics of the four-dimensional universe
in these models. 

We will consider solutions with particular couplings of dilaton to 
the field strengths. In the classical solution of a $p$-brane, 
the coupling to dilaton for field strengths includes the parameter $N$.  
This is specified in string theory as $N=4$. 
Though there are classical solutions 
for other values of coupling, the solutions 
in these models are no longer related to D-branes and M-branes. 
The dilaton coupling is related to the power of the warp factor 
in the metric and ensures the existence of a classical solution of  
$p$-branes in string theory. 
We know that it is not easy to find a time-dependent  
intersecting brane solution under the assumption 
of $N\neq 4$ \cite{Minamitsuji:2011jt}.

As in the case of other dynamical brane systems, 
we assume that the warp factor 
depends not only on the coordinate of the relative 
and overall transverse space but also the worldvolume coordinate. 
In a suitable ansatz for fields, we find that  
the warp factors arise from field strengths 
in the time-dependent background and then a system composed of $n$-branes 
can be characterized by $n$ warp factors arising from $n$ field strengths. 
Unfortunately, there are few solutions in which all harmonic functions 
depend on time except the D-brane bound states \cite{Maeda:2012xb}. 
Many other interesting models contain cosmological solutions, 
as a result of which they are not such close relatives of supergavity 
theories if all warp factors in the metric depend on time 
\cite{Minamitsuji:2010uz}. 
These time-dependent solutions are still inadequate for explaining 
cosmological behavior, such as inflation or 
accelerated expansion of our four-dimensional Universe. 

In the present paper,
we will 
investigate the partially localized intersecting brane solutions,
especially those involving more than two branes 
which have never been discussed yet 
except the delocalized brane systems. 
We will construct various explicit intersecting
dynamical D-brane, M-brane solutions in eleven and ten dimensions. 
We will also clarify the asymptotic spacetime structures of them
in the near-horizon limit,
and show that they describe the static spacetime 
whose metric is written by warped anti-de Sitter (AdS) 
spacetimes with the internal space. 
We will give the complete classification of these solutions,
focusing on their relations, 
and discuss the application of
them to cosmology in Secs.~\ref{sec:three} - \ref{sec:cosmology}. 
Among them,
as the new examples we discuss,
in Sec.~\ref{sec:three} we will focus on 
the partially localized brane solution involving three branes in terms of 
Kaluza-Klein (KK) reduction or T-duality map.
In Sec.~\ref{sec:ma}, we provide brief discussions for intersection
pairings on the massive supergravity theory.
In Sec. \ref{sec:asymptotic}, 
we discuss the behavior of brane intersections involving two branes 
in the near-horizon limits;
the ten- or eleven-dimensional spacetime 
metric is written by AdS or warped AdS spacetime with the internal space. 
In Sec. \ref{sec:cosmology}, we describe how the four-dimensional spacetime 
could be represented in the present model via an appropriate compactification. 
We show that there exists no accelerating expansion of our Universe 
in the Einstein conformal frame, although the accelerating 
expansion of the universe is possible in the non-Einstein frame. 
To illustrate this, we construct cosmological models of the D$p$-D$(p+2)$-NS5 
brane system, which is relevant to type II string theory. 
We give the classification of the D$p$-D$(p+2)$-NS5 brane, 
M-brane, KK monopole systems and their application to cosmology. 
Section \ref{sec:discussions} is devoted to conclusion and remarks.


\section{Dynamical partially localized intersection involving three branes 
system}
\label{sec:three}
In this section, we construct the time-dependent partially localized 
brane solution in ten- and eleven-dimensional theory. We concentrate on  
the dynamical intersection involving the three branes 
system. These are a generalization of the solutions found in 
\cite{Minamitsuji:2011jt}. 

\subsection{Dynamical D3-D5-NS5 system in ten-dimensional IIB theory}
\label{sec:IIB}

We start the cosmological D3-D5-NS5 system in ten-dimensional IIB theory. 
These solutions give AdS in warped ten-dimensional spacetime at 
the near-horizon limit of the intersecting branes. 
We also discuss the time-dependent D2-D6-KK monopole system 
in ten-dimensional IIA theory after performing
a T-duality transformation in D3-D5-NS5 solution. 

In the ten-dimensional IIB supergavity theory, 
the field equations are given by
\Eqrsubl{B:equations:Eq}{
&&R_{MN}=\frac{1}{2}\pd_M\phi \pd_N \phi
+\frac{1}{2\cdot 3!}\e^{\phi} 
\left(3F_{MAB} {F_N}^{AB}-\frac{1}{4}g_{MN} F_{(3)}^2\right)\nn\\
&&~~~~~~+\frac{1}{2\cdot 3!}\e^{-\phi} 
\left(3H_{MAB} {H_N}^{AB}
-\frac{1}{4}g_{MN} H_{(3)}^2\right)
+\frac{1}{4\cdot 4!}F_{MABCD} {F_N}^{ABCD}\,,
   \label{B:Einstein:Eq}\\
&&d\ast d\phi=\frac{1}{2\cdot 3!}
\e^{\phi}F_{(3)}\wedge \ast F_{(3)}
-\frac{1}{2\cdot 3!}\e^{-\phi}H_{(3)}\wedge \ast H_{(3)}\,,
   \label{B:scalar:Eq}\\
&&d\left(\e^{\phi}\ast F_{(3)}\right)+H_{(3)}\wedge\ast F_{(5)}=0\,,
   \label{B:F:Eq}\\
&&d\left(\e^{-\phi}\ast H_{(3)}\right)
 +\ast F_{(5)}\wedge F_{(3)}=0\,,
   \label{B:H:Eq}\\
&&F_{(5)}=\pm\ast F_{(5)}\,,
   \label{B:5:Eq}
}
where $\ast$ is the Hodge operator in the ten-dimensional spacetime, and 
we define 
\Eqrsubl{B:strength:Eq}{
F_{(3)}&=&dC_{(2)}\,,\\
H_{(3)}&=&dB_{(2)}\,,\\
F_{(5)}&=&dC_{(4)}+\frac{1}{2}\left(C_{(2)}\wedge H_{(3)}-
B_{(2)}\wedge F_{(3)}\right)\,.
}
Here $B_{(2)}$, $C_{(2)}$, and $C_{(4)}$ are the NS 2-form, 
RR 2-form, and RR 4-form, respectively.  
For the cosmological D3-D5-NS5 system, 
we look for solutions whose spacetime metrics have the form
\Eqr{
ds^2&=& h_3^{-1/2}(x, y, z)\left[h_5(z)h_{\rm NS}(y)\right]^{-1/4}
\left[q_{\mu\nu}(\Xsp)dx^{\mu}dx^{\nu}
   +h_3(x, y, z)h_{\rm NS}(y)\gamma_{ij}(\Ysp)dy^idy^j\right.\nn\\
  &&\left. +h_5(z)h_{\rm NS}(y)dv^2+h_3(x, y, z)h_5(y)
  u_{ab}(\Zsp)dz^adz^b\right]\,,
   \label{B:metric:Eq}
}
where $q_{\mu\nu}(\Xsp)$, $\gamma_{ij}(\Ysp)$, and $u_{ab}(\Zsp)$ are 
the metrics of 
the three-dimensional spacetime X, of the three-dimensional space Y, 
and of the three-dimensional space Z, which depend only
on the three-dimensional coordinates $x^{\mu}$, on the three-dimensional 
ones $y^i$, and on the three-dimensional ones $z^a$, respectively.

Concerning the other fields, we adopt the following assumptions:
\Eqrsubl{B:fields:Eq}{
C_{(2)}&=&\omega_{(2)}\,,\\
B_{(2)}&=&\tilde{\omega}_{(2)}\,,\\
\e^{2\phi}&=&h_5^{-1}h_{\rm NS}\,,\\
C_{(4)}&=&\omega_{(4)}\pm h_3^{-1}\Omega(\Xsp)\wedge dv\,,
} 
where $C_{(4)}$, $B_{(2)}$, and $C_{(2)}$ are gauge potentials 
for D3-, NS5-, and D5-brane, 
and $\Omega(\Xsp)$ is the volume 3-form. The forms $\omega_{(2)}$, 
$\tilde{\omega}_{(2)}$, and $\omega_{(4)}$ satisfy the relations 
\Eqrsubl{B:forms:Eq}{
&&d\omega_{(2)}=\pd_ah_5\,\ast_{\Zsp}\left(dz^a\right)\wedge dv\,,~~~~~~~
d\tilde{\omega}_{(2)}=\pd_ih_{\rm NS}\,\ast_{\Ysp}\left(dy^i\right)
\wedge dv\,,\\
&&d\omega_{(4)}=\pm h_5^{-1}\pd_ih_3\ast_{\Ysp}\left(dy^i\right)
\pm h_{\rm NS}^{-1}\pd_ah_3\ast_{\Zsp}\left(dz^a\right)\nn\\
&&~~~~~~~~~
+\frac{1}{2}\left[\pd_ah_5\,\tilde{\omega}_{(2)}\wedge
\ast_{\Zsp}\left(dz^a\right)
-\pd_ih_{\rm NS}\,\omega_{(2)}\wedge\ast_{\Ysp}\left(dy^i\right)
\right]\wedge dv\,,
}
where $\ast_{\Ysp}$ and $\ast_{\Zsp}$ are the Hodge operator in the 
Y and Z space. 

\begin{table}[h]
\caption{\baselineskip 14pt
Dynamical D3-D5-NS5 brane system. 
Here $\circ$ denotes the worldvolume coordinate.}
\label{B}
{\scriptsize
\begin{center}
\begin{tabular}{|c|c|c|c|c|c|c|c|c|c|c|}
\hline
&0&1&2&3&4&5&6&7&8&9
\\
\hline
D3 & $\circ$ & $\circ$ & $\circ$ & & &   & $\circ$ &&&
\\
\cline{2-11}
D5 & $\circ$ &$\circ$ & $\circ$ & $\circ$ &$\circ$ & $\circ$ & 
 & && 
\\
\cline{2-11}
NS5 & $\circ$ &$\circ$ &  $\circ$ &&&&& $\circ$ 
& $\circ$ & $\circ$  
\\
\cline{2-11}
$x^N$ & $t$ & $x^1$ & $x^2$ & $y^1$ & $y^2$ & $y^3$ & $v$ & $z^1$
& $z^2$ & $z^3$
\\
\hline
\end{tabular}
\end{center}
}
\label{tableIIB}
\end{table}

\subsubsection{The scalar and gauge field equations}
Under the ansatz given above, we first simplify the field 
equations other than the Einstein equations. 
The assumption (\ref{B:fields:Eq}) implies that $F_{(3)}$ and 
$H_{(3)}$ are closed forms depending on the coordinates of X, Y, and Z. 
In terms of the ansatz (\ref{B:fields:Eq}) for fields, the field equations
for 3-forms are automatically satisfied. 
The Bianchi identity of the 3-forms gives
\Eqrsubl{B:h3:Eq}{
&&h_3(x, y, z)=K_0(x)+K_1(y, z)\,;
~~~h_5\lap_{\Ysp}K_1+h_{\rm NS}\lap_{\Zsp}K_1=0\,,\\
&&\lap_{\rm Y}h_{\rm NS}=0,~~~\lap_{\Zsp}h_5=0\,,
   \label{B:3f:Eq}
 }
where $\lap_{\Ysp}$, $\lap_{\Zsp}$ are the Laplace 
operators on the space of Y and Z, respectively. 
 
From the self-duality requirement for $F_{(5)}$, (\ref{B:5:Eq}), 
it follows that $F_{(5)}$ can be expressed in terms of functions  
$h_3$, $h_5$, and $h_{\rm NS5}$ on ten-dimensional spacetime satisfying 
\eqref{B:3f:Eq}.  

Finally, utilizing the assumption of fields (\ref{B:fields:Eq}) and 
metric (\ref{B:metric:Eq}), 
the scalar field equation can be replaced by 
\Eq{
h_5h_{\rm NS}^{-1}\lap_{\Ysp}h_{\rm NS}
-h_5^{-1}h_{\rm NS}\lap_{\Zsp}h_5=0\,.
}

\subsubsection{The Einstein equations}
\label{sec:Einstein}

Under our ansatz, the Einstein equations become
\Eqrsubl{B:cEinstein:Eq}{
&&\hspace{-0.5cm}R_{\mu\nu}(\Xsp)-h_3^{-1}D_{\mu}D_{\nu}h_3
+\frac{1}{8}h_3^{-1}q_{\mu\nu}(\Xsp)
\left[2\lap_{\Xsp}h_3+h_{\rm NS}^{-1}\left(2h_3^{-1}\lap_{\Ysp}h_3
+h_{\rm NS}^{-1}\lap_{\Ysp}h_{\rm NS}^{-1}\right)\right.\nn\\
&&\left.~~~~~~~~+h_5^{-1}\left(2h_3^{-1}\lap_{\Zsp}h_3
+h_5^{-1}\lap_{\Zsp}h_5^{-1}\right)\right]=0\,,
   \label{B:cEinstein-mn:Eq}\\
&&\hspace{-0.5cm}h_3^{-1}\pd_{\mu}\pd_ih_3=0\,,
   \label{B:cEinstein-mi:Eq}\\
&&\hspace{-0.5cm}h_3^{-1}\pd_{\mu}\pd_ah_3=0\,,
   \label{B:cEinstein-ma:Eq}\\
&&\hspace{-0.5cm}R_{ij}(\Ysp)-\frac{1}{8}\gamma_{ij}(\Ysp)
\left[2h_{\rm NS}\lap_{\Xsp}h_3+\left(2h_3^{-1}\lap_{\Ysp}h_3
-3h_{\rm NS}^{-1}\lap_{\Ysp}h_{\rm NS}^{-1}\right)\right.\nn\\
&&\left.~~~~~~~~+h_5^{-1}h_{\rm NS}\left(2h_3^{-1}\lap_{\Zsp}h_3
-h_5^{-1}\lap_{\Zsp}h_5^{-1}\right)\right]=0\,,
   \label{B:cEinstein-ij:Eq}\\
&&\hspace{-0.5cm}2h_5h_{\rm NS}\lap_{\Xsp}h_3+h_5\left(2h_3^{-1}\lap_{\Ysp}h_3
-3h_{\rm NS}^{-1}\lap_{\Ysp}h_{\rm NS}^{-1}\right)
+h_{\rm NS}\left(2h_3^{-1}\lap_{\Zsp}h_3
-3h_5^{-1}\lap_{\Zsp}h_5^{-1}\right)=0\,,
   \label{B:cEinstein-vv:Eq}\\
&&\hspace{-0.5cm}R_{ab}(\Zsp)-\frac{1}{8}h_5u_{ab}(\Zsp)
\left[2h_3^{-1}\lap_{\Xsp}h_3+h_{\rm NS}^{-1}\left(2h_3^{-1}\lap_{\Ysp}h_3
-h_{\rm NS}^{-1}\lap_{\Ysp}h_{\rm NS}^{-1}\right)\right.\nn\\
&&\left.~~~~~~~~+\left(2h_3^{-1}\lap_{\Zsp}h_3
+3h_5^{-1}\lap_{\Zsp}h_5^{-1}\right)\right]=0\,,
   \label{B:cEinstein-ab:Eq}
 }
where $D_{\mu}$ is the covariant derivative with respect to the metric 
$q_{\mu\nu}(\Xsp)$, and $\lap_{\Ysp}$, $\lap_{\Zsp}$ are the
Laplace operators on the space of Y, Z, 
and $R_{\mu\nu}(\Xsp)$, $R_{ij}(\Ysp)$, and $R_{ab}(\Zsp)$ are the Ricci
tensors for the metrics $q_{\mu\nu}(\Xsp)$, $\gamma_{ij}(\Ysp)$, 
and $u_{ab}(\Zsp)$, respectively.

Equations (\ref{B:cEinstein-mi:Eq}) and (\ref{B:cEinstein-ma:Eq}) 
are consistent with \eqref{B:3f:Eq}.  Taking account of the form 
of $h_3$, the Einstein equations reduce to
\Eqrsubl{B:cEinstein2:Eq}{
&&\hspace{-1cm}R_{\mu\nu}(\Xsp)-h_3^{-1}D_{\mu}D_{\nu}K_0
+\frac{1}{8}h_3^{-1}q_{\mu\nu}(\Xsp)
\left[2\lap_{\Xsp}K_0+h_{\rm NS}^{-1}\left(2h_3^{-1}\lap_{\Ysp}K_1
+h_{\rm NS}^{-1}\lap_{\Ysp}h_{\rm NS}^{-1}\right)\right.\nn\\
&&\left.~~~~~~~~+h_5^{-1}\left(2h_3^{-1}\lap_{\Zsp}K_1
+h_5^{-1}\lap_{\Zsp}h_5^{-1}\right)\right]=0\,,
   \label{B:cEinstein-mn2:Eq}\\
&&\hspace{-1cm}R_{ij}(\Ysp)-\frac{1}{8}\gamma_{ij}(\Ysp)
\left[2h_{\rm NS}\lap_{\Xsp}K_0+\left(2h_3^{-1}\lap_{\Ysp}K_1
-3h_{\rm NS}^{-1}\lap_{\Ysp}h_{\rm NS}^{-1}\right)\right.\nn\\
&&\left.~~~~~~~~+h_5^{-1}h_{\rm NS}\left(2h_3^{-1}\lap_{\Zsp}K_1
-h_5^{-1}\lap_{\Zsp}h_5^{-1}\right)\right]=0\,,
   \label{B:cEinstein-ij2:Eq}\\
&&\hspace{-1cm}
2h_5h_{\rm NS}\lap_{\Xsp}K_0+h_5\left(2h_3^{-1}\lap_{\Ysp}K_1
-3h_{\rm NS}^{-1}\lap_{\Ysp}h_{\rm NS}^{-1}\right)
+h_{\rm NS}\left(2h_3^{-1}\lap_{\Zsp}K_1
-3h_5^{-1}\lap_{\Zsp}h_5^{-1}\right)=0\,,
   \label{B:cEinstein-vv2:Eq}\\
&&\hspace{-1cm}R_{ab}(\Zsp)-\frac{1}{8}h_5u_{ab}(\Zsp)
\left[2h_3^{-1}\lap_{\Xsp}K_0+h_{\rm NS}^{-1}\left(2h_3^{-1}\lap_{\Ysp}K_1
-h_{\rm NS}^{-1}\lap_{\Ysp}h_{\rm NS}^{-1}\right)\right.\nn\\
&&\left.~~~~~~~~+\left(2h_3^{-1}\lap_{\Zsp}K_1
+3h_5^{-1}\lap_{\Zsp}h_5^{-1}\right)\right]=0\,.
   \label{B:cEinstein-ab2:Eq}
 }
These equations are equivalent to \eqref{B:3f:Eq} with 
\Eq{
R_{\mu\nu}(\Xsp)=0\,,~~~~R_{ij}(\Ysp)=0\,,~~~~R_{ab}(\Zsp)=0\,,~~~~
D_{\mu}D_{\nu}K_0=0\,.
   \label{B:Ricci:Eq}
}
The result (\ref{B:Ricci:Eq}) strongly restricts the ten-dimensional metric
because X, Y, Z are required to be Ricci flat. 

Let us consider the case in which the ten-dimensional metric is 
given by 
\Eq{ 
q_{\mu\nu}=\eta_{\mu\nu}\,,~~~~\gamma_{ij}=\delta_{ij}\,,
~~~~u_{ab}=\delta_{ab}\,,
 \label{B:particular:Eq}
 }
where $\eta_{\mu\nu}$ is the three-dimensional 
Minkowski metric, and $\delta_{ij}$ are  
the three-dimensional Euclidean metric, respectively.
If the D3-brane is located at the origin of the Y and Z spaces, 
the solution for field equations is given by 
\Eq{ 
h_3(x, y, z)=c_{\mu}x^{\mu}+\tilde{c}+\frac{M}
{\left(4M_{\rm NS}y+4M_5z\right)^3},~~~~
h_5(z)=\frac{M_5}{z},~~~~h_{\rm NS}(y)=\frac{M_{\rm NS}}{y},
  \label{B:h:Eq}
}
where $c_{\mu}$, $\tilde{c}$, $M_5$, $M_{\rm NS}$, $M$ 
are constant parameters, and $y^2=\delta_{ij}y^iy^j$, 
$z^2=\delta_{ab}z^az^b$. 

\subsubsection{Warped AdS${}_4$ spacetime}
In Sec.~\ref{sec:Einstein}, we have constructed 
the time-dependent 
D3-D5-NS5 brane solution in ten-dimensional IIB theory. 
In this subsubsection, we will present that the solution gives AdS 
in warped ten-dimensional spacetime at the near-horizon limit of the 
intersecting branes below. 

Now we make a coordinate transformation,
\Eq{
y^i=\frac{\xi^i}{4M_{\rm NS}}r\cos^2\alpha\,,~~~~
z^a=\frac{\zeta^a}{4M_5}r\sin^2\alpha\,,
}
where $\xi^i$, $\zeta^a$ coordinates satisfy 
\Eq{
\xi^i\xi_i=1\,,~~~~\zeta^a\zeta_a=1\,,~~~~
d\xi^id\xi_i=d\tilde{\Omega}_{(2)}^2\,,~~~~d\zeta^ad\zeta_a=d\Omega_{(2)}^2\,. 
}
Here, $d\Omega_{(2)}^2$ and $d\tilde{\Omega}_{(2)}^2$
 are the line elements of 2-sphere. 
Thus, in the near-horizon limit $r\rightarrow 0$, where 
the function $K_0$ in $h_3$ can be dropped, 
the metric of the D3-D5-NS5 system becomes
\Eqr{
ds^2&=&\left(\frac{M\sin\alpha\cos\alpha}{4M_5M_{\rm NS}}\right)^{1/2}
\left[\frac{r^4}{M}\eta_{\mu\nu}+\frac{dr^2}{r^2}
+\left(\frac{4M_5M_{\rm NS}}{\sin\alpha\cos\alpha}\right)^2\frac{dv^2}{M}
\right.\nn\\
&&\left.+d\alpha^2+\frac{1}{4}\cos^2\alpha d\tilde{\Omega}_{(2)}^2
+\frac{1}{4}\sin^2\alpha d\Omega_{(2)}^2\right].
}
Hence, the ten-dimensional metric of the D3-D5-NS5 system gives 
a warped product of AdS${}_4$ with an internal 6-space.

\subsection{Dynamical D2-D6-KK monopole solution}
In this subsection, we apply the T-duality formulation developed in string theory to the D3-D5-NS5 solution in order to construct the 
dynamical D2-D6-KK monopole system in IIA theory.
We start from the dynamical D3-D5-NS5 solution in the string frame: 
\Eqrsubl{D2D6K:D3D5N:Eq}{
ds^2&=& h_3^{-1/2}(x, y, z)\left[h_5(z)h_{\rm NS}(y)\right]^{-1/4}
\left[\eta_{\mu\nu}(\Xsp)dx^{\mu}dx^{\nu}
   +h_3(x, y, z)h_{\rm NS}(y)\delta_{ij}(\Ysp)dy^idy^j\right.\nn\\
  &&\left. +h_5(z)h_{\rm NS}(y)dv^2+h_3(x, y, z)h_5(y)
  \delta_{ab}(\Zsp)dz^adz^b\right]\,,
   \label{D2D6K:metric:Eq}\\
C_{(2)}&=&\omega_{(2)}\,,\\
B_{(2)}&=&\tilde{\omega}_{(2)}\,,\\
\e^{2\phi}&=&h_5^{-1}h_{\rm NS}\,,\\
C_{(4)}&=&\omega_{(4)}\pm h_3^{-1}\Omega(\Xsp)\wedge dv\,,
}
where $\eta_{\mu\nu}$ is the four-dimensional 
Minkowski metric, and $\delta_{ij}$ are the three-dimensional 
Euclidean metric, respectively.

The ten-dimensional T-duality map from the type IIB theory to type 
IIA theory is given by
\cite{Bergshoeff:1994cb, Bergshoeff:1995as, Breckenridge:1996tt, Costa:1996zd}
\Eqr{
&&g^{(\rm A)}_{xx}=\frac{1}{g^{(\rm B)}_{xx}}\,,~~~~
g^{(\rm A)}_{\mu\nu}=g^{(\rm B)}_{\mu\nu}
-\frac{g^{(\rm B)}_{x\mu}g^{(\rm B)}_{x\nu}
-B^{(\rm B)}_{x\mu}B^{(\rm B)}_{x\nu}}
{g^{(\rm B)}_{xx}}\,,~~~~
g^{(\rm A)}_{x\mu}=-\frac{B^{(\rm B)}_{x\mu}}{g^{(\rm B)}_{xx}}\,,\nn\\
&&\e^{2\phi_{(\rm A)}}=\frac{\e^{2\phi_{(\rm B)}}}{g^{(\rm B)}_{xx}}\,,~~~~
C_{\mu}=C_{x\mu}+C_{(0)}B_{x\mu}^{(\rm B)}\,,~~~~C_x=-C_{(0)}\,,\nn\\
&&B^{(\rm A)}_{\mu\nu}=B^{(\rm B)}_{\mu\nu}
+2\frac{B^{(\rm B)}_{x[\mu}\,g^{(\rm B)}_{\nu]x}}{g^{(\rm B)}_{xx}}\,,~~~~
B^{(\rm A)}_{x\mu}=-\frac{g^{(\rm B)}_{x\mu}}{g^{(\rm B)}_{xx}}\,,~~~~
C_{x\mu\nu}=C_{\mu\nu}
+2\frac{C_{x[\mu}\,g^{(\rm B)}_{\nu]x}}{g^{(\rm B)}_{xx}}\,,\nn\\
&&C_{\mu\nu\rho}=C_{\mu\nu\rho x}+\frac{3}{2}\left(
C_{x[\mu}\,B^{(\rm B)}_{\nu\rho]}
-B^{(\rm B)}_{x[\mu}\,C_{\nu\rho]}
-4\frac{B^{(\rm B)}_{x[\mu}\,C_{|x|\nu}g^{(\rm B)}_{\rho x}}
{g^{(\rm B)}_{xx}}\right)\,,
   \label{D2D6K:duality:Eq}
}
where $x$ is the coordinate to which the T dualization is applied in the 
D3-D5-NS5 system, and 
$\mu$, $\nu$, and $\rho$ denote the coordinates other than $x$. 

In terms of the T-duality map (\ref{D2D6K:duality:Eq}), 
the D3-D5-NS5 solution (\ref{D2D6K:D3D5N:Eq}) becomes 
\Eqrsubl{D2D6K:solution:Eq}{
ds^2_{(\rm A)}&=& 
     h_2^{1/2}(x, y, z)h_6^{1/2}(z)\left[h_2^{-1}(x, y, z)
     h_6^{-1}(z)\eta_{\mu\nu}(\Xsp)dx^{\mu}dx^{\nu}
     +h_6^{-1}(z)\delta_{ij}(\Ysp)dy^idy^j\right.\nn\\
     & &\left.+h_{\rm K}(z)\delta_{ab}(\Zsp)dz^adz^b
     +h_{\rm K}^{-1}(z)\left(dv+A_adz^a\right)^2\right],
   \label{D2D6K:metric2:Eq}\\
C_{(3)}&=&\pm h_2^{-1}dt\wedge dx^1\wedge dx^2,\\
C_{(1)}&=&\omega_{(1)}\,,\\
\e^{2\phi_{(\rm A)}}&=&h_2^{1/2}h_6^{-3/2}\,,
}
where $A_a$ is the 1-form, and 
$ds^2_{{(\rm A)}}$ is the ten-dimensional metric in the Einstein 
frame, $C_{(3)}$ 
and $C_{(1)}$ are gauge potentials for D2- and D4-branes, 
and $\omega_{(1)}$ satisfies 
\Eq{
d\omega_{(1)}=\pd_{a}h_6\ast_{\Zsp}dz^a\,.
}
Here $\ast_{\Zsp}$ is the Hodge operator in the Z space and 
$h_2$, $h_6$, and $h_{\rm K}$ can be written by
\Eq{ 
h_2(x, y, z)=h_3(x, y, z),~~~~~h_6(z)=h_5(z),~~~~~
h_{\rm K}(y)=h_{\rm NS}(y)\,.
  \label{D2D6K:h:Eq}
  }

\subsection{Dynamical D2-D2-D6 brane solution}
\label{subsec:D226}
Now we construct the dynamical D2-D2-D6 brane solutions in 
ten-dimensional type IIA string theory. 

Let us consider the ten-dimensional spacetime with the metric
\Eqr{
ds^2&=& h_2^{3/8}(t, x, \ell, z)k_2^{3/8}(\ell, z)h_6^{7/8}(z)
\left[-h^{-1}_2(t, x, \ell, z)k_2^{-1}(\ell, z)h_6^{-1}(z)dt^2\right.\nn\\
&&+k_2^{-1}(\ell, z)h_6^{-1}(z)q_{\mu\nu}(\Xsp)dx^{\mu}dx^{\nu}
+h^{-1}_2(t, x, \ell, z)h_6^{-1}(z)\gamma_{ij}(\Ysp)dy^idy^j\nn\\
&&\left.+h_6^{-1}(z)w_{mn}({\rm L})d\ell^{m}d\ell^{n}+u_{ab}(\Zsp)dz^adz^b
\right],
   \label{D2D2D6:metric:Eq}
}
where $q_{\mu\nu}(\Xsp)$ is a two-dimensional metric which
depends only on the two-dimensional coordinates $x^{\mu}$, 
$\gamma_{ij}(\Ysp)$ is a two-dimensional metric which
depends only on the two-dimensional coordinates $y^i$, 
$w_{mn}({\rm L})$ is a two-dimensional metric which
depends only on the two-dimensional coordinates $\ell^m$, 
and finally $u_{ab}(\Zsp)$ is a three-dimensional metric which
depends only on the three-dimensional coordinates $z^a$.  
We further require that the scalar field $\phi$ 
and the form fields satisfy the following conditions:
\Eqrsubl{D2D2D6:fields:Eq}{
\e^{\phi}&=&\left(h_2k_2\right)^{1/4}h_6^{-3/4},\\
F_{(2)}&=&\e^{-3\phi/2}\ast\left[d\left(h_6^{-1}\right)\wedge
  \Omega(\Xsp)\wedge\Omega(\Ysp)\wedge\Omega({\rm L})\right]\,,\\
F_{(4)}&=&d\left(h_2^{-1}\right)\wedge dt\wedge\Omega(\Ysp)
+d\left(k_2^{-1}\right)\wedge dt\wedge\Omega(\Xsp)\,,
      }
where $\Omega(\Xsp)$, $\Omega(\Ysp)$, and $\Omega({\rm L})$ denote 
the volume 2-forms,
\Eqrsubl{D2D2D6:volume:Eq}{
\Omega(\Xsp)&=&\sqrt{q}\,dx^1\wedge dx^2,\\
\Omega(\Ysp)&=&\sqrt{\gamma}\,dy^1\wedge dy^2,\\
\Omega({\rm L})&=&\sqrt{w}\,d\ell^1\wedge d\ell^2,
}
respectively.

\begin{table}[h]
\caption{\baselineskip 14pt
Dynamical D2-D2-D6 brane system in the metric \eqref{D2D2D6:metric:Eq}. 
Here $\circ$ denotes the worldvolume coordinate. }
\label{D2D2D6}
{\scriptsize
\begin{center}
\begin{tabular}{|c|c|c|c|c|c|c|c|c|c|c|}
\hline
&0&1&2&3&4&5&6&7&8&9\\
\hline
D2 & $\circ$ & $\circ$ & $\circ$ &  &  &  &  
&&& \\
\cline{2-11}
D2 & $\circ$ &&&  $\circ$  &   $\circ$ &  &  
&&& \\
\cline{2-11}
D6 & $\circ$ & $\circ$ & $\circ$ & $\circ$ & $\circ$ &  
 $\circ$ & $\circ$ &  &  & \\
\cline{2-11}
$x^N$ & $t$ & $x^1$ & $x^2$ & $y^1$ & $y^2$ & $\ell^1$ & $\ell^2$
& $z^1$ & $z^2$ & $z^3$ \\
\hline
\end{tabular}
\end{center}
}
\label{table_226}
\end{table}

The field equations reduce to  
\Eqrsubl{D2D2D6:Einstein equations:Eq}{
&&R_{\mu\nu}(\Xsp)=0,~~~~R_{ij}(\Ysp)=0,~~~~R_{mn}({\rm L})=0,
~~~~R_{ab}(\Zsp)=0,
   \label{D2D2D6:Ricci:Eq}\\ 
&&h_2(t, x, \ell, z)=K_0(t)+K_1(x, \ell, z)\,,~~~
k_2=k_2(\ell, z)\,,~~~~h_6=h_6(z)\,,\\
&&\pd_t^2K_0=0\,,~~~\lap_{\Zsp}K_1
+h_6\left(\lap_{\rm L}K_1+k_2\lap_{\Xsp}K_1\right)=0, \\
&&\lap_{\Zsp}k_2+h_6\lap_{\rm L}k_2=0,
~~~\lap_{\Zsp}h_6=0\,,
   \label{D2D2D6:warp factor h:Eq}
 }
where $\triangle_{\Xsp}$, $\triangle_{\rm L}$, $\lap_{\Zsp}$ are
the Laplace operators on 
$\Xsp$, L,  $\Zsp$ space, and $R_{\mu\nu}(\Xsp)$, 
$R_{ij}(\Ysp)$, $R_{mn}({\rm L})$, and $R_{ab}(\Zsp)$
are the Ricci tensors constructed from the metrics $q_{\mu\nu}(\Xsp)$, 
$\gamma_{ij}(\Ysp)$, $w_{mn}({\rm L})$, $u_{ab}(\Zsp)$, 
respectively.   
As a special example, we consider the case
\Eq{ 
q_{\mu\nu}=\delta_{\mu\nu},~~~~\gamma_{ij}=\delta_{ij},
~~~~u_{ab}=\delta_{ab}\,,~~~~w_{mn}=\delta_{mn},
 \label{D2D2D6:s-metric:Eq}
 }
where $\delta_{\mu\nu}$, $\delta_{ij}$, $\delta_{mn}$, and $\delta_{ab}$
 are the two-, two-, two-, three-dimensional Euclidean metric.  
In this case, the solution of field equations can be obtained
explicitly as
\Eqrsubl{D2D2D6:solutions1:Eq}{
h_2(t, x, \ell, z)&=&\bar{c}t+\tilde{c}
+M\left[|x^{\mu}-x^{\mu}_0|^2
+\frac{M_2}{|\ell^{m}-\ell^{n}_0|^2+4M_6z}\right],
 \label{D2D2D6:solution-r2:Eq}\\
k_2(\ell, z)&=&\frac{M_2}{\left(|\ell^{m}-\ell^{m}_0|^2+4M_6z\right)^2},
 \label{D2D2D6:solution-s2:Eq}\\
h_6(z)&=&\frac{M_6}{z}\,,
 \label{D2D2D6:solution-n:Eq}
}
where $\bar{c}$, $\tilde{c}$, $M$, $M_2$, and $M_6$ are 
constant parameters, and the constants 
$x_0^{\mu}$, $\ell_0^m$ denote the positions of the branes, 
and $z^2=\delta_{ab}z^az^b$.

\subsection{M2-M2-KK monopole system}
\label{sec:M}

Lifting the dynamical D2-D2-D6 brane solution from ten to eleven 
dimensions, we obtain a time-dependent M2-M2-KK monopole system, 
in which the only non-trivial fields are the metric and the 4-form, given by
\Eqr{
&&ds^2=h^{1/3}_2(t, x, \ell, z)k_2^{1/3}(r)
\left[-h^{-1}_2(t, x, \ell, z)k_2^{-1}(\ell, r)dt^2
+k_2^{-1}(\ell, z)\delta_{\mu\nu}dx^{\mu}dx^{\nu}
+\delta_{mn}d\ell^{m}d\ell^{n}\right.\nn\\
&&\left.~~~~~+h^{-1}_2(t, x, \ell, z)\delta_{ij}dy^idy^j
+h_{\rm K}(z)\delta_{ab}dz^adz^b
+h^{-1}_{\rm K}(z)\left(dv+A_adz^a\right)^2
\right],  
 \label{m2m2K:metric:Eq}
}
where $A_a$ denotes the 1-form, and 
$\delta_{\mu\nu}$, $\delta_{mn}$, $\delta_{ij}$, $\delta_{ab}$ 
are the two-, two-, two-, three-dimensional flat metrics, respectively.
The functions $h_2$, $k_2$, and $h_{\rm K}$ are written by 
\Eqrsubl{m2m2K:solutions1:Eq}{
h_2(t, x, \ell, z)&=&\bar{c}t+\tilde{c}
+ M\left[|x^{\mu}-x^{\mu}_0|^2
+\frac{M_2}{|\ell^{m}-\ell^{m}_0|^2+4M_{\rm K}z}\right],
 \label{m2m2K:solution-r2:Eq}\\
k_2(\ell, z)&=&\frac{M_2}
{\left(|\ell^{m}-\ell^{m}_0|^2+4M_{\rm K}z|\right)^2},
 \label{m2m2K:solution-s2:Eq}\\
h_{\rm K}(z)&=&\frac{M_{\rm K}}{z}\,,
 \label{m2m2K:solution-n:Eq}
}
where $\bar{c}$, $\tilde{c}$, $M$, $M_2$, and $M_{\rm K}$ are 
constant parameters, and the constants $x_0^{\mu}$, $\ell_0^m$
denote the positions of the branes, and $z^2=\delta_{ab}z^az^b$.

\begin{table}[h]
\caption{\baselineskip 14pt
Time-dependent M2-M2 KK monopole in the metric \eqref{m2m2K:metric:Eq}. 
Here $\circ$ denotes the worldvolume coordinate and $\bullet$ 
denotes the fiber coordinate of the KK monopole, respectively.}
\label{M2M2K}
{\scriptsize
\begin{center}
\begin{tabular}{|c|c|c|c|c|c|c|c|c|c|c|c|}
\hline
&0&1&2&3&4&5&6&7&8&9&10\\
\hline
M2 & $\circ$ & $\circ$ &  $\circ$ &&  &  &
&  &  &  & \\
\cline{2-12}
M2 & $\circ$ &&&  $\circ$ &  $\circ$ & & & &  &  &\\ 
\cline{2-12}
KK & $\circ$ & $\circ$ &  $\circ$ & $\circ$ & $\circ$ & $\circ$ &
$\circ$ & $\bullet$ &  &  & \\
\cline{2-12}
$x^N$ & $t$ & $x^1$ & $x^2$ & $y^1$ & $y^2$ & $\ell^1$ & $\ell^2$
& $v$ & $z^1$ & $z^2$ & $z^3$\\
\hline
\end{tabular}
\end{center}
}
\label{table_M2M2K}
\end{table}
The M2-M2-KK monopole solution can be obtained by replacing 
the 5-sphere with a lens space in the eleven-dimensional 
M2-M2 brane system.

\Eqr{
&&ds^2=h^{1/3}_2(t, x, \ell, r)k_2^{1/3}(r)
\left[-h^{-1}_2(t, x, \ell, r)k_2^{-1}(\ell, r)dt^2
+k_2^{-1}(\ell, r)\delta_{\mu\nu}dx^{\mu}dx^{\nu}
+\delta_{mn}d\ell^{m}d\ell^{n}\right.\nn\\
&&\left.~~~~~+h^{-1}_2(t, x, \ell, r)\delta_{ij}dy^idy^j
+h_{\rm K}(r)\delta_{ab}dz^adz^b
+h^{-1}_{\rm K}(r)\left(dv+A_adz^a\right)^2
\right],  
 \label{m2m2K:metric2:Eq}
}
where $A_a$ denotes the 1-form, and $d\Omega_{(2)}^2$ is the metric of 
a 2-sphere. 

\section{Dynamical intersecting solutions in massive IIA theory}
\label{sec:ma}
We have so far the examples of intersecting D$p$-brane ($p\le 6$)  
in the ten-dimensional theory and M-brane in eleven-dimensional 
theory that give rise to warped products 
of AdS with certain internal spaces. 
In this section, we consider dynamical solutions for the 
D4-D8 brane system, which appears in the
ten-dimensional massive type IIA supergravity. 
The bosonic action of the D4-D8 brane system in the 
Einstein frame is given by 
\cite{Youm:1999ti, Cvetic:1999xx, Cvetic:1999un, 
Nastase:2003dd, Brandhuber:1999np}
\Eq{
\hspace{-1cm}
S=\frac{1}{2{\kappa}^2}\int \left(R\ast{\bf 1}
 -\frac{1}{2}d\phi \wedge \ast d\phi 
 -\frac{1}{2\cdot 4!}\e^{\phi/2}F_{(4)}\wedge\ast F_{(4)}
 -\frac{1}{2}\e^{5\phi/2}m^2 \ast{\bf 1}\right)\,,
   \label{D4D8:action:Eq}
   }
where ${\kappa}^2$ is the ten-dimensional gravitational constant, 
$\ast$ is the Hodge dual operator in the ten-dimensional spacetime, and
$m$ is a constant parameter, which is the dual of the 10-form field
strength $F_{(10)}$ in the string frame.

The field equations are
\Eqrsubl{D4D8:field eqs:Eq}{
&&\hspace{-1cm}d\ast d\phi
=\frac{1}{4}\left(5m^2\e^{5\phi/2}\ast{\bf 1}+
 \frac{1}{4!}\e^{\phi/2}F_{(4)}\wedge\ast F_{(4)}\right),
  \label{D4D8:scalar eq:Eq}\\
&&\hspace{-1cm}d(\e^{\phi/2}\ast F_{(4)})=0\,,
  \label{D4D8:gauge eq:Eq}\\
&&\hspace{-1cm}R_{MN}=\frac{1}{2}\pd_M\phi \pd_N\phi 
+ \frac{1}{16}m^2\e^{5\phi/2}g_{MN}
+\frac{1}{2\cdot 4!}\e^{\phi/2}\left(4F_{MABC}{F_N}^{ABC}
-\frac{3}{8}g_{MN}F_{(4)}^2\right).
   \label{D4D8:Einstein:Eq}
  }
It was observed that we can construct a solution 
whose spacetime metric has the form \cite{Binetruy:2007tu}
\Eqrsubl{D4D8:fields:Eq}{
ds^2&=&h^{1/12}\left[h_4^{-3/8}q_{\mu\nu}(\Xsp)dx^{\mu}dx^{\nu}
+h_4^{5/8}(d{r}^2+{r}^2d\Omega_{(4)}^2)\right],
   \label{D4D8:metric:Eq}\\
d\Omega_{(4)}^2&=&
d\alpha^2+\cos^2\alpha d\Omega_{(3)}^2\,,
    \label{D4D8:metric s4:Eq}\\
h_4(x, {r})&=&a_{\mu}x^{\mu}+\frac{c_1}{{r}^{10/3}}+c_2\,,~~~~
h({r}, \alpha)=\frac{3}{2}m{r}\sin\alpha\,,
  \label{D4D8:h:Eq}
}
where $q_{\mu\nu}$ is a five-dimensional metric 
depending only on the coordinates $x^\mu$ of $\Xsp$, and 
$d\Omega_{(3)}^2$ and $d\Omega_{(4)}^2$ denote the line elements 
of 
unit 3- 
and 4-spheres,
and $a_{\mu}$, $c_1$, and $c_2$ are constants, respectively. 
As for the scalar field and the 4-form field strength, 
we can adopt the following forms:
\Eqrsubl{D4D8:fields2:Eq}{
\e^{\phi}&=&h^{-5/6}h_4^{-1/4},
  \label{D4D8:scalar:Eq}\\
F_{(4)}&=&\e^{-\phi/2}\ast \left[d\left(h_4^{-1}\right)
\wedge \Omega(\Xsp)\right],
  \label{D4D8:gauge:Eq}
}
where $\Omega(\Xsp)$ is given by
\Eq{
\Omega(\Xsp)=\sqrt{-q}\,dx^0\wedge dx^1\wedge dx^2 \wedge dx^3\wedge dx^4.
}

If we further define a new coordinate $U$ by ${r}^2=U^3$.
 From Eq.~(\ref{D4D8:h:Eq}),
we see that $h_4$ is a linear function of $x^\mu$. Hence,
keeping the values of these coordinates finite, 
the metric in the limit $U\rightarrow 0$ becomes 
\Eq{
ds^2=c_1^{1/8}\left(\frac{3}{2}m\sin\alpha\right)^{1/12}
\left[c_1^{-1/2}U^2q_{\mu\nu}dx^{\mu}dx^{\nu}
+c_1^{1/2}\left(\frac{9dU^2}{4U^2}+d\Omega_{(4)}^2\right)\right],
   \label{D4D8:metric2:Eq}
} 
while the dilaton is given by 
\Eq{
\e^{\phi}=c_1^{-1/4}\left(\frac{3}{2}m\sin\alpha\right)^{-5/6}.
}
This is a static metric.
In particular, in the case $q_{\mu\nu}$ is a five-dimensional 
Minkowski metric $\eta_{\mu\nu}$, the above ten-dimensional 
metric becomes a warped ${\rm AdS}_6\times {\rm S}^4$ space 
\cite{Brandhuber:1999np, Cvetic:1999un, Behrndt:1999mk, Nunez:2001pt, 
Nastase:2003dd, Binetruy:2007tu}. 
A new supersymmetric solution in the massive IIA theory and 
its holographic dual were found 
recently \cite{Itsios:2012dc, Bergman:2012kr}, 
which can be uplifted to the type 
IIB theory using the non-Abelian T-duality transformation.

It is possible to introduce the KK monopole in the D4-D8 brane system. 
If we add the KK monopole to the D4-D8 brane, in the near-horizon limit 
the metric of S${}^3$ in (\ref{D4D8:metric s4:Eq}) is replaced by 
the lens space:
\Eq{
d\Omega_{(3)}^2=\frac{1}{4}\left[d\Omega_{(2)}^2+\left(dv+\omega\right)^2
\right],
    \label{D4D8:3Dmetric:Eq}
}
where $d\omega=\Omega_{(2)}$ is the volume form of a unit 2-sphere, and 
$d\Omega_{(2)}^2$ is the metric of 2-sphere. The metric 
(\ref{D4D8:3Dmetric:Eq}) is viewed as a U(1) bundle over S${}^2$. 
The ten-dimensional metric in the near-horizon limit is thus written by
\Eqrsubl{D4D8:metrics:Eq}{
ds^2&=&c_1^{1/8}\left(\frac{3}{2}m\sin\alpha\right)^{1/12}
\left[c_1^{-1/2}U^2q_{\mu\nu}dx^{\mu}dx^{\nu}
+c_1^{1/2}\left(\frac{9dU^2}{4U^2}+ds_4^2\right)\right],
   \label{D4D8:metric3:Eq}\\
ds_4^2&=&d\alpha^2+\frac{1}{4}\cos^2\alpha \left[
d\Omega_{(2)}^2+\left(dv+\omega\right)^2\right]\,,
}
where $ds_4^2$ is the line element of four-dimensional space.

If we perform the T-duality on the ten-dimensional metric 
(\ref{D4D8:metric3:Eq}), the solution can be viewed as the 
D5-D7-NS5 brane system:
\Eqrsubl{D5D7N:fields:Eq}{
ds^2&=&\left(h_5h_{\rm NS}\right)^{3/4}h_7
\left[\left(h_5h_7h_{\rm NS}\right)^{-1}\eta_{\mu\nu}(\Xsp)dx^{\mu}dx^{\nu}
+h_7^{-1}\delta_{ij}(\Ysp)dy^idy^j\right.\nn\\
&&\left.+h_5^{-1}dv^2+h_{\rm NS}^{-1}dz^2\right],
   \label{D5D7N:metric:Eq}\\
h_5(t, y, z)&=&c\,t+\tilde{c}
+\frac{M_5}{\left[4M_{\rm NS}|y^i-y^i_0|
+\frac{4M_7}{9}z^{3}\right]^{5/3}},
    \label{D5D7N:h:Eq}\\
h_7(z)&=&M_7\,z,~~~~
h_{\rm NS}(y)=\frac{M_{\rm NS}}{|y^i-y^i_0|},
 \label{D5D7N:h2:Eq}
}
where $\eta_{\mu\nu}(\Xsp)$ is a five-dimensional Minkowski metric, 
$\delta_{ij}(\Ysp)$ is the three-dimensional Euclidean metric, 
$M_5$, $M_7$, $M_{\rm NS}$ are constants, 
the constant $y^i_0$ represents the position of 
the NS5-brane, and the ten-dimensional T-duality map from the type 
IIA theory to type IIB theory is given by
\cite{Bergshoeff:1995as, Breckenridge:1996tt}
\Eqr{
&&g^{(\rm B)}_{xx}=\frac{1}{g^{(\rm A)}_{xx}}\,,~~~~
g^{(\rm B)}_{\mu\nu}=g^{(\rm A)}_{\mu\nu}
-\frac{g^{(\rm A)}_{x\mu}g^{(\rm A)}_{x\nu}
-B^{(\rm A)}_{x\mu}B^{(\rm A)}_{x\nu}}
{g^{(\rm A)}_{xx}}\,,~~~~
g^{(\rm B)}_{x\mu}=-\frac{B^{(\rm A)}_{x\mu}}{g^{(\rm A)}_{xx}}\,,\nn\\
&&\e^{2\phi_{(\rm B)}}=\frac{\e^{2\phi_{(\rm A)}}}{g^{(\rm A)}_{xx}}\,,~~~~
B^{(\rm B)}_{\mu\nu}=B^{(\rm A)}_{\mu\nu}
+2\frac{g^{(\rm A)}_{x[\mu}\,B^{(\rm A)}_{\nu]x}}{g^{(\rm A)}_{xx}},~~~~
B^{(\rm B)}_{x\mu}=-\frac{g^{(\rm A)}_{x\mu}}{g^{(\rm A)}_{xx}},\nn\\
&&C_{\mu\nu}=C_{\mu\nu x}-2C_{[\mu}B_{\nu]x}^{(\rm A)}
+2\frac{g^{(\rm A)}_{x[\mu}\,B^{(\rm A)}_{\nu]x}C_x}{g^{(\rm A)}_{xx}}\,,
~~~~C_{x\mu}=C_{\mu}
-\frac{C^{(\rm A)}_{x}\,g^{(\rm A)}_{x\mu}}{g^{(\rm A)}_{xx}}\,,\nn\\
&&C_{\mu\nu\rho x}=C_{\mu\nu\rho}-\frac{3}{2}\left(
C_{[\mu}\,B^{(\rm A)}_{\nu\rho]}
-\frac{g^{(\rm A)}_{x[\mu}B^{(\rm A)}_{\nu\rho]}\,C_x}
{g^{(\rm A)}_{xx}}
+\frac{g^{(\rm A)}_{x[\mu}\,C_{\nu\rho]x}}
{g^{(\rm A)}_{xx}}\right)\,,~~~~C_{(0)}=-C_x\,,
   \label{D4:duality:Eq}
}
where $x$ is the coordinate to which the T dualization is applied, and 
$\mu$, $\nu$, $\rho$ denote the coordinates other than $x$.

The near-horizon structure of the D5-D7-NS5 brane system can be 
obtained via an appropriate coordinate transformation and 
embedding of AdS${}_6$ in the IIB theory, 
\Eqr{
ds^2&=&c_1\cos^{1/2}\alpha\left[U^2\eta_{\mu\nu}dx^{\mu}dx^{\nu}
+c_2\frac{dU^2}{U^2}\right.\nn\\
&&\left.+c_3\left(
d\alpha^2+\frac{1}{4}\cos^2\alpha d\Omega_{(2)}^2\right)
+c_4\sin^{2/3}\alpha\cos^{-2}\alpha dv^2\right]\,,
}
where $c_i~(i=1, \cdots\,, 4)$ are constants. 
\begin{table}[h]
\caption{\baselineskip 14pt
Dynamical D5-D7-NS5 brane system. 
Here $\circ$ denotes the worldvolume coordinate.}
\label{D5D7N}
{\scriptsize
\begin{center}
\begin{tabular}{|c|c|c|c|c|c|c|c|c|c|c|}
\hline
&0&1&2&3&4&5&6&7&8&9
\\
\hline
D5 & $\circ$ & $\circ$ & $\circ$ & $\circ$ & $\circ$ &   &&& $\circ$ &
\\
\cline{2-11}
D7 & $\circ$ &$\circ$ & $\circ$ & $\circ$ &$\circ$ & $\circ$ & 
 $\circ$ & $\circ$ && 
\\
\cline{2-11}
NS5 & $\circ$ &$\circ$ &  $\circ$ & $\circ$ &$\circ$ &&& 
&& $\circ$  
\\
\cline{2-11}
$x^N$ & $t$ & $x^1$ & $x^2$ & $x^3$ & $x^4$ & $y^1$ & $y^2$ & $y^3$
& $v$ & $z$
\\
\hline
\end{tabular}
\end{center}
}
\label{tableD5D7N}
\end{table}

\section{Asymptotic AdS spacetime in dynamical partially localized brane 
system}
\label{sec:asymptotic}
In this section, we discuss the asymptotic geometries in the construction 
of dynamical partially localized brane solutions. We present the 
asymptotic spacetime for intersection involving a two brane system. 
These solutions give AdS in warped ten- or eleven-dimensional spacetime 
at the near-horizon limit of the intersecting branes.

\subsection{Dynamical D3-brane system}
We present the asymptotic geometries in the D3-brane solution in this 
subsection. We also discuss the cosmological F1-D2 
brane solutions in ten-dimensional IIA theory after performing
a T-duality transformation in the time-dependent D3-wave system. 

\subsubsection{AdS${}_5$ spacetime in D3-KK monopole system}
We first consider the dynamical D3-KK monopole 
solution of type IIB supergravity. 
The field equations can be written as 
\Eqrsubl{D3K:equations:Eq}{
&&R_{MN}=\frac{1}{4\cdot 4!} F_{MABCD} {F_N}^{ABCD},
   \label{D3K:Einstein:Eq}\\
&&d\left[\ast F_{(5)}\right]=0,
   \label{D3K:F:Eq}\\
&&F_{(5)}=\pm\ast F_{(5)}.
   \label{D3K:F2:Eq}
}
Now we will briefly summarize the results for the dynamical 
D3-KK monopole solution 
in type IIB supergravity \cite{Cvetic:2000cj}.

\begin{table}[h]
\caption{\baselineskip 14pt
Dynamical D3-brane and KK monopole system in the metric \eqref{D3K:metric:Eq}. 
Here $\circ$ denotes the worldvolume coordinate and $\bullet$ 
denotes the fiber coordinate of the KK monopole, respectively.}
\label{D3KK}
{\scriptsize
\begin{center}
\begin{tabular}{|c|c|c|c|c|c|c|c|c|c|c|}
\hline
&0&1&2&3&4&5&6&7&8&9\\
\hline
D3 & $\circ$ & $\circ$ & $\circ$ & $\circ$ &  &  &
&  &  &  \\
\cline{2-11}
KK & $\circ$ & $\circ$ & $\circ$ & $\circ$ & $\circ$ & $\circ$ 
& $\bullet$ & &  &  \\ 
\cline{2-11}
$x^N$ & $t$ & $x^1$ & $x^2$ & $x^3$ & $y^1$ & $y^2$ & $v$
& $z^1$ & $z^2$ & $z^3$ \\
\hline
\end{tabular}
\end{center}
}
\label{table_A}
\end{table}

The ansatz of the dynamical D3-brane system with a single RR 5-form 
$F_{(5)}$ is given by 
\Eqrsubl{D3K:field:Eq}{
&&ds^2=h^{-1/2}(x, y, r)q_{\mu\nu}(\Xsp)dx^{\mu}dx^{\nu} 
  +h^{1/2}(x, y, r)\left[\gamma_{ij}(\Ysp)dy^idy^j\right.\nn\\
&& \left. ~~~~~+h_{\rm K}(r)u_{ab}(\Zsp)dz^adz^b+h_{\rm K}^{-1}(r)
  \left(dv+A_adz^a\right)^2\right],  
 \label{D3K:metric:Eq}\\
&&u_{ab}(\Zsp)dz^adz^b=dr^2+r^2w_{mn}(\Zsp')dp^mdp^n\,,\\
&&F_{(5)}=\left(1\pm\ast\right)d\left(h^{-1}\right)\wedge\Omega(\Xsp),
  \label{D3K:gauge:Eq}
}
where $A_a$ is the 1-form, and 
$q_{\mu\nu}(\Xsp)$, $\gamma_{ij}(\Ysp)$, $w_{mn}(\Zsp')$, and 
$u_{ab}(\Zsp)$ are the metrics of the 
four-dimensional spacetime X,  of the two-dimensional space Y, 
of the two-dimensional space $\Zsp'$, 
and of the three-dimensional space Z, which depend only
on the four-dimensional coordinates $x^{\mu}$, 
on the two-dimensional ones $y^i$, on the two-dimensional ones $p^m$, 
and on the three-dimensional ones $z^a$, and 
the volume 4-form 
$\Omega(\Xsp)$ is given by
\Eq{
\Omega(\Xsp)=\sqrt{-q}\,dt\wedge dx^1\wedge dx^2\wedge dx^3\,.
    \label{D3K:volume:Eq}
}

Under the ansatz for fields (\ref{D3K:field:Eq}), 
the Einstein equations lead to 
\Eqrsubl{D3K:cEinstein:Eq}{
&&R_{\mu\nu}(\Xsp)-h^{-1}D_{\mu}D_{\nu} h
+\frac{1}{4}q_{\mu\nu}h^{-2}\left[h\lap_{\Xsp}h
+\triangle_{\Ysp}h
+h_{\rm K}^{-1}\left(\pd_r^2h+\frac{2}{r}\pd_rh\right)\right]
=0,
 \label{D3K:cEinstein-mu:Eq}\\
&&h^{-1}\pd_{\mu}\pd_i h=0,
 \label{D3K:cEinstein-mi:Eq}\\
 &&h^{-1}\pd_{\mu}\pd_r h=0,
 \label{D3K:cEinstein-mr:Eq}\\
&&R_{ij}(\Ysp)-\frac{1}{4}\gamma_{ij}h^{-1}\left[h\lap_{\Xsp}h
+\triangle_{\Ysp} h
+h_{\rm K}^{-1}\left(\pd_r^2h+\frac{2}{r}\pd_rh\right)\right]=0
 \label{D3K:cEinstein-ij:Eq},\\
&&h_{\rm K}\lap_{\Xsp}h+h_{\rm K}h^{-1}\left[\lap_{\Ysp}h
+h_{\rm K}^{-1}\left(\pd_r^2h+\frac{2}{r}\pd_rh\right)\right]
+2h_{\rm K}^{-1}\left(\pd_r^2h_{\rm K}+\frac{2}{r}\pd_rh_{\rm K}\right)=0,
 \label{D3K:cEinstein-rr:Eq}\\
&&R_{mn}(\Zsp')-w_{mn}(\Zsp')-\frac{1}{4}h_{\rm K}
\left(r^2w_{mn}+h_{\rm K}^{-2}A_mA_n\right)\left[\lap_{\Xsp}h
\right.\nn\\
&&\left.~~~~~~
+h^{-1}\left\{\lap_{\Ysp}h
+h_{\rm K}^{-1}\left(\pd_r^2h+\frac{2}{r}\pd_rh\right)\right\}
+2h_{\rm K}^{-1}\left(\pd_r^2h+\frac{2}{r}\pd_rh\right)\right]\nn\\
&&~~~~~~-\frac{1}{4}h^{-1}_{\rm K}
\left(r^2w_{mn}-h_{\rm K}^{-2}A_mA_n\right)
\left(\pd_r^2h_{\rm K}+\frac{2}{r}\pd_rh_{\rm K}\right)=0,
\label{D3K:cEinstein-ab:Eq}\\
&&h_{\rm K}^{-1}\left[\lap_{\Xsp}h+h^{-1}\left\{\lap_{\Ysp}h
+h_{\rm K}^{-1}\left(\pd_r^2h+\frac{2}{r}\pd_rh\right)\right\}
-2h_{\rm K}^{-2}\left(\pd_r^2h_{\rm K}
+\frac{2}{r}\pd_rh_{\rm K}\right)\right]=0,
 \label{D3K:cEinstein-vv:Eq}
}
where we used the condition 
$dh_{\rm K}=\ast_{\Zsp}dA_{(1)}$, and 
$D_{\mu}$ is the covariant derivative constructed from the metric 
$q_{\mu\nu}(\Xsp)$, and $\lap_{\Ysp}$ denotes the
Laplace operators on the space of Y, 
and $R_{\mu\nu}(\Xsp)$, $R_{ij}(\Ysp)$ and $R_{mn}(\Zsp')$ are the Ricci
tensors with respect to the metrics $q_{\mu\nu}(\Xsp)$, $\gamma_{ij}(\Ysp)$ 
and $w_{mn}(\Zsp')$, respectively.
From Eqs.~(\ref{D3K:cEinstein-mi:Eq}) and (\ref{D3K:cEinstein-mr:Eq}),  
the function $h$ becomes 
\Eq{
h(x, y, r)= h_0(x)+h_1(y, r).
  \label{D3K:warp:Eq}
}
With this form of $h$, the Einstein equations 
(\ref{D3K:cEinstein:Eq}) are rewritten as
\Eqrsubl{D3K:cEinstein2:Eq}{
&&\hspace{-0.9cm}R_{\mu\nu}(\Xsp)-h^{-1}D_{\mu}D_{\nu} h_0
+\frac{1}{4}q_{\mu\nu}h^{-2}\left[h\lap_{\Xsp}h_0
+\triangle_{\Ysp}h_1
+h_{\rm K}^{-1}\left(\pd_r^2h_1+\frac{2}{r}\pd_rh_1\right)\right]
=0,
 \label{D3K:cEinstein-mu2:Eq}\\
&&\hspace{-0.9cm}R_{ij}(\Ysp)-\frac{1}{4}\gamma_{ij}h^{-1}\left[h\lap_{\Xsp}h_0
+\triangle_{\Ysp} h_1
+h_{\rm K}^{-1}\left(\pd_r^2h_1+\frac{2}{r}\pd_rh_1\right)\right]=0
 \label{D3K:cEinstein-ij2:Eq},\\
&&\hspace{-0.9cm}h_{\rm K}\lap_{\Xsp}h_0+h_{\rm K}h^{-1}\left[\lap_{\Ysp}h_1
+h_{\rm K}^{-1}\left(\pd_r^2h_1+\frac{2}{r}\pd_rh_1\right)\right]
+2h_{\rm K}^{-1}\left(\pd_r^2h_{\rm K}+\frac{2}{r}\pd_rh_{\rm K}\right)=0,
 \label{D3K:cEinstein-rr2:Eq}\\
&&\hspace{-0.9cm}R_{mn}(\Zsp')-w_{mn}(\Zsp')-\frac{1}{4}
\left(h_{\rm K}r^2w_{mn}+h_{\rm K}^{-1}A_mA_n\right)
\left[\lap_{\Xsp}h_0\right.\nn\\
&&\left.~~~~~~+h^{-1}\left\{\lap_{\Ysp}h_1
+h_{\rm K}^{-1}\left(\pd_r^2h_1+\frac{2}{r}\pd_rh_1\right)\right\}
+2h_{\rm K}^{-1}\left(\pd_r^2h_{\rm K}
+\frac{2}{r}\pd_rh_{\rm K}\right)\right]=0\,,
\label{D3K:cEinstein-ab2:Eq}\\
&&\hspace{-0.9cm}
h_{\rm K}^{-1}\left[\lap_{\Xsp}h_0+h^{-1}\left\{\lap_{\Ysp}h_1
+h_{\rm K}^{-1}\left(\pd_r^2h_1+\frac{2}{r}\pd_rh_1\right)\right\}
-2h_{\rm K}^{-1}\left(\pd_r^2h_{\rm K}
+\frac{2}{r}\pd_rh_{\rm K}\right)\right]=0\,.
 \label{D3K:cEinstein-vv2:Eq}
}
Let us next consider the gauge field equations. 
Under the ansatz for fields (\ref{D3K:field:Eq}),
the equation of motion for the gauge field~(\ref{D3K:F:Eq}) gives
\Eqr{
\left[h_{\rm K}\lap_{\Ysp}h_1
+\left(\pd_r^2h_1+\frac{2}{r}\pd_rh_1\right)\right]
\,\Omega(\Ysp)\wedge dr\wedge\Omega(\Zsp')\wedge dv=0,
 }
where we have used (\ref{D3K:warp:Eq}), and 
the volume 2-forms $\Omega(\Ysp)$, 
$\Omega(\Zsp')$ are defined by
\Eqrsubl{D3K:volume2:Eq}{
\Omega(\Ysp)&=&\sqrt{\gamma}\,dy^1\wedge dy^2,\\
\Omega(\Zsp')&=&\sqrt{w}\,dp^1\wedge dp^2.
}
Hence, the gauge field equation reduces to
\Eq{
\lap_{\Ysp}h_1+h_{\rm K}^{-1}\left(\pd_r^2h_1+\frac{2}{r}\pd_rh_1\right)=0.
   \label{D3K:h1:Eq}
}

Let us go back to the Einstein equations (\ref{D3K:cEinstein2:Eq}).
If $F_{(5)}=0$, the function $h_1$ becomes trivial, and  
 the internal space is no longer warped~\cite{Kodama:2005cz}.
On the other hand, for $F_{(5)}\ne 0$, the first term in 
Eq.~(\ref{D3K:cEinstein-mu2:Eq}) depends only on $x$,
whereas the rest on $x$ as well as $y$ and $r$.
Thus, Eqs.~(\ref{D3K:cEinstein2:Eq}) together with (\ref{D3K:h1:Eq}) give
\Eqrsubl{D3K:Einstein2:Eq}{
&&\hspace{-1cm}R_{\mu\nu}(\Xsp)=0,~~~~R_{ij}(\Ysp)=0,
~~~~R_{mn}(\Zsp')=w_{mn}(\Zsp'),
   \label{D3K:Ricci:Eq}\\
&&\hspace{-1cm}
h(x, z)=h_0(x)+h_1(y, r)\,;~~D_{\mu}D_{\nu}h_0=0,~~~~
\lap_{\Ysp}h_1+h_{\rm K}^{-1}\left(\pd_r^2h_1+\frac{2}{r}\pd_rh_1\right)=0,
   \label{D3K:warp2:Eq}\\
&&\hspace{-1cm}\pd_r^2h_{\rm K}+\frac{2}{r}\pd_rh_{\rm K}=0.
   \label{D3K:K:Eq} 
 }

Now we set the ten-dimensional metric: 
\Eq{ 
q_{\mu\nu}=\eta_{\mu\nu}\,,\quad \gamma_{ij}=\delta_{ij}\,,
\quad w_{mn}(\Zsp')dp^mdp^n=d\Omega_{(2)}^2\,,
 \label{D3K:special metric:Eq}
 }
where $\eta_{\mu\nu}$ is the four-dimensional 
Minkowski metric, and $\delta_{ij}$ is 
the two-dimensional Euclidean metric, and $d\Omega_{(2)}^2$ is the metric of 
the two-dimensional sphere. 
In this case, the solution for $h$ and $h_{\rm K}$ can be obtained
explicitly as
\Eqrsubl{D3K:solutions:Eq}{ 
h(x, y, r)&=&c_{\mu}x^{\mu}+\bar{c}
+\sum_{l}\frac{M_l}{\left(|y^i-y^i_l|^2+4M_{\rm K}r\right)^2},
 \label{D3K:solution1:Eq}\\
h_{\rm K}(r)&=&\frac{M_{\rm K}}{r},
 \label{D3K:solution2:Eq}
}
where $c_{\mu}$, $\bar{c}$, $M_l$, and $M_{\rm K}$ are constant parameters,
and the constant $y^i_l$ denotes the position of the brane. 

If the D3-brane is located at the origin of Y space, we have 
\Eq{
h(x, y, r)=c_{\mu}x^{\mu}+\bar{c}+\frac{M}{\left(y^2+4M_{\rm K}r\right)^2},
 \label{D3K:solution3:Eq}
} 
where $y^2=\delta_{ij}y^iy^j$\,. 
If we use the following coordinate transformation, 
\Eqr{
y^1=\zeta \cos\psi\cos\theta,~~~~y^2=\zeta \cos\psi\sin\theta,~~~~
r=\frac{1}{4}M_{\rm K}^{-1}\zeta^2\sin^2\psi\,,
}
the ten-dimensional metric becomes 
\Eqr{
ds^2&=&h^{-1/2}(x, \zeta)q_{\mu\nu}(\Xsp)dx^{\mu}dx^{\nu} 
  +h^{1/2}(x, \zeta)\left[d\zeta^2+\zeta^2\left\{
  d\psi^2+\cos^2\psi d\theta^2\right.\right.\nn\\
  &&\left.\left.+\frac{1}{4}\sin^2\theta\left(d\Omega_{(2)}^2
  +M_{\rm K}^{-2}\left(dv+A_adz^a\right)^2\right)\right\}\right], 
  \label{D3K:metric2:Eq} 
}
where $h$ is given by 
\Eqr{
h(x, \zeta)=c_{\mu}x^{\mu}+\bar{c}+\frac{M}{\zeta^4}\,.
 \label{D3K:solution4:Eq}
}
Then the metric \eqref{D3K:metric2:Eq} in the limit $\zeta\rightarrow 0$ gives 
AdS${}_5$ spacetime with internal space: 
\Eqr{
ds^2&=&M^{-1/2}\zeta^2q_{\mu\nu}(\Xsp)dx^{\mu}dx^{\nu} 
  +M^{1/2}\frac{d\zeta^2}{\zeta^2}+M^{1/2}\left[
  d\psi^2+\cos^2\psi d\theta^2\right.\nn\\
  &&\left.+\frac{1}{4}\sin^2\theta\left\{
  d\Omega_{(2)}^2
  +M^{-2}_{\rm K}\left(dv+A_adz^a\right)^2\right\}\right]. 
  \label{D3K:metric3:Eq} 
}
The solution (\ref{D3K:metric3:Eq}) in the type IIB supergravity 
theory enters in this particular string theory
in much the same way that static spacetime physics 
(with general relativity as the near-horizon limit) 
arises in conventional string theory.

\subsubsection{Warped AdS${}_3$ spacetime in D3-wave system}
\label{sec:BTZ}
We present the time-dependent D3-brane and wave solution. 
The field equations are given by
\Eqrsubl{D3w:field equations:Eq}{
&&\hspace{-1cm}R_{MN}=\frac{1}{4\cdot 4!}F_{M_2\cdots A_5} 
{F_N}^{A_2\cdots A_5}\,,
   \label{D3w:Einstein:Eq}\\
&&\hspace{-1cm}dF_{(5)}=0\,,~~~~~F_{(5)}=\ast F_{(5)}\,,
   \label{D3w:gauge:Eq}
}
where $\ast$ is the Hodge operator in the ten-dimensional spacetime.
We assume that the ten-dimensional metric takes the form
\Eqr{
ds^2&=&h^{1/2}(t, z)\left[-h^{-1}(t, z)h_{\rm W}^{-1}(t, y, z)dt^2
+h^{-1}(t, z)h_{\rm W}(t, y, z)
\left\{\left(h_{\rm W}^{-1}(t, y, z)-1\right)dt+dx\right\}^2\right.\nn\\
&&\left.
+h^{-1}(t, z)\gamma_{ij}(\Ysp)dy^idy^j+u_{ab}(\Zsp)dz^adz^b\right], 
 \label{D3w:metric:Eq}
}
where $\gamma_{ij}(\Ysp)$, $u_{ab}(\Zsp)$ are the metrics depending only on 
$y^i$, $z^a$ coordinates of dimensions two, six, respectively.

We also assume that the gauge field 
strength $F_{\left(5\right)}$ is given by
\Eqr{
F_{\left(5\right)}&=&(1\pm\ast) d\left[h^{-1}(t, z)\wedge dt
\wedge dx\wedge\Omega(\Ysp)\right],
  \label{D3w:ansatz for gauge:Eq}
}
where $\Omega(\Ysp)$ denotes the volume 
2-form 
\Eqr{
\Omega(\Ysp)=\sqrt{\gamma}\,dy^1\wedge dy^2\,.
   \label{D3w:volume:Eq}
}
Here, $\gamma$ is the determinant of the metric $\gamma_{ij}$.

\begin{table}[h]
\caption{\baselineskip 14pt
Dynamical D3-brane and wave system in the metric \eqref{D3w:metric:Eq}. 
Here $\circ$ denotes the worldvolume coordinate and 
$\star$ denotes the wave coordinate, respectively.}
\label{D3pp}
{\scriptsize
\begin{center}
\begin{tabular}{|c|c|c|c|c|c|c|c|c|c|c|}
\hline
&0&1&2&3&4&5&6&7&8&9\\
\hline
D3 & $\circ$ & $\circ$ & $\circ$ & $\circ$ &  &  &
&  &  &  \\
\cline{2-11}
W & $\circ$ & $\star$ & & & & & & &  &  \\ 
\cline{2-11}
$x^N$ & $t$ & $x$ & $y^1$ & $y^2$ & $z^1$ & $z^2$ & $z^3$
& $z^4$ & $z^5$ & $z^6$ \\
\hline
\end{tabular}
\end{center}
}
\label{table_B}
\end{table}
Let us first consider the Einstein equation (\ref{D3w:Einstein:Eq}). 
In terms of the ansatz for fields (\ref{D3w:metric:Eq}) and 
(\ref{D3w:ansatz for gauge:Eq}), 
the Einstein equations are written by
\Eqrsubl{D3w:cEinstein:Eq}{
&&\hspace{-0.5cm}
-\frac{1}{2}\left[\left(2-h_{\rm W}\right)h_{\rm W}-4\right]h^{-1}\pd_t^2h
+(2-h_{\rm W})\pd_t^2h_{\rm W}+\lap_{\Ysp}h_{\rm W}+h^{-1}\lap_{\Zsp}h_{\rm W}
+\frac{1}{2}(2-h_{\rm W})h^{-2}\lap_{\Zsp}h\nn\\
&&~~~~+\frac{1}{2}(2-h_{\rm W})\pd_th_{\rm W}\pd_t\ln h=0\,,
 \label{D3w:cEinstein-tt:Eq}\\
&&\hspace{-0.5cm}\pd_t\pd_i h_{\rm W}=0\,,
 \label{D3w:cEinstein-ti:Eq}\\
&&\hspace{-0.5cm}\pd_t\pd_a h_{\rm W}+h^{-1}\pd_t\pd_a h=0,
 \label{D3w:cEinstein-ta:Eq}\\
&&\hspace{-0.5cm}-\frac{1}{2}h_{\rm W}^2h^{-1}\pd_t^2h
+h_{\rm W}\pd_t^2h_{\rm W}-\lap_{\Ysp}h_{\rm W}-h^{-1}\lap_{\Zsp}h_{\rm W}
+\frac{1}{2}h_{\rm w}h^{-2}\lap_{\Zsp}h\nn\\
&&~~~~+\frac{1}{2}h_{\rm W}\pd_t h_{\rm W}\pd_t\ln h=0\,,
 \label{D3w:cEinstein-xx:Eq}\\
&&\hspace{-0.5cm}
R_{ij}(\Ysp)-\frac{1}{4}h^{-1}\gamma_{ij}\left(h_{\rm W}\pd_t^2h
+\pd_th_{\rm W}\pd_th\right)
+\frac{1}{4}h^{-2}\gamma_{ij}\lap_{\Zsp}h=0\,,
 \label{D3w:cEinstein-ij:Eq}\\
&&\hspace{-0.5cm}
R_{ab}(\Zsp)-\frac{1}{4}hh_{\rm W}u_{ab}\left(h^{-1}
\pd_t^2h+\pd_t\ln h_{\rm W}\pd_t\ln h\right)-
\frac{1}{4}u_{ab}h^{-1}\lap_{\Zsp}h=0\,,
  \label{D3w:cEinstein-ab:Eq}
}
where the Laplace operators on 
$\Ysp$, $\Zsp$ space are defined by 
$\triangle_{\Ysp}$, $\lap_{\Zsp}$, and
$R_{ij}(\Ysp)$ and $R_{ab}(\Zsp)$ denote the Ricci tensors
constructed from the metrics $\gamma_{ij}(\Ysp)$, $u_{ab}(\Zsp)$, 
respectively.
From Eqs.~(\ref{D3w:cEinstein-ti:Eq}) and 
(\ref{D3w:cEinstein-ta:Eq}), the function $h$ can be expressed as 
\Eqrsubl{D3w:warp:Eq}{
&&h= h_0(t)+h_1(z),~~~~h_{\rm W}=h_{\rm W}(y, z)\,,~~~~~~{\rm For}~~
\pd_{t}h_{\rm W}=0\,,
  \label{D3w:warp1:Eq}\\
&&h= h(z),~~~~h_{\rm W}= k_0(t)+k_1(y, z)\,,~~~~~~{\rm For}~~
\pd_th=0.
  \label{D3w:warp2:Eq}  
}

Next we consider the gauge field equations (\ref{D3w:gauge:Eq}).
Using the ansatz (\ref{D3w:ansatz for gauge:Eq}), we have 
\Eq{
d\left[\pd_a h\left(\ast_{\Zsp}dz^a\right)\right]
=0,
  \label{D3w:gauge2:Eq}
 }
where $\ast_{\Zsp}$ 
denotes the Hodge operator on $\Zsp$. 
Equations (\ref{D3w:gauge2:Eq}) reduces to
\Eq{
\lap_{\Zsp}h=0,~~~\pd_t\pd_a h=0.
    \label{D3w:gauge3:Eq}
}
This is consistent with the Einstein equations (\ref{D3w:warp:Eq}). 
Setting $\pd_th=0$,  
the Einstein equations (\ref{D3w:cEinstein:Eq}) are thus rewritten as
\Eqrsubl{D3w:c2Einstein:Eq}{
&&
(2-h_{\rm W})\pd_t^2h_{\rm W}+\lap_{\Ysp}h_{\rm W}+h^{-1}\lap_{\Zsp}h_{\rm W}
+\frac{1}{2}(2-h_{\rm W})h^{-2}\lap_{\Zsp}h=0\,,
 \label{D3w:c2Einstein-tt:Eq}\\
&&h_{\rm W}\pd_t^2h_{\rm W}-\lap_{\Ysp}h_{\rm W}-h^{-1}\lap_{\Zsp}h_{\rm W}
+\frac{1}{2}h_{\rm W}h^{-2}\lap_{\Zsp}h=0\,,
 \label{D3w:c2Einstein-xx:Eq}\\
&&R_{ij}(\Ysp)+\frac{1}{4}h^{-2}\gamma_{ij}\lap_{\Zsp}h=0\,,
 \label{D3w:c2Einstein-ij:Eq}\\
&&R_{ab}(\Zsp)-\frac{1}{4}u_{ab}h^{-1}\lap_{\Zsp}h=0\,.
  \label{D3w:c2Einstein-ab:Eq}
}
Hence, the field equations reduce to
\Eqrsubl{D3w:fields2:Eq}{
&&R_{ij}(\Ysp)=0,~~~~R_{ab}(\Zsp)=0,
   \label{D3w:Ricci2:Eq}\\
&&h=h(z),~~~~h_{\rm W}=k_0(t)+k_1(y, z),
   \label{D3w:h2:Eq}\\
&&\pd_t^2k_0=0, 
~~~h\lap_{\Ysp}k_1+\triangle_{\Zsp}k_1=0\,.
   \label{D3w:warp2-1:Eq}
 }
For $F_{\left(5\right)}=0$, the function $h_1$ becomes trivial. 

If we choose $\pd_th_{\rm W}=0$, we have 
\Eqrsubl{D3w:fields3:Eq}{
&&R_{ij}(\Ysp)=0,~~~~R_{ab}(\Zsp)=0,
   \label{D3w:Ricci:Eq}\\
&&h=h_0(t)+h_1(z),
   \label{D3w:h:Eq}\\
&&\pd_t^2h_0=0,~~~\triangle_{\Zsp}h_1=0,
   \label{D3w:warp1-1:Eq}\\
&&h\lap_{\Ysp}h_{\rm W}+\triangle_{\Zsp}h_{\rm W}=0.
   \label{D3w:warp1-2:Eq}
 }
Here we focus on the case
\Eq{
\gamma_{ij}=\delta_{ij}\,,~~~u_{ab}=\delta_{ab}\,,~~~~\pd_th=0\,,
 \label{D3w:flat:Eq}
 }
where $\delta_{ij}$, $\delta_{ab}$ are
the two- and six-dimensional Euclidean metrics,
respectively. 
Then, the solution for $h$ and $h_{\rm W}$ can be written by 
\Eqrsubl{D3w:solutions1:Eq}{
h_{\rm W}(t, z)&=&c\,t+\bar{c}
+\sum_{l}\frac{M_l}{\left[|y^i-y^i_0|^2
+M|z^a-z^a_0|^{-2}\right]^{-1}},
 \label{D3w:solution-r:Eq}\\
h(z)&=&\frac{M}{|z^a-z^a_0|^4},
 \label{D3w:solution-s:Eq}
}
where $c$, $\bar{c}$, $M_l$, and $M$ are constant parameters,
and the constants $y_0^i$ and $z_0^a$ denote the positions of the branes.
If the D3-brane is located at the origin of Y and Z spaces, 
the solution is given by 
\Eqrsubl{D3w:solutions2:Eq}{
h_{\rm W}(t, z)&=&c\,t+\bar{c}+\frac{M_{\rm W}}{\left(|y^i|^2
+M|z^a|^{-2}\right)^{-1}},
 \label{D3w:solution-r2:Eq}\\
h(z)&=&\frac{M}{|z^a|^4}\,,
 \label{D3w:solution-s2:Eq}
}
where $M_{\rm W}$ is constant. 
Now we introduce coordinates 
\Eq{
y^1=\frac{1}{r}\cos\theta\cos\alpha\,,~~~~
y^2=\frac{1}{r}\sin\theta\cos\alpha\,,~~~~
z^a=\frac{rM^{1/2}}{\sin\alpha}\mu^a\,,
   \label{D3w:coordinate:Eq}
}
where $\mu^a$ is defined as 
\Eq{
\mu_a\mu^a=1\,,~~~~~~d\Omega^2_{(5)}=d\mu_ad\mu^a\,.
}
Here $d\Omega^2_{(5)}$ is the line element of the unit 5-sphere. 
In terms of \eqref{D3w:solutions2:Eq}
the metric of the D3-wave system becomes
\Eqrsubl{D3w:metric2:Eq}{
ds^2&=&M^{1/2}\sin^{-2}\alpha\left[ds^2_{\rm AdS_3}
+d\alpha^2+\cos^2\alpha d\theta^2
+\sin^2\alpha d\Omega^2_{(5)}\right]\,,\\
ds^2_{\rm AdS}&=&-r^2h_{\rm W}^{-1}dt^2
+r^2h_{\rm W}\left[(h_{\rm W}^{-1}-1)dt+dx\right]^2+r^{-2}dr^2\,,
    \label{D3w:BTZ:Eq}\\
h_{\rm W}&=&c\,t+\bar{c}+\frac{M_{\rm W}}{r^2}\,.
}
Since the metric (\ref{D3w:metric2:Eq}) is the extremal BTZ black hole, 
which is locally AdS${}_3$, the dynamical 
D3-wave system is a warped product of AdS${}_3$ with a  
7-sphere. In terms of the coordinate transformation 
$\tan(\alpha/2)=\e^{\rho}$, we have the AdS${}_5$ metric:
\Eqr{
ds_5^2&=&\sin^{-2}\alpha\left[ds^2_{\rm AdS_3}
+d\alpha^2+\cos^2\alpha d\theta^2\right]\nn\\
&=&d\rho^2+\sinh^2\rho\,d\theta^2+\cosh^2\rho\,ds^2_{\rm AdS_3}\,.
}
Then the ten-dimensional metric can be also written by the 
product of AdS${}_5$ with a 5-sphere \cite{Cvetic:2000cj}. 

If we apply T-duality in the $x$ direction of the ten-dimensional spacetime, 
the D3-brane and wave become D2-brane, F1 string, respectively. 
Then we can obtain the solution for dynamical F1-D2 brane followed by 
a T-duality map in the time-dependent D3-wave system. 
We will show how to obtain the dynamical
F1-D2 brane solution via 
the T-duality in the next subsection. 

%
\subsubsection{Warped AdS${}_2$ spacetime in the F1-D2 brane system}

Now we consider the dynamical F1-D2 brane solution 
in ten-dimensional IIA theory. 
We assume that the ten-dimensional metric takes the form
\Eqr{
ds^2&=&h_{\rm F}^{1/4}(t, y, z)h^{3/8}(z)\left[-h_{\rm F}^{-1}(t, y, z)
h^{-1}(z)dt^2+h_{\rm F}^{-1}(t, y, z)dx^2\right.\nn\\
&&\left.+h^{-1}(z)\gamma_{ij}(\Ysp)dy^idy^j
+u_{ab}(\Zsp)dz^adz^b\right], 
 \label{F1D2:metric:Eq}
}
where $\gamma_{ij}(\Ysp)$ is the two-dimensional metric depending only on 
the coordinates $y^i$ of Y, and $u_{ab}(\Zsp)$ is the six-dimensional 
metric depending only on the coordinates $z^a$ of Z.

We also assume that the scalar and gauge field 
strengths $H_{\left(3\right)}$, $F_{\left(4\right)}$ are given by
\Eqrsubl{F1D2:fields:Eq}{
\e^{\phi}&=&h_{\rm F}^{-1/2}h^{1/4}\,,\\
H_{\left(3\right)}&=&d\left(h_{\rm F}^{-1}\right)\wedge dt\wedge dx\,,
  \label{F1D2:gauge1:Eq}\\
F_{\left(4\right)}&=&d\left(h^{-1}\right)\wedge dt\wedge \Omega(\Ysp),
  \label{F1D2:gauge2:Eq}
}
where $\Omega(\Ysp)$ is the volume 2-form, 
\Eqr{
\Omega(\Ysp)=\sqrt{\gamma}\,dy^1\wedge dy^2\,.
   \label{F1D2:volume:Eq}
}
Here, $\gamma$ is the determinant of the metric $\gamma_{ij}$.

\begin{table}[h]
\caption{\baselineskip 14pt
Dynamical F1-D2 brane system in the metric \eqref{F1D2:metric:Eq}. 
Here $\circ$ denotes the worldvolume coordinate.}
\label{F1D2}
{\scriptsize
\begin{center}
\begin{tabular}{|c|c|c|c|c|c|c|c|c|c|c|}
\hline
&0&1&2&3&4&5&6&7&8&9\\
\hline
F1 & $\circ$ & & & $\circ$ &  &  &
&  &  &  \\
\cline{2-11}
D2 & $\circ$ & $\circ$ & $\circ$ & & & & & &  &  \\ 
\cline{2-11}
$x^N$ & $t$ & $y^1$ & $y^2$ & $x$ & $z^1$ & $z^2$ & $z^3$
& $z^4$ & $z^5$ & $z^6$ \\
\hline
\end{tabular}
\end{center}
}
\label{table_F1D2}
\end{table}
By using the ansatz for fields \eqref{F1D2:metric:Eq} and 
\eqref{F1D2:fields:Eq}, we get 
\Eqrsubl{F1D2:solution1:Eq}{
&&R_{ij}(\Ysp)=0,~~~~R_{ab}(\Zsp)=0,
   \label{F1D2:Ricci:Eq}\\
&&h_{\rm F}=h_0(t)+h_1(y, z),~~~
\pd_t^2h_0=0\,, ~~~
h\lap_{\Ysp}h_1+\triangle_{\Zsp}h_1=0\,,~~~\triangle_{\Zsp}h=0\,,
   \label{F1D2:warp1-2:Eq}
 }
where the Laplace operators on 
$\Ysp$, $\Zsp$ spaces are defined by $\triangle_{\Ysp}$, $\lap_{\Zsp}$, and
$R_{ij}(\Ysp)$ and $R_{ab}(\Zsp)$ denote the Ricci tensors
constructed from the metrics $\gamma_{ij}(\Ysp)$, $u_{ab}(\Zsp)$, 
respectively. 
Let us consider the case 
\Eq{
\gamma_{ij}=\delta_{ij}\,,~~~u_{ab}=\delta_{ab}\,,
 \label{F1D2:flat:Eq}
 }
where $\delta_{ij}$, $\delta_{ab}$ are
the two- and six-dimensional Euclidean metrics,
respectively. 
The solution of $h$ and $h_{\rm F}$ can be expressed as 
\Eqrsubl{F1D2:solutions1:Eq}{
h_{\rm F}(t, y, z)&=&\bar{c}t+\tilde{c}
+\sum_{l}M_l\left[|y^i-y^i_l|^2
+M|z^a-z^a_0|^{-2}\right],
 \label{F1D2:solution-r:Eq}\\
h(z)&=&\frac{M}{|z^a-z^a_0|^{4}},
 \label{F1D2:solution-s:Eq}
}
where $\bar{c}$, $\tilde{c}$, $M_l$, and $M$ are constant parameters,
and the constants $y_l^{i}$ and $z_0^a$ denote the positions of the branes.
If we consider the case where F1-brane is located at the origin of the 
Y, Z spaces and use a coordinate transformation (\ref{D3w:coordinate:Eq}), 
we have 
\Eqrsubl{F1D2:solutions2:Eq}{
h_{\rm F}(t, r)&=&\bar{c}t+\tilde{c}+\frac{M_{\rm F}}{r^2},
 \label{F1D2:solution-r2:Eq}\\
h(r)&=&\frac{\sin^4\alpha}{Mr^4}\,,
 \label{F1D2:solution-s2:Eq}
}
where $M_{\rm F}$ is constant.
In the near-horizon limit $r\rightarrow 0$, the metric becomes 
\Eqr{
ds^2&=&M_{\rm F}^{1/4}M^{5/8}(\sin\alpha)^{-5/2}
\left[-\frac{r^4}{M_{\rm F}}dt^2+\frac{dr^2}{r^2}+d\alpha^2
+\cos^2\alpha d\theta^2\right.\nn\\
&&\left.+\sin^2\alpha d\Omega_{(5)}^2
+\left(M_{\rm F}M\right)^{-1}\sin^4\alpha\left(dy^1\right)^2\right].
    \label{F1D2:metric2:Eq}
}
Then, the ten-dimensional metric is a warped product of the 
AdS${}_2$ with an eight-dimensional internal space. 

We can construct the solution of the F1-D2 brane system 
(\ref{F1D2:metric:Eq}) in terms of the T-duality map 
in the D3-wave solution. 
We start from the dynamical D3-wave solution in the string frame 
in the type IIB theory, 
\Eqrsubl{F1D2:metricD3:Eq}{
ds^2_{({\rm B})}&=&h^{-1/2}(z)\left[-dt^2+dx^2
+\left\{h_{\rm W}(t, y, z)-1\right\}\left(dt-dx\right)^2\right.\nn\\
&&\left.+\delta_{ij}(\Ysp)dy^idy^j+h(z)\delta_{ab}(\Zsp)dz^adz^b\right], 
 \label{F1D2:metric3:Eq}\\
C_{(4)}&=&h^{-1}(z)dt\wedge dx\wedge dy^1\wedge dy^2+\omega_{(4)}\,,
}
where the warp factors $h$ and $h_{\rm W}$ are given by
\Eqrsubl{F1D2:solutions-D3p:Eq}{
h_{\rm W}(t, y, z)&=&At+B+\sum_l\frac{M_l}{\left(|y^i-y^i_l|^2
+M|z^a-z^a_0|^{-2}\right)^{-1}},
 \label{F1D2:solution-D3p1:Eq}\\
h(z)&=&\frac{M}{|z^a-z^a_0|^4}\,.
 \label{F1D2:solution-D3p2:Eq}
}
Here, the constant $y^i_l$ and $z^a_0$ denote 
the positions of the branes. 

The 4-form $\omega_{(4)}$ satisfies 
\Eq{
d\omega_{(4)}=\pm \pd_ah\ast_{\Zsp}\left(dz^a\right)\,.
}
Here $\ast\Zsp$ denotes the Hodge operator on Z.
Now we will obtain the dynamical solution of a D3-pp after 
we apply T-duality in the $x$ direction of the ten-dimensional spacetime. 
In terms of the T-duality map (\ref{D2D6K:duality:Eq}), 
the D3-wave solution (\ref{F1D2:metric3:Eq}) becomes 
\Eqrsubl{F1D2:solution2:Eq}{
ds^2_{{(\rm A)}}&=&h_{\rm F}^{1/4}h_2^{3/8}\left[-(h_{\rm F}h_2)^{-1}dt^2
+h_{\rm F}^{-1}dx^2+h_2^{-1}\delta_{ij}(\Ysp)dy^idy^j
+\delta_{ab}(\Zsp)dz^adz^b\right], 
 \label{F1D2:metric4:Eq}\\
\e^{\phi_{(\rm A)}}&=&h_{\rm F}^{-1/2}h_2^{1/4}\,,\\
B_{\left(2\right)}&=&h_{\rm F}^{-1} dt\wedge dx\,,
  \label{F1D2:gauge3:Eq}\\
C_{\left(3\right)}&=&h_2^{-1}dt\wedge dy^1\wedge dy^2,
  \label{F1D2:gauge4:Eq}
}
where $ds^2_{{(\rm A)}}$ is the ten-dimensional metric in the Einstein 
frame, and $h$ and $h_{\rm F}$ can be written by
\Eqrsubl{F1D2:solutions3:Eq}{
h_{\rm F}(t, y, z)&=&h_{\rm W}(t, y, z),
 \label{F1D2:solution3-r:Eq}\\
h_2(z)&=&h(z)\,.
 \label{F1D2:solution3-s:Eq}
}
Finally we obtain the solutions (\ref{F1D2:metric:Eq}) and 
(\ref{F1D2:fields:Eq}) which 
are derived from the dynamical D3-wave 
solution via T-duality.

\subsection{AdS spacetime from M2-brane solutions}
\label{sec:M2}

In this subsection, we discuss asymptotic spacetime of the M2-brane 
in the eleven-dimensional theory. 
We also present the dynamical D-brane solutions 
in ten dimensions in terms of dimensional reduction of 
the eleven-dimensional theory.

\subsubsection{AdS${}_4$ spacetime in M2-brane and KK monopole system}
The eleven-dimensional action 
which contains 
the metric $g_{MN}$ and 4-form field strength $F_{(4)}$
is given by
\Eq{
S=\frac{1}{2\kappa^2}\int \left[R\ast{\bf 1}
 -\frac{1}{2\cdot 4!}F_{(4)}\wedge\ast F_{(4)}\right],
\label{m:action:Eq}
}
where $\kappa^2$ is the eleven-dimensional gravitational constant,
$\ast$ is the Hodge operator in the eleven-dimensional space-time. 
The field strength $F_{(4)}$ is given by the 3-form gauge potential 
\Eq{
F_{(4)}=dC_{(3)}\,.}

The field equations are given by
\Eqrsubl{m:field equations:Eq}{
&&\hspace{-1cm}R_{MN}=\frac{1}{2\cdot 4!}\left[4F_{MABC} {F_N}^{ABC}
-\frac{1}{2}g_{MN} F^2_{(4)}\right],
   \label{m:Einstein:Eq}\\
&&\hspace{-1cm}d\left[\ast F_{(4)}\right]=0\,,~~~~~dF_{(4)}=0\,.
   \label{m:gauge:Eq}
}

We present the solution involving the M2-brane and KK monopole system. 
We take the eleven-dimensional metric to be 
\cite{Cvetic:1999xx}
\Eqrsubl{M2N:metric:Eq}{
ds^2&=&h_2^{-2/3}(x, y, z)q_{\mu\nu}(\Xsp)dx^{\mu}dx^{\nu}
+h_2^{1/3}(x, y, z)\left[\gamma_{ij}(\Ysp)dy^idy^j\right.\nn\\
&&\left.+h_{\rm K}(z)u_{ab}(\Zsp)dz^adz^b
+h_{\rm K}^{-1}(z)\left(dv+A_adz^a\right)^2\right], 
 \label{M2N:metric1:Eq}\\
&&\hspace{-1.4cm}u_{ab}(\Zsp)dz^adz^b=dr^2+r^2w_{mn}(\Zsp')dp^mdp^n\,,
 \label{M2N:metric2:Eq}
}
where $A_a$ denotes the 1-form, and $q_{\mu\nu}(\Xsp)$, $\gamma_{ij}(\Ysp)$, 
$u_{ab}(\Zsp)$, and $w_{mn}(\Zsp')$ are, respectively,
the three-, four-, three-, two-dimensional metrics of the X, Y, Z, $\Zsp'$ 
spaces, and X, Y, Z, $\Zsp'$ are the three-, four-, three-, two-dimensional 
space with coordinates $x^{\mu}$, $y^i$, $z^a$, $p^m$, respectively.

The 4-form gauge field strength $F_{\left(4\right)}$ is assumed to be
\Eqr{
F_{\left(4\right)}&=&d\left[h_2^{-1}(x, y, z)\right]\wedge\Omega(\Xsp),
  \label{M2N:ansatz:Eq}
}
where the volume 3-form $\Omega(\Xsp)$ is given by 
\Eqr{
\Omega(\Xsp)=\sqrt{-q}\,dx^0\wedge dx^1\wedge dx^2\,.
   \label{M2N:volume:Eq}
}
Here, $q$ is the determinant of the metric $q_{\mu\nu}$.

\begin{table}[h]
\caption{\baselineskip 14pt
Dynamical M2-brane and KK monopole in the metric \eqref{M2N:metric1:Eq}. 
Here $\circ$ denotes the worldvolume coordinate and $\bullet$ 
denotes the fiber coordinate of the KK monopole, respectively.}
\label{M2N}
{\scriptsize
\begin{center}
\begin{tabular}{|c|c|c|c|c|c|c|c|c|c|c|c|}
\hline
&0&1&2&3&4&5&6&7&8&9&10\\
\hline
M2 & $\circ$ & $\circ$ &  $\circ$ &&  &  &
&  &  &  & \\
\cline{2-12}
KK & $\circ$ & $\circ$ & $\circ$ & $\circ$ & $\circ$ & $\circ$ & $\circ$
& $\bullet$ &  &  &\\ 
\cline{2-12}
$x^N$ & $t$ & $x^1$ & $x^2$ & $y^1$ & $y^2$ & $y^3$ & $y^4$
& $v$ & $z^1$ & $z^2$ & $z^3$\\
\hline
\end{tabular}
\end{center}
}
\label{table_M2N}
\end{table}

By using the ansatz for fields \eqref{M2N:metric:Eq} and  
\eqref{M2N:ansatz:Eq}, and the condition 
$dh_{\rm K}=\ast_{\Zsp}dA$, 
the field equations give 
\Eqrsubl{M2N:fields:Eq}{
&&R_{\mu\nu}(\Xsp)=0,~~~~R_{ij}(\Ysp)=0,~~~~R_{ab}(\Zsp)=0,~~~~
R_{mn}(\Zsp')=w_{mn}(\Zsp')\,,
   \label{M2N:Ricci:Eq}\\
&&h_2(x, y, z)=h_0(x)+h_1(y, z)\,,
   \label{M2N:h:Eq}\\
&&D_{\mu}D_{\nu}h_0=0, ~~~~h_{\rm K}\lap_{\Ysp}h_1+\lap_{\Zsp}h_1=0\,,~~~
\lap_{\Zsp}h_{\rm K}=0\,,
   \label{M2N:warp1-2:Eq}
 }
where $D_{\mu}$ is the covariant derivative constructed from 
the metric $q_{\mu\nu}$, and $\triangle_{\Ysp}$, $\lap_{\Zsp}$ denote 
the Laplace operators on $\Ysp$, $\Zsp$ space, and $R_{\mu\nu}(\Xsp)$, 
$R_{ij}(\Ysp)$, and $R_{ab}(\Zsp)$ are the Ricci tensors
with respect to the metrics $q_{\mu\nu}(\Xsp)$, 
$\gamma_{ij}(\Ysp)$, $u_{ab}(\Zsp)$, respectively. 

We set the eleven-dimensional metric: 
\Eq{
q_{\mu\nu}=\eta_{\mu\nu}\,,~~~\gamma_{ij}=\delta_{ij}\,,
~~~u_{ab}=\delta_{ab}\,,
 \label{M2N:Minkowski:Eq}
 }
where $\eta_{\mu\nu}$ denotes the three-dimensional
Minkowski metric and $\delta_{ij}$, $\delta_{ab}$ are
the four- and three-dimensional Euclidean metrics, respectively. 
Under the eleven-dimensional metric (\ref{M2N:Minkowski:Eq}), 
we have 
\Eqrsubl{M2N:solutions:Eq}{
h_2(x, y, r)&=&c_{\mu}x^{\mu}+\tilde{c}
+\frac{M_2}{\left(y^2+4M_{\rm K}r\right)^3},
 \label{M2N:solution-r:Eq}\\
h_{\rm K}(r)&=&\frac{M_{\rm K}}{r}\,,
 \label{M2N:solution-s:Eq}
}
where $c_{\mu}$, $\tilde{c}$, $M_2$, and $M_{\rm K}$ 
are constants, and $y^2=\delta_{ij}y^iy^j$.

Let us consider 
a coordinate transformation 
\Eq{
y^i=\zeta\xi^i\cos\alpha\,,~~~~
r=\frac{1}{4}M_{\rm K}^{-1}\zeta^2\sin^2\alpha\,,
  \label{M2N:coordinate:Eq}
}
where $\xi_i\xi^i=1$. In terms of \eqref{M2N:coordinate:Eq}, 
the eleven-dimensional metric \eqref{M2N:metric:Eq} becomes 
\Eq{
ds^2=h_2^{-2/3}\eta_{\mu\nu}(\Xsp)dx^{\mu}dx^{\nu}
+h_2^{1/3}\left[d\zeta^2+\zeta^2ds^2(\Msp_7)\right],
   \label{M2N:metric3:Eq}
}
where $h_2$ and $ds^2(\Msp_7)$ are given by
\Eqrsubl{M2N:h2:Eq}{
h_2&=&c_{\mu}x^{\mu}+\tilde{c}+\frac{M_2}{\zeta^6}\,,~~~~~~
h_{\rm K}(r)=\frac{4M_{\rm K}^2}{\zeta^2\sin^2\alpha}\,,\\
ds^2(\Msp_7)&=&d\alpha^2+\cos^2\alpha d\Omega_{(3)}^2
+\frac{1}{4}\sin^2\alpha
\left[d\Omega_{(2)}^2+M_{\rm K}^{-2}\left(dv+A_adz^a\right)^2\right].
}
Here $d\Omega^2_{(2)}$ and $d\Omega^2_{(3)}=d\xi_id\xi^i$
are the line element of the unit 2- and 3-sphere, respectively. 
In the near-horizon limit $\zeta\rightarrow 0$, the metric becomes 
AdS${}_4\times\Msp_7$.
After performing 
Kaluza-Klein reduction on the $v$ coordinate, 
the solution (\ref{M2N:solutions:Eq}) gives 
a dynamical D2-D6 brane. 
We will discuss the dynamical D2-D6 brane 
solution in the next subsection.

\subsubsection{AdS${}_4$ spacetime in the D2-D6 brane system}
\label{subsec:D2D6}
Now we discuss the dynamical D2-D6 brane solutions 
in ten-dimensional 
type IIA string theory. 
The action for IIA theory in the Einstein 
frame can be written as 
\Eq{
S=\frac{1}{2\kappa^2}\int \left(R\ast{\bf 1}
 -\frac{1}{2}d\phi \wedge \ast d\phi
 -\frac{1}{2\cdot 2!}\e^{3\phi/2}F_{(2)}\wedge\ast F_{(2)}
 -\frac{1}{2\cdot 4!}\e^{\phi/2}F_{(4)}\wedge\ast F_{(4)}\right),
\label{IIA:action:Eq}
}
where $\kappa^2$ is the ten-dimensional gravitational constant, 
$\ast$ is the Hodge dual operator 
in the ten-dimensional spacetime, and $F_{(2)}$, $F_{(4)}$ are 
RR 2-form, RR 4-form field strengths, respectively.
The expectation values of fermionic fields are assumed to be zero.

After variations with respect to the metric, the scalar field, 
and the gauge field, the field equations are given by
\Eqrsubl{IIA:equations:Eq}{
&&\hspace{-1.2cm}R_{MN}=\frac{1}{2}\pd_M\phi \pd_N \phi
+\frac{1}{2\cdot 2!}\e^{3\phi/2} 
\left(2F_{MA} {F_N}^{A}-\frac{1}{8}g_{MN} F_{(2)}^2\right)\nn\\
&&+\frac{1}{2\cdot 4!}\e^{\phi/2} 
\left(4F_{MABC} {F_N}^{ABC}
-\frac{3}{8}g_{MN} F_{(4)}^2\right),
   \label{IIA:Einstein:Eq}\\
&&\hspace{-1.2cm}d\ast d\phi=\frac{3}{4\cdot 2!}
\e^{3\phi/2}F_{(2)}\wedge \ast F_{(2)}
+\frac{1}{4\cdot 4!}\e^{\phi/2}F_{(4)}\wedge \ast F_{(4)},
   \label{IIA:scalar:Eq}\\
&&\hspace{-1.2cm}d\left(\e^{3\phi/2}\ast F_{(2)}\right)=0,
   \label{IIA:2f:Eq}\\
&&\hspace{-1.2cm}
d\left(\e^{\phi/2}\ast F_{(4)}\right)=0.
   \label{IIA:4f:Eq}
}
The ten-dimensional metric is assumed to be
\Eqr{
ds^2&=& h_2^{3/8}(x, y, z)h_6^{7/8}(z)\left[h_2^{-1}(x, y, z)
     h_6^{-1}(z)q_{\mu\nu}(\Xsp)dx^{\mu}dx^{\nu}
     +h_6^{-1}(z)\gamma_{ij}(\Ysp)dy^idy^j\right.\nn\\
     & &\left.+u_{ab}(\Zsp)dz^adz^b\right],
   \label{D2D6:metric:Eq}
}
where $q_{\mu\nu}(\Xsp)$ denotes the three-dimensional metric depending 
only on the coordinates $x^{\mu}$ of X, 
$\gamma_{ij}(\Ysp)$ denotes the four-dimensional metric depending 
only on the coordinates $y^i$ of Y, and $u_{ab}(\Zsp)$ denotes the 
three-dimensional metric depending 
only on the coordinates $z^a$ of Z, respectively.

The metric form \Eqref{D2D6:metric:Eq} is 
a straightforward generalization of the case of a static D2-D6 brane
system with a dilaton coupling \cite{Cvetic:2000cj}.
Furthermore, we assume that the scalar field $\phi$ 
and the gauge field strengths are given by
\Eqrsubl{D2D6:fields:Eq}{
\e^{\phi}&=&h_2^{1/4}h_6^{-3/4},\\
F_{(2)}&=&\e^{-3\phi/2}\ast\left[d\left(h_6^{-1}\right)\wedge
  \Omega(\Xsp)\wedge\Omega(\Ysp)\right]\,,\\
F_{(4)}&=&d\left(h_2^{-1}\right)\wedge\Omega(\Xsp)\,,
      }
where $\Omega(\Xsp)$ and $\Omega(\Ysp)$ denote 
the volume 3- and 4-forms,
\Eqrsubl{D2D6:volume:Eq}{
\Omega(\Xsp)&=&\sqrt{-q}\,dx^0\wedge dx^1\wedge dx^2,\\
\Omega(\Ysp)&=&\sqrt{\gamma}\,dy^1\wedge dy^2\wedge dy^3\wedge dy^4,
}
respectively.

\begin{table}[h]
\caption{\baselineskip 14pt
Dynamical D2-D6 brane system in the metric \eqref{D2D6:metric:Eq}. 
Here $\circ$ denotes the worldvolume coordinate. }
\label{D2D6}
{\scriptsize
\begin{center}
\begin{tabular}{|c|c|c|c|c|c|c|c|c|c|c|}
\hline
&0&1&2&3&4&5&6&7&8&9\\
\hline
D2 & $\circ$ & $\circ$ & $\circ$ &  &  &  &  
&&& \\
\cline{2-11}
D6 & $\circ$ & $\circ$ & $\circ$ & $\circ$ & $\circ$ &  
 $\circ$ & $\circ$ &  &  & \\
\cline{2-11}
$x^N$ & $t$ & $x^1$ & $x^2$ & $y^1$ & $y^2$ & $y^3$ & $y^4$
& $z^1$ & $z^2$ & $z^3$ \\
\hline
\end{tabular}
\end{center}
}
\label{table_26}
\end{table}

The field equations reduce to  
\Eqrsubl{D2D6:Einstein equations:Eq}{
&&R_{\mu\nu}(\Xsp)=0,~~~~R_{ij}(\Ysp)=0,~~~~R_{ab}(\Zsp)=0,
   \label{D2D6:Ricci:Eq}\\ 
&&h_2(x, y, z)=K(x)+L(y, z)\,,~~~~h_6=h_6(z);~~\\
&&D_{\mu}D_{\nu}K=0\,,~~~\lap_{\Zsp}L(y, z)+h_6\lap_{\Ysp}L(y, z)=0, ~~~~
\lap_{\Zsp}h_6=0\,, 
   \label{D2D6:warp factor h:Eq}
 }      
where $D_{\mu}$ is the covariant derivative with respect to
the metric $q_{\mu\nu}$, and $\triangle_{\Ysp}$, $\lap_{\Zsp}$ are
the Laplace operators on 
$\Ysp$, $\Zsp$ space, and $R_{\mu\nu}(\Xsp)$, 
$R_{ij}(\Ysp)$, and $R_{ab}(\Zsp)$ are the Ricci tensors
constructed from the metrics $q_{\mu\nu}(\Xsp)$, 
$\gamma_{ij}(\Ysp)$, $u_{ab}(\Zsp)$, respectively. 

Let us consider the case
\Eq{ 
q_{\mu\nu}=\eta_{\mu\nu},~~~~\gamma_{ij}=\delta_{ij},
~~~~u_{ab}=\delta_{ab}\,,
 \label{D2D6:s-metric:Eq}
 }
where $\eta_{\mu\nu}$ is the three-dimensional 
Minkowski metric, and $\delta_{ij}$, $\delta_{ab}$ are  
the four-dimensional, three-dimensional Euclidean metric.  
In this case, the solution for $h_2$ and $h_6$ can be obtained
explicitly as
\Eq{ 
h_2(x, z)=c_{\mu}x^{\mu}+\tilde{c}+\sum_l\frac{M_l}
{\left(|y^i-y^i_l|^2+4M_6|z^a-z^a_0|\right)^3},~~~~
h_6(z)=\frac{M_6}{|z^a-z^a_0|},
 \label{D2D6:h:Eq}
}
where $c_{\mu}$, $\tilde{c}$, $M_6$, $M_l$, $y^i_l$, and 
$z^a_0$ are constant parameters. 

Now we investigate the geometrical properties of 
the D2-D6 brane metric \eqref{D2D6:s-metric:Eq}. 
We consider the case where the D2-brane is 
located at the origin of the Y space and use the coordinate 
\Eq{ 
\delta_{ab}dz^adz^b=dr^2+r^2d\Omega^2_{(2)}\,,
 \label{D2D6:s-metric2:Eq}
 }
where $d\Omega^2_{(2)}$ is the line element of the two-dimensional 
unit 2-sphere. Then we have
\Eq{ 
h_2(x, r)=c_{\mu}x^{\mu}+\tilde{c}+\frac{M_2}
{\left(y^2+4M_6r\right)^3},~~~~
h_6(r)=\frac{M_6}{r}\,,
 \label{D2D6:h2:Eq}
}
where $y^2=\delta_{ij}y^iy^j$, and $M_2$ is constant. 

In terms of 
a coordinate transformation 
\Eq{
y^i=\zeta\xi^i\cos\alpha\,,~~~~
r=\frac{1}{4}M_6^{-1}\zeta^2\sin^2\alpha\,,
  \label{D2D6:coordinate:Eq}
}
the eleven-dimensional metric \eqref{D2D6:metric:Eq} becomes 
\Eqrsubl{D2D6:metric2:Eq}{
ds^2&=& \left(\frac{\zeta\sin\alpha}{2M_6}\right)^{1/4}
h_2^{3/8}(x, \zeta)\left[h_2^{-1}(x, \zeta)
     \eta_{\mu\nu}(\Xsp)dx^{\mu}dx^{\nu}\right.\nn\\
&&\left.     +h_2(x, \zeta)\left\{d\zeta^2+\zeta^2\left(d\alpha^2+
\cos^2\alpha d\Omega^2_{(3)}
     +\frac{1}{4}\sin^2\alpha d\Omega^2_{(2)}
\right)\right\}\right],
   \label{D2D6:metric2-1:Eq}\\
h_2(x, \zeta)&=&A_{\mu}x^{\mu}+B+\frac{M_2}{\zeta^6},~~~~
h_6(\zeta)=\frac{4M_6^2}{\zeta^2\sin^2\alpha},~~~~
}
where $\xi_i\xi^i=1$, 
and $d\Omega^2_{(3)}$ is the line element of 3-spheres. 
Then, the metric \eqref{D2D6:metric2-1:Eq} 
in the limit $\zeta\rightarrow 0$ reads
\Eqr{
\hspace{-0.2cm}
ds^2=M_2^{3/8}\left(\frac{\sin\alpha}{2M_{6}}\right)^{\frac{1}{4}}
\left[\left(\frac{\zeta^4}{M_2}\right)\eta_{\mu\nu}(\Xsp)dx^{\mu}dx^{\nu}
+\frac{d\zeta^2}{\zeta^2}+d\alpha^2+\cos^2\alpha d\Omega^2_{(3)}
     +\frac{\sin^2\alpha}{4} d\Omega^2_{(2)}\right].
   \label{D2D6:metric3:Eq}
}
Hence, the ten-dimensional 
metric for $\alpha=0$ and $\gamma_{ij}=\delta_{ij}$ 
becomes a warped ${\rm AdS}_4\times {\rm S}^6$ spacetime.

\subsubsection{AdS${}_3$ spacetime in the M2-wave system}
We discuss the dynamical M2-brane and wave solution. 
We adopt the following ansatz for the eleven-dimensional metric:
\Eqr{
ds^2&=&h_2^{-2/3}(z)\left[-h_{\rm W}^{-1}(t, y, z)dt^2
+h_{\rm W}(t, y, z)\left\{\left(h_{\rm W}^{-1}(t, y, z)-1\right)dt
+dx\right\}^2+dy^2\right.\nn\\
&&\left.+h_2(z)u_{ab}(\Zsp)dz^adz^b\right], 
 \label{M2p:metric:Eq}
}
where $u_{ab}(\Zsp)$ is an eight-dimensional metric which
depends only on the eight-dimensional coordinates $z^a$. 

We further require that the 4-form field satisfies the
following condition:
\Eqr{
F_{\left(4\right)}&=& d\left[h_2^{-1}(z)\right]\wedge dt
\wedge dx\wedge dy\,.
  \label{M2p:ansatz:Eq}
}

\begin{table}[h]
\caption{\baselineskip 14pt
Dynamical M2-brane and wave system in the metric \eqref{M2p:metric:Eq}. 
Here $\circ$ denotes the worldvolume coordinate and $\star$ 
denotes the wave coordinate, respectively.}
\label{M2p}
{\scriptsize
\begin{center}
\begin{tabular}{|c|c|c|c|c|c|c|c|c|c|c|c|}
\hline
&0&1&2&3&4&5&6&7&8&9&10\\
\hline
M2 & $\circ$ & $\circ$ &  $\circ$ &&  &  &
&  &  &  & \\
\cline{2-12}
W & $\circ$ & $\star$ & & & & &
& &  &  &\\ 
\cline{2-12}
$x^N$ & $t$ & $x$ & $y$ & $z^1$ & $z^2$ & $z^3$ & $z^4$
& $z^5$ & $z^6$ & $z^7$ & $z^8$\\
\hline
\end{tabular}
\end{center}
}
\label{table_M2p}
\end{table}

If we use ansatz for fields \eqref{M2p:metric:Eq} and 
\eqref{M2p:ansatz:Eq}, the field equations give 
\Eqrsubl{M2p:fields:Eq}{
&&R_{ab}(\Zsp)=0,
   \label{M2p:Ricci:Eq}\\
&&h_{\rm W}=h_0(t)+h_1(y, z)\,,~~~
\pd_t^2h_0=0,~~~h_2\pd_y^2h_1+\triangle_{\Zsp}h_1=0\,,~~~
\triangle_{\Zsp}h_2=0\,,
   \label{M2p:warp1:Eq}
 }
where $\lap_{\Zsp}$ denotes 
the Laplace operator on $\Zsp$ space, and 
$R_{ab}(\Zsp)$ is the Ricci tensor with respect to 
the metric $u_{ab}(\Zsp)$. 
Now we consider the case
\Eq{
u_{ab}dz^adz^b=\delta_{ab}dz^adz^b=dr^2+r^2d\Omega_{(7)}^2\,,
 \label{M2p:flat metric:Eq}
 }
where $\delta_{ab}$ is 
the eight-dimensional Euclidean metrics, and $d\Omega_{(7)}^2$ 
is the line element of a unit 7-sphere, respectively. 
Substituting the eleven-dimensional metric (\ref{M2p:flat metric:Eq}) 
into the field equations (\ref{M2p:fields:Eq}), we have 
\Eqrsubl{M2p:solutions1:Eq}{
h_{\rm W}(t, y, r)&=&\bar{c}t+\tilde{c}
+M_{\rm W}\left(y^2+\frac{M}{4r^4}\right)\,,
 \label{M2p:solution-r:Eq}\\
h_2(r)&=&\frac{M}{r^6},
 \label{M2p:solution-s:Eq}
}
where $\bar{c}$, $\tilde{c}$, $M_{\rm W}$, and $M$ are constants. 
If we introduce a coordinate transformation
\Eq{
y=\frac{\cos\alpha}{\zeta}\,,~~~~
r^2=\frac{\zeta M^{1/2}}{2\sin\alpha}\,,
   \label{M2p:coordinate:Eq}
}
the metric of the M2-wave system
becomes
\Eqrsubl{M2p:metric2:Eq}{
ds^2&=&\frac{M^{1/3}}{4\sin^{2}\alpha}\left(ds^2_{\rm AdS_3}
+d\alpha^2\right)
+M^{1/3}d\Omega^2_{(7)}\,,\\
ds^2_{\rm AdS}&=&-\zeta^2h_{\rm W}^{-1}dt^2
+\zeta^2h_{\rm W}\left[(h_{\rm W}^{-1}-1)dt+dx\right]^2+\zeta^{-2}d\zeta^2\,,
    \label{M2p:BTZ:Eq}\\
h_{\rm W}&=&At+B+\frac{M_{\rm W}}{\zeta^2}\,.
}
The metric \eqref{M2p:metric2:Eq} is the extremal BTZ black hole, 
which is locally AdS${}_3$. 
The dynamical M2-wave system 
is a warped product of AdS${}_3$ with an 8-sphere. 
If we take the coordinate transformation $\tan(\alpha/2)=\e^{\rho}$, 
the four-dimensional line element can be written by AdS${}_4$ spacetime: 
\Eq{
ds^2_4=\frac{1}{\sin^{2}\alpha}\left(ds^2_{\rm AdS_3}
+d\alpha^2\right)=d\rho^2+\cosh^2\rho\,ds^2_{\rm AdS_3}\,.
}
This is a foliation of AdS${}_3$ \cite{Cvetic:2000cj}. 
Hence, the eleven-dimensional metric becomes AdS${}_4\times{\rm S}^7$. 

\subsubsection{AdS${}_2$ spacetime in the F1-D0 brane system}
In this subsubsection, we discuss the time-dependent F1-D0 brane solution. 
We assume that the ten-dimensional spacetime has the metric
\Eqr{
ds^2=h_{\rm F}^{-3/4}(z)h^{-7/8}(t, y, z)\left[-dt^2
+h(t, y, z)dy^2+h_{\rm F}(z)h(t, y, z)u_{ab}(\Zsp)dz^adz^b\right], 
 \label{F1D0:metric:Eq}
}
where $u_{ab}(\Zsp)$ is an eight-dimensional metric which
depends only on the eight-dimensional coordinates $z^a$. 

Concerning the other fields, we adopt the following assumptions: 
\Eqrsubl{F1D0:ansatz:Eq}{
\e^{\phi}&=&h_{\rm F}^{-1/2}h^{3/4}\,,
  \label{F1D0:scalar:Eq}\\
F_{\left(2\right)}&=&
d\left[h^{-1}(t, y, z)\right]\wedge dt,
  \label{F1D0:gauge:Eq}\\
H_{\left(3\right)}&=&d\left[h_{\rm F}^{-1}(z)\right]\wedge dt
\wedge dy\,.
  \label{F1D0:gauge2:Eq}
}

\begin{table}[h]
\caption{\baselineskip 14pt
Dynamical F1-D0 brane system in the metric \eqref{F1D0:metric:Eq}. 
Here $\circ$ denotes the worldvolume coordinate.}
\label{F1D0}
{\scriptsize
\begin{center}
\begin{tabular}{|c|c|c|c|c|c|c|c|c|c|c|}
\hline
&0&1&2&3&4&5&6&7&8&9\\
\hline
F1 & $\circ$ & $\circ$ & & &  &  &
&  &  &  \\
\cline{2-11}
D0 & $\circ$ && & & & & & &  &  \\ 
\cline{2-11}
$x^N$ & $t$ & $y$ & $z^1$ & $z^2$ & $z^3$ & $z^4$ & $z^5$
& $z^6$ & $z^7$ & $z^8$ \\
\hline
\end{tabular}
\end{center}
}
\label{table_F1D0}
\end{table}

Then, the field equations reduce to
\Eqrsubl{F1D0:fields:Eq}{
&&R_{ab}(\Zsp)=0,
   \label{F1D0:Ricci:Eq}\\
&&h=h_0(t)+h_1(y, z),~~~h_{\rm F}=h_{\rm F}(z)\,,
   \label{F1D0:h:Eq}\\
&&\pd_t^2h_0=0\,, ~~~
h_{\rm F}\pd_y^2h_1+\triangle_{\Zsp}h_1=0\,,~~~
\triangle_{\Zsp}h_{\rm F}=0\,,
   \label{F1D0:warp1-2:Eq}
 }
where $\lap_{\Zsp}$ is the Laplace operators on 
$\Zsp$ space, and $R_{ab}(\Zsp)$ 
denotes 
the Ricci tensor constructed from the metric $u_{ab}(\Zsp)$, respectively. 
We assume that the metric of Z space is given by 
\Eq{
u_{ab}=\delta_{ab}\,,
 \label{F1D0:flat:Eq}
 }
where $\delta_{ab}$ is the eight-dimensional Euclidean metrics.  
Thus, the solution of $h$ and $h_{\rm F}$ can be expressed as 
\Eqrsubl{F1D0:solutions1:Eq}{
h(t, y, z)&=&\bar{c}t+\tilde{c}
+\sum_{l}M_l\left[\left(y-y_l\right)^2
+\frac{M_{\rm F}}{4|z^a-z^a_0|^{4}}\right],
 \label{F1D0:solution-r:Eq}\\
h_{\rm F}(z)&=&\frac{M_{\rm F}}{|z^a-z^a_0|^{6}},
 \label{F1D0:solution-s:Eq}
}
where $\bar{c}$, $\tilde{c}$, $M_l$, and $M_{\rm F}$ are constants,
and $y_l$ and $z_0^a$ are also constant parameters. 

We consider the case where the D0-brane is 
located at the origin of the Y space and use the coordinate 
\Eq{ 
\delta_{ab}dz^adz^b=dr^2+r^2d\Omega^2_{(7)}\,,
 \label{F1D0:s-metric2:Eq}
 }
where $d\Omega^2_{(7)}$ is the line element of the seven-dimensional 
unit 7-sphere. Then we have
\Eq{ 
h(x, r)=A_{\mu}x^{\mu}+B+{M}
\left(y^2+\frac{M_{\rm F}}{4r^4}\right),~~~~
h_{\rm F}(r)=\frac{M_{\rm F}}{r^6}\,,
 \label{F1D0:h2:Eq}
}
where $M$ is constant. 

If we use a coordinate transformation 
\Eq{
y=\frac{\cos\alpha}{\zeta}\,,~~~~
r^2=\frac{\zeta M_{\rm F}^{1/2}}{2\sin\alpha}\,,
  \label{F1D0:coordinate:Eq}
}
the metric of the near-horizon region is written by
\Eqr{
ds^2&=&8^{-3/4}M^{1/8}M_{\rm F}^{3/8}(\sin\alpha)^{-9/4}
\left(-\frac{\zeta^4}{M}dt^2+\frac{d\zeta^2}{\zeta^2}+d\alpha^2
+4\sin^2\alpha d\Omega_{(7)}^2\right)\,,
    \label{F1D0:metric2:Eq}
}
where $d\Omega_{(7)}^2$ is the line element of a unit 7-sphere.
This is a metric of the warped product of AdS${}_2$ with an 8-sphere. 

We can obtain the M2-M2 brane system in terms of 
lifting the F1-D2 brane (\ref{F1D2:solutions2:Eq}) 
back to the eleven-dimensional theory. We will discuss the dynamical 
M2-M2 brane solution in the next subsubsection. 
%
%
\subsubsection{AdS${}_2$ spacetime in the M2-M2 brane system}

In this subsubsection, 
we discuss the dynamical intersecting M2-M2 brane solution.  
We adopt the metric ansatz for the eleven-dimensional spacetime, 
\Eqr{
ds^2&=&h^{1/3}_2(t, x, z)k_2^{1/3}(z)
\left[-h_2^{-1}(t, x, z)k_2^{-1}(z)dt^2
+k_2^{-1}(z)q_{\mu\nu}(\Xsp)dx^{\mu}dx^{\nu}\right.\nn\\
&&\left.+h_2^{-1}(t, x, z)\gamma_{ij}(\Ysp)dy^idy^j
+u_{ab}(\Zsp)dz^adz^b\right], 
 \label{M2M2:metric:Eq}
}
where $q_{\mu\nu}(\Xsp)$, $\gamma_{ij}(\Ysp)$, and 
$u_{ab}(\Zsp)$ are the metrics of the 
two-dimensional spacetime X, of the two-dimensional space Y, 
and of the six-dimensional space Z, which depend only
on the two-dimensional coordinates $x^{\mu}$, 
on the two-dimensional ones $y^i$, and on 
the four-dimensional ones $z^a$, respectively.

The 4-form gauge field strength $F_{\left(4\right)}$ is assumed to be
\Eqr{
F_{\left(4\right)}= d\left[h_2^{-1}(t, x, z)
\,dt\wedge \Omega(\Ysp)+k_2^{-1}(z)
\,dt\wedge \Omega(\Xsp)\right],
  \label{M2M2:ansatz:Eq}
}
where the volume 2-forms $\Omega(\Xsp)$ and $\Omega(\Ysp)$ are defined by   
\Eqrsubl{M2M2:volume:Eq}{
\Omega(\Xsp)&=&\sqrt{q}\,dx^1\wedge dx^2\,,
   \label{M2M2:volume y1:Eq}\\
\Omega(\Ysp)&=&\sqrt{\gamma}\,dy^1\wedge dy^2\,.
   \label{M2M2:volume y2:Eq}
}
Here, $q$ and $\gamma$ are the determinants of the metric 
$q_{\mu\nu}$, and $\gamma_{ij}$, respectively.

\begin{table}[h]
\caption{\baselineskip 14pt
Dynamical M2-M2 brane system in the metric \eqref{M2M2:metric:Eq}. 
Here $\circ$ denotes the worldvolume coordinate.}
\label{M2M2}
{\scriptsize
\begin{center}
\begin{tabular}{|c|c|c|c|c|c|c|c|c|c|c|c|}
\hline
&0&1&2&3&4&5&6&7&8&9&10\\
\hline
M2 & $\circ$ & $\circ$ &  $\circ$ &&  &  &
&  &  &  & \\
\cline{2-12}
M2 & $\circ$ &&&  $\circ$ &  $\circ$ & & & &  &  &\\ 
\cline{2-12}
$x^N$ & $t$ & $x^1$ & $x^2$ & $y^1$ & $y^2$ & $z^1$ & $z^2$
& $z^3$ & $z^4$ & $z^5$ & $z^6$\\
\hline
\end{tabular}
\end{center}
}
\label{table_M2M2}
\end{table}

Under the assumption for the metric \eqref{M2M2:metric:Eq} and field strength 
\eqref{M2M2:ansatz:Eq}, the field equations give 
\Eqrsubl{M2M2:solution:Eq}{
&&R_{\mu\nu}(\Xsp)=0,~~~~R_{ij}(\Ysp)=0,~~~~
R_{ab}(\Zsp)=0,
   \label{M2M2:Ricci:Eq}\\
&&h_2=h_0(t)+h_1(x, z)\,, ~~~ 
\pd_t^2h_0=0, ~~~k_2\lap_{\Xsp}h_1+\triangle_{\Zsp}h_1=0\,,
~~~\triangle_{\Zsp}k_2=0\,,
   \label{M2M2:warp:Eq}
 }
where the Laplace operators on the space of 
$\Xsp$, $\Zsp$ are defined by $\triangle_{\Xsp}$, $\lap_{\Zsp}$, and  
$R_{\mu\nu}(\Xsp)$, $R_{ij}(\Ysp)$, and $R_{ab}(\Zsp)$
are the Ricci tensors with respect to the metrics 
$q_{\mu\nu}(\Xsp)$, $\gamma_{ij}(\Ysp)$, $u_{ab}(\Zsp)$, 
respectively. We set the ten-dimensional metric: 
\Eq{
q_{\mu\nu}=\delta_{\mu\nu}\,,~~~\gamma_{ij}=\delta_{ij}\,,~~~
u_{ab}=\delta_{ab}\,,
 \label{M2M2:flat metric:Eq}
 }
where $\delta_{\mu\nu}$, $\delta_{ij}$, $\delta_{ab}$ are
the two-, two-, six-dimensional Euclidean metrics,
respectively. 
The ten-dimensional field equations (\ref{M2M2:solution:Eq}) give 
\Eqrsubl{M2M2:solutions:Eq}{
h_2(t, x, z)&=&\bar{c}t+\tilde{c}
+\sum_{l}M_l\left[|x^{\mu}-x^{\mu}_l|^2
+M_2|z^a-z^a_0|^{-2}\right],
 \label{M2M2:solutions1-1:Eq}\\
k_2(z)&=&\frac{M_2}{|z^a-z^a_0|^4},
 \label{M2M2:solutions1-2:Eq}
}
where $\bar{c}$, $\tilde{c}$, $M_l$, and $M_2$ are constants,
and the constants $x_l^{\mu}$ and $z_0^a$ represent
 the positions of the branes. 

Let us consider the case where the M2-brane is located at the 
origin of the X and Z spaces. 
If we use the coordinate transformation 
\Eq{
x^1=\frac{1}{r}\cos\theta\cos\alpha\,,~~~~
x^2=\frac{1}{r}\sin\theta\cos\alpha\,,~~~~
z^a=\frac{rM_2^{1/2}}{\sin\alpha}\mu^a\,,
   \label{M2M2:coordinate:Eq}
}
the harmonic functions take the following form:
\Eqrsubl{M2M2:solutions2:Eq}{
h_2(t, r)&=&\bar{c}t+\tilde{c}+\frac{M}{r^2},
 \label{M2M2:solution-r2:Eq}\\
k_2(r)&=&\frac{\sin^4\alpha}{M_2\,r^4}\,,
 \label{M2M2:solution-s2:Eq}
}
where $M$ is constant, and $\mu^a$ is defined as 
\Eq{
\mu_a\mu^a=1\,,~~~~~~d\Omega^2_{(5)}=d\mu_ad\mu^a\,.
}
Here $d\Omega^2_{(5)}$ is the line element of the unit 5-sphere. 

The eleven-dimensional metric in the 
the near-horizon limit $r\rightarrow 0$ is thus written by
\Eqr{
ds^2&=&M^{1/3}M_2^{2/3}(\sin\theta)^{-8/3}
\left[-\frac{r^4}{M}dt^2+\frac{dr^2}{r^2}+d\alpha^2
+\cos^2\theta d\theta^2\right.\nn\\
&&\left.+\sin^2\theta d\Omega_{(5)}^2
+\left(MM_2\right)^{-1}\sin^4\theta\left\{\left(dy^1\right)^2
+\left(dy^2\right)^2\right\}\right]\,,
    \label{M2M2:metric2:Eq}
}
where $d\Omega_{(5)}^2$ is the line element of 
a unit 5-sphere. The near-horizon geometry of this system 
is a warped product of AdS${}_2$ with a 7-sphere and a 2-torus.

\subsection{AdS spacetime from M5-brane solutions}
\label{sec:M5}
In this subsection, we discuss asymptotic geometries for 
the dynamical intersecting brane solutions 
including M5-waves and KK monopole in eleven dimensions. 
The dimensional reduction of these generates the cosmological D-brane 
solutions in the ten-dimensional supergravity theories. We also 
briefly discuss these objects. 

\subsubsection{AdS${}_7$ spacetime in M5-brane and KK monopole system}
In this subsubsection, we construct the time-dependent 
M5-brane and KK monopole solution. 
The eleven-dimensional metric is assumed to be 
\cite{Cvetic:1999xx}
\Eqrsubl{M5K:metric:Eq}{
&&ds^2=h_5^{-1/3}(x, y, z)q_{\mu\nu}(\Xsp)dx^{\mu}dx^{\nu}
+h_5^{2/3}(x, y, z)\left[dy^2\right.\nn\\
&&\left.~~~~+h_{\rm K}(z)u_{ab}(\Zsp)dz^adz^b
+h_{\rm K}^{-1}(z)\left(dv+A_adz^a\right)^2\right], 
 \label{M5K:metric1:Eq}\\
&&u_{ab}(\Zsp)dz^adz^b=dr^2+r^2w_{mn}(\Zsp')dp^mdp^n\,,
}
where $A_a$ is the 1-form, and 
$q_{\mu\nu}(\Xsp)$ is a six-dimensional metric which
depends only on the six-dimensional coordinates $x^{\mu}$, 
$w_{mn}(\Zsp')$ is a two-dimensional metric which
depends only on the two-dimensional coordinates $p^m$, 
and $u_{ab}(\Zsp)$ is a three-dimensional metric which
depends only on the three-dimensional coordinates $z^a$, respectively.

We assume the form of 4-form gauge field 
strength $F_{\left(4\right)}$: 
\Eqr{
F_{\left(4\right)}&=&\ast d\left[h_5^{-1}(x, y, z)
\wedge\Omega(\Xsp)\right],
  \label{M5K:ansatz:Eq}
}
where $\Omega(\Xsp)$ is given by 
\Eqr{
\Omega(\Xsp)=\sqrt{-q}\,dx^0\wedge dx^1\wedge \cdots \wedge dx^5\,.
   \label{M5K:volume:Eq}
}
Here, $q$ is the determinant of the metric $q_{\mu\nu}$.

\begin{table}[h]
\caption{\baselineskip 14pt
Dynamical M5-brane and KK monopole system in the metric 
\eqref{M5K:metric1:Eq}. 
Here $\circ$ denotes the worldvolume coordinate and $\bullet$ 
denotes the fiber coordinate of the KK monopole, respectively.}
\label{M5K}
{\scriptsize
\begin{center}
\begin{tabular}{|c|c|c|c|c|c|c|c|c|c|c|c|}
\hline
&0&1&2&3&4&5&6&7&8&9&10\\
\hline
M5 & $\circ$ & $\circ$ & $\circ$ & $\circ$ &  $\circ$  &  $\circ$  &
&  &  &  & \\
\cline{2-12}
KK & $\circ$ & $\circ$ & $\circ$ & $\circ$ & $\circ$ & $\circ$ & $\circ$
& $\bullet$ &  &  &\\ 
\cline{2-12}
$x^N$ & $t$ & $x^1$ & $x^2$ & $x^3$ & $x^4$ & $x^5$ & $y$
& $v$ & $z^1$ & $z^2$ & $z^3$\\
\hline
\end{tabular}
\end{center}
}
\label{table_M5K}
\end{table}

With ansatz for fields \eqref{M5K:metric:Eq} and 
\eqref{M5K:ansatz:Eq}, the field equations reduce to 
\Eqrsubl{M5K:field:Eq}{
&&R_{\mu\nu}(\Xsp)=0,~~~~R_{ab}(\Zsp)=0,~~~~R_{mn}(\Zsp')=w_{mn}(\Zsp'),
   \label{M5K:Ricci:Eq}\\
&&h_5=h_0(x)+h_1(y, z),~~~~dh_{\rm K}=\ast_{\Zsp}dA_{(1)}\,,
   \label{M5K:h:Eq}\\
&&D_{\mu}D_{\nu}h_0=0, ~~~
h_{\rm K}\pd_y^2h_1+\triangle_{\Zsp}h_1=0\,,~~~\triangle_{\Zsp}h_{\rm K}=0\,,
   \label{M5K:warp1-2:Eq}
 }
where $A_{(1)}$ denotes the 1-form, and 
$D_{\mu}$ is the covariant derivative with respect to
the metric $q_{\mu\nu}$, and $\lap_{\Zsp}$ is 
the Laplace operator on $\Zsp$ space, and $R_{\mu\nu}(\Xsp)$, 
 $R_{ab}(\Zsp)$, $R_{mn}(\Zsp')$ are the Ricci tensors 
with respect ot the metrics $q_{\mu\nu}(\Xsp)$, $u_{ab}(\Zsp)$, 
 $w_{mn}(\Zsp')$, respectively. 

To see
the solutions more explicitly, we consider the case of 
\Eq{
q_{\mu\nu}=\eta_{\mu\nu}\,,
~~~~~u_{ab}dz^adz^b=\delta_{ab}dz^adz^b=dr^2+r^2d\Omega_{(2)}^2\,,
 \label{M5K:flat metric:Eq}
 }
where $\eta_{\mu\nu}$ is the six-dimensional
Minkowski metric and $\delta_{ab}$ is
the three-dimensional Euclidean metric, and $d\Omega_{(2)}^2$ is the line 
element of 2-sphere, respectively.
Then, the solution for $h_5$ and $h_{\rm K}$ can be written as
\Eqrsubl{M5K:solutions:Eq}{
h_5(x, y, r)&=&c_{\mu}x^{\mu}+\tilde{c}
+\sum_{l}\frac{M_l}{\left(|y-y_l|^2
+4M_{\rm K}r\right)^{3/2}},
 \label{M5K:solution-r:Eq}\\
h_{\rm K}(z)&=&\frac{M_{\rm K}}{r},
 \label{M5K:solution-s:Eq}
}
where $c_{\mu}$, $\tilde{c}$, $M_l$, and $M_{\rm K}$ 
are constant parameters, 
and the constant $y_l$ represents the position of the brane.

Now we consider the case where the M5-brane is located at the origin 
of the $y$ and make a change of coordinates, 
\Eq{
y=\zeta\cos\alpha\,,~~~~
r=\frac{1}{4}M_{\rm K}^{-1}\zeta^2\sin^2\alpha\,.
  \label{M5K:coordinate:Eq}
}
In terms of \eqref{M5K:coordinate:Eq} and 
$y_\ell=0$, the eleven-dimensional metric \eqref{M5K:metric:Eq} reads 
\Eq{
ds^2=h_5^{-2/3}\eta_{\mu\nu}(\Xsp)dx^{\mu}dx^{\nu}
+h_5^{-2/3}\left[d\zeta^2+\zeta^2ds^2(\Msp_4)\right],
   \label{M5K:metric3:Eq}
}
where $h_5$ and $ds^2(\Msp_4)$ are given by
\Eqrsubl{M5K:h2:Eq}{
h_5&=&c_{\mu}x^{\mu}+\tilde{c}+\frac{M_5}{\zeta^3}\,,\\
ds^2(\Msp_4)&=&d\alpha^2+\frac{1}{4}\sin^2\alpha
\left[d\Omega_{(2)}^2+\left(dv+A_{a}dz^a\right)^2\right],
}
where $M_5$ is constant. 
In the near-horizon limit $\zeta\rightarrow 0$, the metric becomes 
AdS${}_7\times\Msp_4$.

\subsubsection{AdS${}_7$ spacetime in the D6-NS5 brane system}

Now we construct the dynamical D6-NS5 branes system. 
In this subsubsection, 
we look for solutions whose ten-dimensional metrics have the form
\Eqrsubl{D6N:metric:Eq}{
&&ds^2=h^{7/8}(z)h_{\rm NS}^{3/4}(x, y, z)\left[
h^{-1}(z)h_{\rm NS}^{-1}(x, y, z)q_{\mu\nu}(\Xsp)
dx^{\mu}dx^{\nu}+h^{-1}(z)dy^2\right.\nn\\
&&\left.~~~~~~+u_{ab}(\Zsp)dz^adz^b\right], 
 \label{D6N:metric1:Eq}\\
&&u_{ab}(\Zsp)dz^adz^b=dr^2+r^2w_{mn}(\Zsp')dp^mdp^n\,,
}
where $q_{\mu\nu}(\Xsp)$ is the metric of a six-dimensional spacetime, 
$w_{mn}(\Zsp')$ is the metric of a two-dimensional spacetime, and 
$u_{ab}(\Zsp)$ is a three-dimensional metric. The metrics 
$q_{\mu\nu}(\Xsp)$, $w_{mn}(\Zsp')$, and $u_{ab}(\Zsp)$ depend on the 
six-dimensional coordinate $x^{\mu}$, two-dimensional coordinates $p^m$, 
and three-dimensional coordinates $z^a$, respectively. 

We also assume that other fields are the function of time 
\Eqrsubl{D6N:ansatz:Eq}{
\e^{\phi}&=&h_{\rm NS}^{1/2}\,h^{-3/4}\,,
  \label{D6N:scalar:Eq}\\
H_{\left(3\right)}&=&\e^{\phi}\ast d\left[h_{\rm NS}^{-1}(x, y, z)
\,\Omega(\Xsp)\right],
  \label{D6N:gauge:Eq}\\
F_{\left(2\right)}&=&\e^{-3\phi/2}\ast d\left[h^{-1}(z)\Omega(\Xsp)\wedge dy\right]\,,
  \label{D6N:gauge2:Eq}
}
where $\Omega(\Xsp)$ denotes the volume 6-form
\Eqr{
\Omega(\Xsp)&=&\sqrt{-q}\,dx^0\wedge dx^1 \wedge\cdots\wedge dx^5\,.
   \label{D6N:volume x:Eq}
}
Here, $q$ 
is the determinant of the metric $q_{\mu\nu}$.

\begin{table}[h]
\caption{\baselineskip 14pt
Dynamical D6-NS5 brane system in the metric \eqref{D6N:metric1:Eq}. 
Here $\circ$ denotes the worldvolume coordinate. }
\label{D6N}
{\scriptsize
\begin{center}
\begin{tabular}{|c|c|c|c|c|c|c|c|c|c|c|}
\hline
&0&1&2&3&4&5&6&7&8&9\\
\hline
D6 & $\circ$ & $\circ$ & $\circ$ & $\circ$ & $\circ$ &  $\circ$  &  
$\circ$ &&& \\
\cline{2-11}
NS5 & $\circ$ & $\circ$ & $\circ$ & $\circ$ & $\circ$ &  
 $\circ$ & &  & & \\
\cline{2-11}
$x^N$ & $t$ & $x^1$ & $x^2$ & $x^3$ & $x^4$ & $x^5$ & $y$
& $z^1$ & $z^2$ & $z^3$ \\
\hline
\end{tabular}
\end{center}
}
\label{table_D6N}
\end{table}

From the ansatz for fields \eqref{D6N:metric:Eq} and 
\eqref{D6N:ansatz:Eq}, we get 
\Eqrsubl{D6N:solution1:Eq}{
&&R_{\mu\nu}(\Xsp)=0,~~~~R_{ab}(\Zsp)=0,~~~~R_{mn}(\Zsp')=w_{mn}(\Zsp'),
   \label{D6N:Ricci:Eq}\\
&&h_{\rm NS}=h_0(x)+h_1(y, r),~~~
   \label{D6N:h:Eq}
D_{\mu}D_{\nu}h_0=0\,,~~~
h\pd_y^2h_1+\triangle_{\Zsp}h_1=0\,,
~~~\triangle_{\Zsp}h=0\,,
 }
where $D_{\mu}$ is the covariant derivative constructed by the metric 
$q_{\mu\nu}$, and  
$\lap_{\Zsp}$ is 
the Laplace operator on $\Zsp$ space, 
and $R_{\mu\nu}(\Xsp)$, $R_{ab}(\Zsp)$, and
$R_{mn}(\Zsp')$ are the Ricci tensors
with respect to the metrics $q_{\mu\nu}(\Xsp)$, 
$u_{ab}(\Zsp)$, $w_{mn}(\Zsp')$, respectively. 
Let us consider the case
\Eq{
q_{\mu\nu}=\eta_{\mu\nu}\,,~~~
u_{ab}=\delta_{ab}=\delta_{ab}dz^adz^b=dr^2+r^2d\Omega_{(2)}^2\,,
 \label{D6N:flat metric:Eq}
 }
where $\eta_{\mu\nu}$ is the six-dimensional
Minkowski metric and $\delta_{ab}$ is the 
three-dimensional Euclidean metric, and $d\Omega_{(2)}^2$ is the line 
element of 2-sphere, respectively. 
The solution for $h$ and $h_{\rm NS}$ can be
obtained explicitly as
\Eqrsubl{D6N:solutions1:Eq}{
h_{\rm NS}(x, y, r)&=&c_{\mu}x^{\mu}+\tilde{c}
+\sum_{l}\frac{M_l}{\left[|y-y_l|^2
+4Mr\right]^{\frac{3}{2}}},
 \label{D6N:solution-r:Eq}\\
h(r)&=&\frac{M}{r},
 \label{D6N:solution-s:Eq}
}
where $c_{\mu}$, $\tilde{c}$, $M_l$, and $M$ are constant parameters,
and the constant $y_l$ denotes the position of the brane. 
The solutions (\ref{D6N:metric:Eq}) and (\ref{D6N:solutions1:Eq}) 
can be obtained by the dimensional reduction on the direction 
of $v$ in the solution (\ref{M5K:metric:Eq}). 
If we consider the case where the D6-brane is located at the origin of the 
Y, Z spaces and use the coordinate transformation, 
\Eq{
y=\zeta\cos\alpha\,,~~~~
r=\frac{1}{4}M^{-1}\zeta^2\sin^2\alpha\,,
  \label{D6N:coordinate:Eq}
}
we have 
\Eqrsubl{D6N:solutions2:Eq}{
h_{\rm NS}(x, \zeta)&=&c_{\mu}x^{\mu}+\tilde{c}+\frac{M_5}{\zeta^3},
 \label{D6N:solution-r2:Eq}\\
h(\zeta)&=&\frac{4M^2}{\zeta^2\sin^2\alpha}\,.
 \label{D6N:solution-s2:Eq}
}

In the near-horizon limit $\zeta\rightarrow 0$, the metric becomes 
\Eq{
ds^2=M^{3/4}\left(2M_5\right)^{-1/4}(\sin\alpha)^{1/4}
\left[-\frac{\zeta}{M}\eta_{\mu\nu}(\Xsp)dx^{\mu}dx^{\nu}
+\frac{d\zeta^2}{\zeta^2}
+d\alpha^2+\frac{1}{4}\sin^2\alpha d\Omega_{(2)}^2\right]\,,
    \label{D6N:metric2:Eq}
}
where $d\Omega_{(2)}^2$ is the line element of a unit 2-sphere. 
Then, the ten-dimensional metric is a warped product of the 
AdS${}_7$ with a three-dimensional internal space. 

We can construct the solution (\ref{D6N:solutions2:Eq}) 
after dimensional reduction on the fiber coordinate $v$ in the M5-KK monopole 
(\ref{M5K:solutions:Eq}).
For $c_{\mu}=0$, the function (\ref{D6N:solution-r:Eq}) 
is the consistent with the static 
D6-NS5 solution \cite{Itzhaki:1998uz}.

\subsubsection{AdS${}_3$ spacetime in the M5-brane and wave system}

We construct the dynamical M5-brane and wave system. 
In this subsubsection, we take the following metric ansatz:
\Eqr{
ds^2&=&h_5^{2/3}(z)\left[-h_5^{-1}(z)h_{\rm W}^{-1}(t, y, z)dt^2
+h_5^{-1}(z)h_{\rm W}(t, y, z)
\left\{\left(h_{\rm W}^{-1}(t, y, z)-1\right)dt+dx\right\}^2\right.\nn\\
&&\left.+h_5^{-1}(z)\gamma_{ij}(\Ysp)dy^idy^j+u_{ab}(\Zsp)dz^adz^b\right], 
 \label{M5p:metric:Eq}
}
where $\gamma_{ij}(\Ysp)$ is a four-dimensional metric depending only 
on the four-dimensional coordinates $y^i$, 
and $u_{ab}(\Zsp)$ is a five-dimensional metric which
depends only on the five-dimensional coordinates $z^a$. 

We now take the following ansatz for the gauge field 
strength $F_{\left(4\right)}$:
\Eqr{
F_{\left(4\right)}&=&\ast d\left[h_5^{-1}(z)\wedge dt
\wedge dx\wedge\Omega(\Ysp)\right],
  \label{M5p:ansatz:Eq}
}
where $\Omega(\Ysp)$ is the volume 4-form on Y space:
\Eqr{
\Omega(\Ysp)=\sqrt{\gamma}\,dy^1\wedge dy^2\wedge dy^3\wedge dy^4\,.
   \label{M5p:volume:Eq}
}
Here, $\gamma$ is the determinant of the metric $\gamma_{ij}$.

\begin{table}[h]
\caption{\baselineskip 14pt
Dynamical M5-brane and wave system in the metric \eqref{M5p:metric:Eq}. 
Here $\circ$ denotes the worldvolume coordinate and $\star$ 
denotes the wave coordinate, respectively.}
\label{M5p}
{\scriptsize
\begin{center}
\begin{tabular}{|c|c|c|c|c|c|c|c|c|c|c|c|}
\hline
&0&1&2&3&4&5&6&7&8&9&10\\
\hline
M5 & $\circ$ & $\circ$ &  $\circ$ & $\circ$ & $\circ$ & $\circ$ &
&  &  &  & \\
\cline{2-12}
W & $\circ$ & $\star$ & & & & &
& &  &  &\\ 
\cline{2-12}
$x^N$ & $t$ & $x$ & $y^1$ & $y^2$ & $y^3$ & $y^4$ & $z^1$
& $z^2$ & $z^3$ & $z^4$ & $z^5$\\
\hline
\end{tabular}
\end{center}
}
\label{table_M5p}
\end{table}

By using the ansatz for fields \eqref{M5p:metric:Eq} and 
\eqref{M5p:ansatz:Eq}, the field equations reduce to 
\Eqrsubl{M5p:solution1:Eq}{
&&R_{ij}(\Ysp)=0,~~~~R_{ab}(\Zsp)=0,
   \label{M5p:Ricci:Eq}\\
&&h_{\rm W}=h_0(t)+h_1(y, z)\,,~~~
\pd_t^2h_0=0,~~~h_5\lap_{\Ysp}h_1+\triangle_{\Zsp}h_1=0\,,~~~
\triangle_{\Zsp}h_5=0\,,
   \label{M5p:warp1:Eq}
 }
where the Laplace operators on 
$\Ysp$, $\Zsp$ spaces are defined by 
$\triangle_{\Ysp}$, $\lap_{\Zsp}$, and 
$R_{ij}(\Ysp)$, $R_{ab}(\Zsp)$ are the Ricci tensors with 
respect to the metrics $\gamma_{ij}(\Ysp)$, $u_{ab}(\Zsp)$, 
respectively. 
We take the metric on Y and Z space to be 
\Eq{
\gamma_{ij}=\delta_{ij}\,,
~~~u_{ab}dz^adz^b=\delta_{ab}dz^adz^b=dr^2+r^2d\Omega_{(4)}^2\,,
 \label{M5p:flat metric:Eq}
 }
where $\delta_{ij}$, $\delta_{ab}$ are
the four-dimensional Euclidean metrics, and 
$d\Omega_{(4)}^2$ is the line element of a unit 4-sphere, respectively. 
Then we can get the solution for $h_5$ and $h_{\rm W}$, 
\Eqrsubl{M5p:solutions1:Eq}{
h_{\rm W}(t, y, r)&=&\bar{c}t+\tilde{c}
+M_{\rm W}\left(y^2+\frac{4M}{r}\right),
 \label{M5p:solution-r:Eq}\\
h_5(r)&=&\frac{M}{r^3},
 \label{M5p:solution-s:Eq}
}
where $y^2=\delta_{ij}y^iy^j$\,, and 
$\bar{c}$, $\tilde{c}$, $M_{\rm W}$, and $M$ are constant parameters.

If we introduce coordinate transformation
\Eq{
y^i=\frac{\mu^i}{\zeta}\cos\alpha\,,~~~~
r=\frac{4\zeta^2 M}{\sin^2\alpha}\,,
   \label{M5p:coordinate:Eq}
}
the metric of the M5-brane and wave system becomes
\Eqrsubl{M5p:metric2:Eq}{
ds^2&=&\frac{4M^{2/3}}{\sin^{2}\alpha}\left(ds^2_{\rm AdS}
+d\alpha^2+\cos^2\alpha d\Omega^2_{(3)}\right)
+M^{2/3}d\Omega^2_{(4)}\,,\\
ds^2_{\rm AdS}&=&-\zeta^2h_{\rm W}^{-1}dt^2
+\zeta^2h_{\rm W}\left[(h_{\rm W}^{-1}-1)dt+dx\right]^2+\zeta^{-2}d\zeta^2\,,
    \label{M5p:BTZ:Eq}\\
h_{\rm W}&=&At+B+\frac{M_{\rm W}}{\zeta^2}\,,
}
where $\mu^i\mu_i=1$, and the line element of the 3-sphere is defined 
by $d\mu^id\mu_i=d\Omega^2_{(3)}$. 
The metric \eqref{M5p:metric2:Eq} is the extremal BTZ black hole, 
which is locally AdS${}_3$. 
The dynamical M5-wave system 
is a warped product of AdS${}_3$ with the internal spaces. 
However, under the coordinate transformation $\tan(\alpha/2)=\e^{\rho}$, 
the seven-dimensional line element can be written by  
\Eq{
ds^2_7=\frac{1}{\sin^{2}\alpha}\left[ds^2_{\rm AdS_3}
+d\alpha^2+\cos^2\alpha\,d\Omega_{(3)}^2\right]
=d\rho^2+\sinh^2\rho\,d\Omega_{(3)}^2+\cosh^2\rho\,ds^2_{\rm AdS_3}\,.
}
Since a foliation of AdS${}_3$ and 3-sphere can be expressed as
AdS${}_7$ spacetime, the eleven-dimensional metric becomes 
AdS${}_7\times{\rm S}^4$ \cite{Cvetic:2000cj}. 

If we perform the dimensional reduction of the solution 
(\ref{M5p:solutions1:Eq}) on the coordinate $x$, the M5-wave 
becomes a D0-D4 brane system. We will discuss the dynamical D0-D4 brane
solution in the next subsubsection.

\subsubsection{AdS${}_2$ spacetime in the D0-D4 brane system}
In this subsubsection, we discuss the dynamical solution of 
the D0-D4 brane system.  
We assume that the ten-dimensional metric is written by 
\Eqr{
ds^2&=&h^{1/8}(t, y, z)h_{4}^{5/8}(z)\left[
-h^{-1}(t, y, z)h_{4}^{-1}(z)dt^2
+h_4^{-1}(z)\gamma_{ij}(\Ysp)dy^idy^j
\right.\nn\\
&&\left.+u_{ab}(\Zsp)dz^adz^b\right], 
 \label{D0D4:metric:Eq}
}
where $\gamma_{ij}(\Ysp)$, $u_{ab}(\Zsp)$
are the metrics of the four-dimensional spacetime Y, 
of the
five-dimensional space Z, which depend only
on the four-dimensional coordinates $y^i$, 
on the five-dimensional ones $z^a$, respectively.

\begin{table}[h]
\caption{\baselineskip 14pt
Dynamical D0-D4 brane system. 
Here $\circ$ denotes the worldvolume coordinate. }
\label{twoM}
{\scriptsize
\begin{center}
\begin{tabular}{|c|c|c|c|c|c|c|c|c|c|c|}
\hline
&0&1&2&3&4&5&6&7&8&9
\\
\hline
D0 & $\circ$ &  & &   &   &   &&&&
\\
\cline{2-11}
D4 & $\circ$ &$\circ$ &  $\circ$ & $\circ$ &$\circ$ && 
&  && 
\\
\cline{2-11}
$x^N$ & $t$ & $y^1$ & $y^2$ & $y^3$ & $y^4$ & $z^1$ & $z^2$ & $z^3$
& $z^4$ & $z^5$
\\
\hline
\end{tabular}
\end{center}
}
\label{tableD0D4}
\end{table}

The scalar field $\phi$ and the gauge field 
strengths $F_{\left(2\right)}$, $F_{\left(4\right)}$ are assumed to be 
\Eqrsubl{D0D4:ansatz:Eq}{
\e^{\phi}&=&h^{3/4}h_{4}^{-1/4}\,,
  \label{D0D4:ansatz for scalar:Eq}\\
F_{\left(2\right)}&=&d\left[h^{-1}(x, y, z)\right]
\wedge dt\,,
  \label{D0D4:ansatz for gauge:Eq}\\
F_{\left(4\right)}&=&\e^{-\phi/2}\ast d\left[h_{4}^{-1}(z)
dt\wedge\Omega(\Ysp)\right],
  \label{D0D4:ansatz for gauge2:Eq}
}
where the volume 4-form $\Omega(\Ysp)$ is given by 
\Eqr{
\Omega(\Ysp)=\sqrt{\gamma}\,dy^1\wedge dy^2\wedge dy^3\wedge dy^4\,.
   \label{D0D4:volume y:Eq}
}
Here, $\gamma$ are the determinant of the metric $\gamma_{ij}$.

Under the ansatz for fields \eqref{D0D4:metric:Eq} and 
\eqref{D0D4:ansatz:Eq}, the field equations reduce to 
\Eqrsubl{D0D4:solution1:Eq}{
&&\hspace{-0.5cm}
R_{ij}(\Ysp)=0,~~~~R_{ab}(\Zsp)=0,
   \label{D0D4:Ricci:Eq}\\
&&\hspace{-0.5cm}h=K_0(t)+K_1(y, z),~~~
\pd_t^2K_0=0, ~~~h_{4}\lap_{\Ysp}K_1+\triangle_{\Zsp}K_1=0\,,~~~
\triangle_{\Zsp}h_{4}=0\,,
   \label{D0D4:warp:Eq}
 }
where  $\triangle_{\Ysp}$, $\lap_{\Zsp}$ are
the Laplace operators on 
$\Ysp$, $\Zsp$ spaces, and 
$R_{ij}(\Ysp)$ and $R_{ab}(\Zsp)$ are the Ricci tensors 
constructed from the metrics 
$\gamma_{ij}(\Ysp)$, $u_{ab}(\Zsp)$, respectively. 

Now we consider the case
\Eq{
\gamma_{ij}=\delta_{ij}\,,~~~u_{ab}=\delta_{ab}\,,
 \label{D0D4:flat metric:Eq}
 }
where $\delta_{ij}$, $\delta_{ab}$ are
the four- and five-dimensional Euclidean metrics,
respectively. 
Then, the functions $h_0$ and $h_{4}$ are 
\Eqrsubl{D0D4:solutions1:Eq}{
h(t, y, z)&=&c\,t+\tilde{c}
+\sum_{l}M_l\left[|y^i-y^i_l|^2
+4M_4|z^a-z^a_0|^{-1}\right],
 \label{D0D4:solution-r:Eq}\\
h_{4}(z)&=&\frac{M_4}{|z^a-z^a_0|^{3}},
 \label{D0D4:solution-s:Eq}
}
where $c$, $\tilde{c}$, $M_l$, and $M_4$ are constant parameters,
and $y^i_l$ and $z^a_0$ are also constants.
Now we use the metric 
\Eqr{
u_{ab}dz^adz^b=\delta_{ab}dz^adz^b=dr^2+r^2d\Omega_{(4)}^2\,,
\label{D0D4:flat metric2:Eq}
}
where 
$d\Omega_{(4)}^2$ is the line element of a unit 4-sphere, respectively. 
The forms of 
$h$ and $h_{4}$ are thus 
replaced 
with
\Eqrsubl{D0D4:solutions2:Eq}{
h(t, y, z)&=&c\,t+\tilde{c}
+M\left(y^2+\frac{4M_4}{r}\right),
 \label{D0D4:solution-r2:Eq}\\
h_{4}(z)&=&\frac{M_4}{r^3}\,,
 \label{D0D4:solution-s2:Eq}
} 
where $y^2=\delta_{ij}y^iy^j$, 
and the D0-brane is located at the origin of the Y space, and 
$M$ is constant. 
In terms of Eqs.(\ref{D0D4:flat metric2:Eq}) and (\ref{D0D4:solutions2:Eq}) 
and coordinate transformation 
\Eq{
y^i=\frac{\mu^i}{\zeta}\cos\alpha\,,~~~~
r=\frac{4\zeta^2M_4}{\sin^2\alpha}\,,
   \label{D0D4:coordinate:Eq}
}
the metric in the near-horizon limit is written by
\Eqr{
ds^2&=&\frac{2^{9/4}M^{1/8}M_4^{3/4}}{(\sin\alpha)^{9/4}}
\left(-\frac{\zeta^4}{M}dt^2+\frac{d\zeta^2}{\zeta^2}+d\alpha^2
+\cos^2\alpha d\Omega_{(3)}^2
+\frac{1}{4}\sin^2\alpha d\Omega_{(4)}^2\right)\,,
    \label{D0D4:metric3:Eq}
}
where $\mu^i\mu_i=1$, and the line element of the 3-sphere is defined 
by $d\mu^id\mu_i=d\Omega^2_{(3)}$. 
Thus, the ten-dimensional metric (\ref{D0D4:metric:Eq}) 
describes that the dynamical D0-D4 brane system 
is a warped product of AdS${}_2$ with the internal spaces.

\section{Four-dimensional cosmology}
\label{sec:cosmology}

Now we discuss the application of 
the time-dependent solutions to study
the cosmology. 
In this section, we construct the cosmological model 
from the systems involving three intersecting branes,
i.e., D$p$-D$(p+2)$-NS5 systems 
and provide the brief discussions for other brane systems.
Since our Universe is isotropic and homogeneous, 
we assume that $(p+1)$-dimensional spacetime is an 
isotropic and homogeneous three-space
in the Friedmann-Robertson-Walker universe. 
In the following, the $p$-dimensional metric $q_{\mu\nu}(\Xsp)$ 
is assumed to be Minkowski spacetime. 
We also drop the coordinate dependence on $\Xsp$ 
space except for the time coordinate.

Let us consider the case of the D$p$-D$(p+2)$-NS5 system to 
apply the time-dependent solution to the cosmological models. 
We assume that three space dimensions of the universe can 
be a part of the D$p$-D$(p+2)$-NS5 brane.  
The same branes have to contain three spatial dimensions 
because the four-dimensional universe is isotropic and homogeneous.
Then, the ten-dimensional metric in general can be written by
\Eq{
ds^2=-hdt^2+ds^2(\tilde{\Xsp})+ds^2(\Ysp)+
ds^2({\rm W})+ds^2(\Zsp),
   \label{FRW:metric:Eq}
}
where 
\Eqrsubl{FRW:metric1:Eq}{
ds^2(\tilde{\Xsp})&\equiv&h\delta_{mn}(\tilde{\Xsp})dx^mdx^n,\\
ds^2(\Ysp)&\equiv&h^{\frac{p+1}{8}}_p(t, y, z)
h_{p+2}^{-\frac{5-p}{8}}(z)h_5(z)\gamma_{ij}(\Ysp)dy^idy^j,\\
ds^2({\rm W})&\equiv&h^{-\frac{7-p}{8}}_p(t, y, z)
h_{p+2}^{\frac{p+3}{8}}(z)h_5(z)dw^2,\\
ds^2(\Zsp)&\equiv&h_p^{\frac{p+1}{8}}(t, y, z)
h_{p+2}^{\frac{p+3}{8}}(z)u_{ab}(\Zsp)dz^adz^b,\\
h&\equiv&h^{-\frac{7-p}{8}}_p(t, y, z)
h_{p+2}^{-\frac{5-p}{8}}(z)h_5^{-1/4}(z).
 }
Here, $\delta_{mn}(\tilde{\Xsp})$ is the $(p-1)$-dimensional Euclidean
metric, $\gamma_{ij}(\Ysp)$ and $u_{ab}(\Zsp)$ are the 
three-, $(6-p)$-dimensional metrics, respectively, and $x^m$ denotes 
the coordinate of the $(p-1)$-dimensional Euclid space $\tilde{\Xsp}$, 
$y^i$, $z^a$ are three-, $(6-p)$-dimensional coordinates. 
We also assume $h_{p}=h_{p}(t, y, z)$, 
$h_{p+2}=h_{p+2}(z)$ and $h_5=h_5(z)$. 

Setting $h_p=At+h_1(y, z)$, the ten-dimensional metric
(\ref{FRW:metric1:Eq}) can be written as
\Eqr{
ds^2&=&h_{p+2}^{-\frac{5-p}{8}}h_5^{-\frac{1}{4}}
\left[1+\left(\frac{\tau}{\tau_0}\right)^{-\frac{16}{9+p}}
h_1\right]^{-\frac{7-p}{8}}\nn\\
&&\hspace{-1cm}\times
\left[-d\tau^2+\left(\frac{\tau}{\tau_0}\right)^{\frac{2(p-7)}{9+p}}
\left\{\delta_{PQ}(\tilde{\Xsp})d\theta^Pd\theta^Q
+h_{p+2}h_5dw^2\right\}\right.\nn\\
&&\left.\hspace{-1cm}
+\left\{1+\left(\frac{\tau}{\tau_0}\right)^{-\frac{16}{9+p}}h_1\right\}
\left(\frac{\tau}{\tau_0}\right)^{\frac{2(p+1)}{9+p}}
\left\{h_5\gamma_{ij}(\Ysp)dy^idy^j+h_{p+2}
u_{ab}(\Zsp)dz^adz^b\right\}\right]\,,
   \label{FRW:metric2:Eq}
}
where the cosmic time $\tau$ is defined by
\Eq{
\frac{\tau}{\tau_0}=\left(At\right)^{\frac{p+9}{16}},~~~~\tau_0=
\frac{16}{(p+9)A}.
}
Our three space is given by a three-dimensional subspace in 
$\tilde{\Xsp}$ or Z. However, the power exponent $\lambda$ of the scale 
factor is given by 
\Eq{
\lambda(\tilde{\Xsp})=\frac{p-7}{9+p}\,,~~~~\lambda(\Zsp)=\frac{p+1}{p+9}\,.
}
Since we have obtained $\lambda\le 1$ for $0\le p\le 8$, these
solutions do not lead to accelerated expansion. 
Taking the coordinate transformation 
$\tau=\bar{\tau_{\rm c}}-\bar{\tau}$, we have
\Eqr{
&&\hspace{-1.3cm}ds^2=h_{p+2}^{-\frac{5-p}{8}}h_5^{-\frac{1}{4}}
\left[1+\left(\frac{\bar{\tau}_{\rm c}-\bar{\tau}}
{\tau_0}\right)^{-\frac{16}{9+p}}h_1\right]^{-\frac{7-p}{8}}\nn\\
&&\hspace{-0.8cm}\times\left[-d\bar{\tau}^2
+\left(\frac{\bar{\tau}_{\rm c}-\bar{\tau}}{\tau_0}\right)^{\frac{2(7-p)}{9+p}}
\left\{\delta_{PQ}(\tilde{\Xsp})d\theta^Pd\theta^Q
+h_{p+2}h_5dw^2\right\}\right.\nn\\
&&\left.\hspace{-0.8cm}
+\left\{1+\left(\frac{\bar{\tau}_{\rm c}-\bar{\tau}}{\tau_0}\right)
^{-\frac{16}{9+p}}h_1\right\}
\left(\frac{\bar{\tau}_{\rm c}-\bar{\tau}}{\tau_0}\right)^{\frac{2(p+1)}{9+p}}
\left\{h_5\gamma_{ij}(\Ysp)dy^idy^j+h_{p+2}
u_{ab}(\Zsp)dz^adz^b\right\}\right],
  \label{FRW:ac-metric:Eq}
}
where $\bar{\tau}_{\rm c}$ is a constant. 
For $\bar{\tau}<\bar{\tau}_{\rm c}$ we have accelerated
the expansion of $\tilde{\Xsp}$ spacetime. 
Since the $(p+1)$-dimensional spacetime $\tilde{\Xsp}$ 
is not described by the Einstein-frame metric, 
we consider the cosmic expansion in the Einstein frame. 
In order to discuss the dynamics in the Einstein
frame, we compactify the internal space and perform the 
conformal transformation to make the non-Einstein conformal 
frame into the Einstein frame. 

Now we discuss the cosmological dynamics in the lower-dimensional 
effective theories. 
We compactify $d(\equiv d_{\tilde{\Xsp}}+d_{\Ysp}+d_{\rm W}+d_{\Zsp})$ 
dimensions to the 
$(10-d)$-dimensional universe, where $d_{\tilde{\Xsp}}$, $d_{\Ysp}$, 
$d_{\rm W}$, and $d_{\Zsp}$ denote the compactified dimensions with 
respect to the $\tilde{\Xsp}$, $\Ysp$, W, and $\Zsp$ spaces.
The ten-dimensional metric (\ref{FRW:metric:Eq}) is written by
\Eq{
ds^2=ds^2(\Msp)+ds^2(\Nsp)\,,
   \label{FRW:metric3:Eq}
}
where our Universe is described by the $(10-d)$-dimensional metric 
$ds^2(\Msp)$ and $ds^2(\Nsp)$ is a metric of compactified dimensions.

In order to discuss the dynamics of a $(10-d)$-dimensional universe 
in the Einstein frame, we use the conformal transformation
\Eq{
ds^2(\Msp)=h_p^{B_p}h_{p+2}^{B_{p+2}}h_5^{B_5}ds^2(\bar{\Msp})\,,
}
where $B_p$, $B_{p+2}$, and $B_5$ are expressed as 
\Eqr{
B_p&=&\frac{-(p+1)d+8(d_{\tilde{\Xsp}}+d_{\rm W})}{8(8-d)},~~~~~~
B_{p+2}=\frac{-(p+3)d+8(d_{\tilde{\Xsp}}+d_{\Ysp})}{8(8-d)},\nn\\
B_5&=&\frac{-3d+4(d_{\Ysp}+d_{\Zsp})}{4(8-d)}.
   \label{FRW:power:Eq}
}
Then, the $(10-d)$-dimensional metric in the Einstein frame can be 
written by 
\Eqr{
ds^2(\bar{\Msp})&=&h_p^{B_p'}h_{p+2}^{B_{p+2}'}h_5^{B_5'}
\left[-dt^2+\delta_{P'Q'}
(\tilde{\Xsp}')d\theta^{P'}d\theta^{Q'}
+h_ph_5\gamma_{k'l'}({\Ysp}')dy^{k'}dy^{l'}\right.\nn\\
&&\left.+h_{p+2}h_5dw^2
+h_ph_{p+2}u_{a'b'}({\Zsp}')dz^{a'}dz^{b'}\right],
  \label{FRW:metric-E:Eq}
}
where $\tilde{\Xsp}'$, ${\Ysp}'$, ${\rm W}'$, and ${\Zsp}'$ denote
the $(p-1-d_{\tilde{\Xsp}})$-, $(3-d_{\Ysp})$-, $(1-d_{\rm W})$-, and
$(6-p-d_{\Zsp})$-dimensional spaces, and $B_p'$, $B_{p+2}'$, and $B_5'$ 
are defined by
\Eq{
B_p'=-B_p-\frac{1}{8}(7-p),~~~~B_{p+2}'=-B_{p+2}-\frac{1}{8}(5-p),
~~~~B_5'=B_5-\frac{1}{4}.
}

If we use $h_p=At+h_1$, the $(10-d)$-dimensional 
metric (\ref{FRW:metric-E:Eq}) can be expressed as 
\Eqr{
ds^2(\bar{\Msp})&=&h_{p+2}^{B_{p+2}'}h_5^{B_5'}
\left[1+\left(\frac{\tau}{\tau_0}\right)
^{-\frac{2}{B_p'+2}}h_1\right]^{B_p'}\left[-d\tau^2\right.\nn\\
&&\hspace{-1cm}+
\left(\frac{\tau}{\tau_0}\right)^{\frac{2B_p'}{B_p'+2}}\left\{
\delta_{P'Q'}(\tilde{\Xsp}')d\theta^{P'}d\theta^{Q'}
+h_{p+2}h_5dw^2\right\}
+\left\{1+\left(\frac{\tau}{\tau_0}\right)^
{-\frac{2}{B_p'+2}}h_1\right\}\nn\\
&&\left.\hspace{-1cm}\times
\left(\frac{\tau}{\tau_0}\right)^{\frac{2(B_p'+1)}{B_p'+2}}
\left\{h_5\gamma_{k'l'}({\Ysp}')dy^{k'}dy^{l'}
+h_{p+2}u_{a'b'}({\Zsp}')dz^{a'}dz^{b'}\right\}\right],
  \label{FRW:metric-E2:Eq}
}
where we introduce a cosmic time by
\Eq{
\frac{\tau}{\tau_0}=\left(At\right)^{(B_p'+2)/2},~~~~\tau_0=
\frac{2}{\left(B_p'+2\right)A}.
}

In tables \ref{table_1}-\ref{table_3}, we summarize the cosmological solutions 
 derived from the dynamical brane systems which we have discussed 
 in Sec.\ref{sec:three}. 
 We present the power exponent of the scale factors of our Universe. 
The scale factor of the $(10-d)$-dimensional universe is written by 
$a(\tilde{\Msp})\propto \tau^{\lambda(\tilde{\Msp})}$,
where $\tilde{\Msp}$ is the spatial part of the spacetime $\Msp$, 
and $\lambda(\tilde{\Msp})$ is the power exponent 
of the $\tilde{\Msp}$ space in Jordan frame, and 
$a_{\rm E}(\tilde{\Msp})\propto \tau^{\lambda_{\rm E}(\tilde{\Msp})}$ 
denotes the scale factor of the $\tilde{\Msp}$ space 
in Einstein frame, respectively. 
The expansion law of these brane systems is not complicated because 
the time dependence in the metric comes from only one
brane in the intersections. The solutions we have found do not give 
accelerated expansion of our Universe.

These are the similar results to the case of the other partially 
localized and delocalized intersecting brane solutions. 
For the solutions (\ref{FRW:metric-E2:Eq}), 
we have listed the power exponents of the scale factor
for the intersection involving two branes of 
the ten- or eleven-dimensional supergravity theories 
in the tables in \cite{Minamitsuji:2010kb, Minamitsuji:2011jt}.

The maximum value of $\lambda(\tilde{\rm M})$ is 3/7 in the case of the 
D5-D7-NS5 brane system for the ten-dimensional theory, 
and 2/5 in the case of M2-, M5- , and M2-M2 brane systems with KK monopole 
for the eleven-dimensional model. Although we find the exact
time-dependent brane solution, the power exponent of our scale factor may be 
too small to explain our expanding Universe. Furthermore, in order to 
discuss a de Sitter solution in an intersecting brane system, 
one may need additional ingredients such as a cosmological constant, 
which will be discussed in \cite{Uzawa:2012}. 
The power exponents in the four-dimensional Einstein frame 
depends on how we compactify the internal space in the higher-dimensional 
theory. We list the power exponent of the fastest expansion of 
our four-dimensional Universe in the Einstein frame  
in Table \ref{table_4}.
The fastest expanding case in the Jordan frame
has the power $\lambda(\tilde{\Msp})<1/2$. Then we cannot obtain 
any cosmological model which gives a realistic expansion law.

\section{Conclusion}
  \label{sec:discussions}

In this paper, we have derived the time-dependent solutions 
corresponding to intersecting brane systems. 
We have obtained the dynamical partially localized intersecting branes
in the eleven- or ten-dimensional supergravity models and 
applied them to the four-dimensional cosmology.  
Our solutions constructed using the T-duality map between the 
type IIA and IIB supergravity theories had the similar forms 
to the known dynamical intersecting brane solutions.
These solutions were obtained by replacing a constant $c$ in the warp factor 
$h=c+ h_1$ of a supersymmetric static solution with a linear 
function of the time. This feature is shared by 
 a wide class of supersymmetric solutions beyond examples 
considered in this paper. In the case of intersecting branes, 
the field equations normally indicate that time-dependent solutions 
in supergravity can be found if only one harmonic function in the
metric depends on time. It is not easy to find the solutions of 
the intersecting brane where more than two warp factors depend 
on both time as well as relative or overall transverse space 
coordinates.

We then constructed cosmological models from those solutions 
by smearing some dimensions and compactifying the internal space.
Unfortunately, the powers of the scale factors 
were so small that the dynamical intersecting brane solutions 
could not provide realistic expansions. 
The solutions in the original higher-dimensional theory 
show that the power of the scale factor for the four-dimensional universe 
is proportional to the number of the overlapping dimension $p$. 
If the four-dimensional our Universe stays in the transverse 
space to the time-dependent D$p$-brane, 
the fastest expansion of our Universe can be 
constructed by the dynamical intersection involving D5-brane systems 
in ten-dimensional IIB theory. We can construct the cosmological 
brane solutions involving D6-, D7-, or D8-branes. The 
power of the scale factor for transverse space of these brane system 
is larger than that of the D5-brane. However, these intersecting brane 
systems have the transverse spaces whose dimensions are less than three. 
Then we cannot obtain a homogeneous and isotropic universe as long as 
the transverse space to the intersecting brane is assumed to be a part 
of the four-dimensional universe. On the other hand, we assume that 
our Universe stays along the direction of the overall world volume space 
$\tilde{\Xsp}$. The power of the scale factor of 
$\tilde{\Xsp}$ becomes negative value if $p<7$. In this case, 
although we could find dynamical intersecting brane solutions in which 
the time-dependent warp factor allows the accelerated expansion
in the non-Einstein conformal frame in the higher-dimensional spacetime,
the solutions presented here could not give 
accelerated expansion in the Einstein frame. 

In the lower-dimensional effective theory involving three branes, 
the fastest expansion power is $6/13$, which is found in the case of 
the D5-D7-NS5 brane for the 
nine-dimensional effective theory. Then we have to introduce matter fields 
on the background to construct a realistic cosmological model 
such as inflation, matter- or radiation-dominated universe. 
The properties we have found would also give a clue to investigate 
cosmological brane solutions in a more complicated setup, such as more than 
four intersection of $p$-brane in the eleven- ten-dimensional 
supergravity theory. 

Although the present solutions illustrated in this paper provided 
neither realistic cosmological model, 
the features of the intersecting branes or the method to construct them 
could open new directions to investigate how to obtain not only a realistic
dynamics of branes but also an appropriate higher-dimensional cosmological 
model.

\section*{Acknowledgments}
We would like to thank E. O. Colgain and G. W. Gibbons for valuable comments. 
The work of M.M. was supported by a Yukawa fellowship and 
by a Grant-in-Aid for Young Scientists (B) of JSPS Research,
under Contract No. 24740162. 
K.U. would like to also thank M. Cvetic for discussions and 
her warm hospitality at the University of Pennsylvania.





\begin{table}[p]
\caption{\baselineskip 14pt
Intersecting D-branes with NS-brane. We show the D3-D5-NS5 brane and 
D5-D7-NS5 brane cases which can be applied to cosmology. 
We compactify $d(\equiv d_{\tilde{\Xsp}}+d_{\Ysp}+d_{\rm W}+d_{\Zsp})$ 
dimensions to the $(10-d)$-dimensional Universe,  
where $d_{\tilde{\Xsp}}$, $d_{\Ysp}$, $d_{\rm W}$, and $d_{\Zsp}$ 
denote the compactified dimensions with
respect to the $\tilde{\Xsp}$, $\Ysp$, W, and $\Zsp$ spaces. 
"TD" shows which brane is time dependent.
$\lambda(\tilde{\Msp})$, $\lambda_{\rm E}(\tilde{\Msp})$ are power exponents 
of the cosmic time for the scale factor of the space 
$\tilde{\Msp}$ in Jordan, Einstein frame, respectively.  
}
{\scriptsize
\begin{center}
\begin{tabular}{|c||c|c|c|c|c|c|c|c|c|c|c||c||c|c|c|}
\hline
Branes&&0&1&2&3&4&5&6&7&8&9& TD & $\tilde{\Msp}$ & $\lambda(\tilde{\Msp})$
& $\lambda_{\rm E}(\tilde{\Msp})$
\\
\hline
&D3 & $\circ$ & $\circ$ & $\circ$ &&&& $\circ$ &&&& $\surd$
&  $\tilde{\Xsp}$ \& W & $\lambda(\tilde{\Xsp})=-1/3$&
$\lambda_{\rm E}(\tilde{\Xsp})=\frac{d_{\Ysp}+d_{\Zsp}-4}
{12-2d_{\tilde{\Xsp}}-d_{\Ysp}-2d_{\rm W}-d_{\Zsp}}$
\\
\cline{3-12}
D3-D5-NS5&D5 & $\circ$ &$\circ$& $\circ$ & $\circ$ & $\circ$ & $\circ$ &  &
&&&& & $\lambda({\rm W})=-1/3$ &
 $\lambda_{\rm E}({\rm W})=\frac{d_{\Ysp}+d_{\Zsp}-4}
 {12-2d_{\tilde{\Xsp}}-d_{\Ysp}-2d_{\rm W}-d_{\Zsp}}$
\\
\cline{3-12}
&NS5 & $\circ$ &$\circ$& $\circ$ &  &  &  &  &
 $\circ$  & $\circ$  & $\circ$ && $\Ysp$ \& $\Zsp$ &  $\lambda(\Ysp)$=1/3 &
$\lambda_{\rm E}(\Ysp)=\frac{4-d_{\tilde{\Xsp}}-d_{\rm W}}
{12-2d_{\tilde{\Xsp}}-d_{\Ysp}-2d_{\rm W}-d_{\Zsp}}$
\\
\cline{3-12}
&$x^N$ & $t$ & $x^1$ & $x^2$ & $y^1$ & $y^2$ & $y^3$ & $w$
& $z^1$ & $z^2$ & $z^3$ &
&& $\lambda(\Zsp)$=1/3
 & $\lambda_{\rm E}(\Zsp)=\frac{4-d_{\tilde{\Xsp}}-d_{\rm W}}
 {12-2d_{\tilde{\Xsp}}-d_{\Ysp}-2d_{\rm W}-d_{\Zsp}}$
\\
\hline
\hline
&D5 & $\circ$ & $\circ$ & $\circ$ & $\circ$ & $\circ$ &&&& $\circ$ && $\surd$
& $\tilde{\Xsp}$ \& W  & $\lambda(\tilde{\Xsp})=-1/7$&
$\lambda_{\rm E}(\tilde{\Xsp})=\frac{d_{\Ysp}+d_{\Zsp}-2}
{14-2d_{\tilde{\Xsp}}-d_{\Ysp}-2d_{\rm W}-d_{\Zsp}}$
\\
\cline{3-12}
D5-D7-NS5&D7 & $\circ$ &$\circ$& $\circ$ & $\circ$ & $\circ$ & $\circ$ & 
$\circ$ & $\circ$  &  &&&  & $\lambda({\rm W})=-1/7$ &
 $\lambda_{\rm E}({\rm W})=\frac{d_{\Ysp}+d_{\Zsp}-2}
 {14-2d_{\tilde{\Xsp}}-d_{\Ysp}-2d_{\rm W}-d_{\Zsp}}$
\\
\cline{3-12}
&NS5 & $\circ$ &$\circ$& $\circ$ & $\circ$ & $\circ$ && &
 & & $\circ$  && $\Ysp$ \& $\Zsp$ & $\lambda(\Ysp)$=3/7 &
 $\lambda_{\rm E}(\Ysp)=\frac{6-d_{\tilde{\Xsp}}-d_{\rm W}}
 {14-2d_{\tilde{\Xsp}}-d_{\Ysp}-2d_{\rm W}-d_{\Zsp}}$
\\
\cline{3-12}
&$x^N$ & $t$ & $x^1$ & $x^2$ & $x^3$ & $x^4$ & $y^1$ & $y^2$
& $y^3$ & $w$ & $z$ &
&& $\lambda(\Zsp)$=3/7 & 
$\lambda_{\rm E}(\Zsp)=\frac{6-d_{\tilde{\Xsp}}-d_{\rm W}}
{14-2d_{\tilde{\Xsp}}-d_{\Ysp}-2d_{\rm W}-d_{\Zsp}}$
\\
\hline
\end{tabular}
\end{center}
}
\label{table_1}
\end{table}

\begin{table}[p]
\caption{\baselineskip 14pt
Dynamical M2-brane with KK monopole system in eleven-dimensional 
supergravity theory. 
We compactify $d(\equiv d_{\tilde{\Xsp}}+d_{\Ysp}+d_{\rm U}
+d_{\rm V}+d_{\Zsp})$ dimensions to the $(11-d)$-dimensional Universe,  
where $d_{\tilde{\Xsp}}$, $d_{\Ysp}$, $d_{\rm U}$, $d_{\rm V}$, and $d_{\Zsp}$ 
denote the compactified dimensions with
respect to the $\tilde{\Xsp}$, $\Ysp$, U, V, and $\Zsp$ spaces. 
"TD" denotes which brane (or KK monopole) is time dependent.
$\lambda(\tilde{\Msp})$, $\lambda_{\rm E}(\tilde{\Msp})$ are power exponents 
of the cosmic time for the scale factor of the space
$\tilde{\Msp}$ in Jordan, Einstein frame, respectively.}
{\scriptsize
\begin{center}
\begin{tabular}{|c||c|c|c|c|c|c|c|c|c|c|c|c||c||c|c|c|}
\hline
Case&&0&1&2&3&4&5&6&7&8&9&10& TD & $\tilde{\Msp}$ & $\lambda(\tilde{\Msp})$
& $\lambda_{\rm E}(\tilde{\Msp})$
\\
\hline
&M2 & $\circ$ & $\circ$ & $\circ$ & & & &&&&& 
& $\surd$ &   & $\lambda(\Ysp)=1/4$ &
$\lambda_{\rm E}(\Ysp)=\frac{3-d_{\tilde{\Xsp}}}
{12-2d_{\tilde{\Xsp}}-d_{\Ysp}-d_{\rm U}-d_{\rm V}-d_{\Zsp}}$
\\
\cline{3-13}
M2-M2&M2 & $\circ$ & &  & $\circ$ & $\circ$ & &&&&& 
& &  $\Ysp$ \& U  & $\lambda({\rm U})=1/4$ &
$\lambda_{\rm E}({\rm U})=\frac{3-d_{\tilde{\Xsp}}}
{12-2d_{\tilde{\Xsp}}-d_{\Ysp}-d_{\rm U}-d_{\rm V}-d_{\Zsp}}$
\\
\cline{3-13}
-KK&KK & $\circ$ &$\circ$ &  $\circ$ & $\circ$ &$\circ$ & $\circ$& $\circ$ 
&  &$A_1$& $A_2$ & $A_3$ &
& \& V \& Z  & $\lambda({\rm V})$=1/4 &
$\lambda_{\rm E}({\rm V})=\frac{3-d_{\tilde{\Xsp}}}
{12-2d_{\tilde{\Xsp}}-d_{\Ysp}-d_{\rm U}-d_{\rm V}-d_{\Zsp}}$ 
\\
\cline{3-13}
&$x^N$ & $t$ & $x^1$ & $x^2$ & $y^1$ & $y^2$ & $u^1$ & $u^2$ & $v$
& $z^1$ & $z^2$ & $z^3$ &
& & $\lambda(\Zsp)$=1/4 & 
$\lambda_{\rm E}(\Zsp)=\frac{3-d_{\tilde{\Xsp}}}
{12-2d_{\tilde{\Xsp}}-d_{\Ysp}-d_{\rm U}-d_{\rm V}-d_{\Zsp}}$ 
\\
\hline
\end{tabular}
\end{center}
}
\label{table_2}
\end{table}

\begin{table}[p]
\caption{\baselineskip 14pt
Brane configuration for the D2-D6 with KK monopole system 
and construction of our universe. 
We compactify $d(\equiv d_{\tilde{\Xsp}}+d_{\Ysp}
+d_{\rm V}+d_{\Zsp})$ 
dimensions to the $(10-d)$-dimensional Universe,  
where $d_{\tilde{\Xsp}}$, $d_{\Ysp}$, $d_{\rm V}$, and $d_{\Zsp}$ 
denote the compactified dimensions with
respect to the $\tilde{\Xsp}$, $\Ysp$, V, and $\Zsp$ spaces. 
"TD" shows which brane (or KK monopole) is time dependent.
$\lambda(\tilde{\Msp})$, $\lambda_{\rm E}(\tilde{\Msp})$ are power exponents 
of the cosmic time for the scale factor of the space
$\tilde{\Msp}$ in Jordan, Einstein frame, respectively.
}
\label{two f-1}
{\scriptsize
\begin{center}
\begin{tabular}{|c||c|c|c|c|c|c|c|c|c|c|c||c||c|c|c|}
\hline
Branes&&0&1&2&3&4&5&6&7&8&9& TD & $\tilde{\Msp}$ & $\lambda(\tilde{\Msp})$
& $\lambda_{\rm E}(\tilde{\Msp})$
\\
\hline
&D2 & $\circ$ & $\circ$ & $\circ$ & &&&&&&& $\surd$
& & $\lambda(\Ysp)$=3/11&
$\lambda_{\rm E}(\Ysp)=\frac{3-d_{\tilde{\Xsp}}}
{11-2d_{\tilde{\Xsp}}-d_{\Ysp}-d_{\rm V}-d_{\Zsp}}$
\\
\cline{3-12}
D2-D6-KK&D6 & $\circ$ &$\circ$& $\circ$ & $\circ$ & $\circ$ & $\circ$ & 
 $\circ$ &&   &  && $\Ysp$ \& V \& Z & $\lambda({\rm V})$=3/11 &
 $\lambda_{\rm E}({\rm V})=\frac{3-d_{\tilde{\Xsp}}}
{11-2d_{\tilde{\Xsp}}-d_{\Ysp}-d_{\rm V}-d_{\Zsp}}$
\\
\cline{3-12}
&KK & $\circ$ &$\circ$& $\circ$ & $A_1$ & $A_2$ & $A_3$ &  &
 $\circ$& $\circ$ & $\circ$ && & $\lambda({\rm Z})$=3/11 & 
 $\lambda_{\rm E}({\rm Z})=\frac{3-d_{\tilde{\Xsp}}}
{11-2d_{\tilde{\Xsp}}-d_{\Ysp}-d_{\rm V}-d_{\Zsp}}$
\\
\cline{3-12}
&$x^N$ & $t$ & $x^1$ & $x^2$ & $y^1$ & $y^2$ & $y^3$ & $v$
& $z^1$ & $z^2$ & $z^3$ &
&& & 
\\
\hline
\end{tabular}
\end{center}
}
\label{table_3}
\end{table}

\begin{table}[p]
\caption{\baselineskip 14pt
Intersection involving three branes. Here we show the 
maximum value of the power exponent in the Einstein frame for 
dynamical brane and KK monopole in ten- or eleven-dimensional supergravity 
theory. ``TD" in the table denotes which brane is time dependent.
}
\begin{center}
{\scriptsize
\begin{tabular}{|c|c|c||c|c||c|}
\hline
Branes& TD &dim$(\Msp)$  &$\bar{\Msp}$
& $d$ &
$\lambda_{\rm E}(\bar{\Msp})$
\\
\hline
\hline
D3-D5-NS5&D3& 7& $\tilde{\Xsp}$ \& Y \& W \& Z & 
$(d_{\tilde{\Xsp}}, d_{\Ysp}, d_{\rm W}, d_{\Zsp})=(0, 1, 0, 2)$ &4/9
\\
\cline{2-6}
&D3&7& $\tilde{\tilde{\Xsp}}$ \& Y \& W \& Z & 
$(d_{\tilde{\Xsp}}, d_{\Ysp}, d_{\rm W}, d_{\Zsp})=(0, 2, 0, 1)$ &4/9 
\\
\hline
D5-D7-NS5&D5& 9& $\tilde{\Xsp}$ \& Y \& W \& Z & 
$(d_{\tilde{\Xsp}}, d_{\Ysp}, d_{\rm W}, d_{\Zsp})=(0, 1, 0, 0)$ &6/13
\\
\hline
M2-M2-KK&M2&6& $\tilde{\Xsp}$ \& Y \& U \& V \& Z& 
$(d_{\tilde{\Xsp}}, d_{\Ysp}, d_{\rm U}, d_{\rm V}, d_{\Zsp})
=(0, 1, 2, 0, 2)$ &3/7 
\\
\hline
D2-D6-KK&D2&6& $\tilde{\Xsp}$ \& Y \& V \& U & $(d_{\tilde{\Xsp}}, d_{\Ysp}, 
d_{\rm V}, d_{\Zsp})=(0, 2, 0, 2)$ & 3/7 
\\
\hline
\end{tabular}
}
\label{table_4}
\end{center}
\end{table}


\begin{thebibliography}{99}

\bibitem{Lu:1996jk}
  H.~Lu, S.~Mukherji, C.~N.~Pope and K.~W.~Xu,
  ``Cosmological solutions in string theories,''
  Phys.\ Rev.\  D {\bf 55} (1997) 7926
  [arXiv:hep-th/9610107].

\bibitem{Behrndt:2003cx}
  K.~Behrndt and M.~Cvetic,
  ``Time-dependent backgrounds from supergravity with gauged non-compact
  $R$-symmetry,''
  Class.\ Quant.\ Grav.\  {\bf 20} (2003) 4177
  [arXiv:hep-th/0303266].

\bibitem{Gibbons:2005rt}
  G.~W.~Gibbons, H.~Lu and C.~N.~Pope,
  ``Brane worlds in collision,''
  Phys.\ Rev.\ Lett.\  {\bf 94} (2005) 131602
  [arXiv:hep-th/0501117].
  
\bibitem{Chen:2005jp}
  W.~Chen, Z.~W.~Chong, G.~W.~Gibbons, H.~Lu and C.~N.~Pope,
  ``Horava-Witten stability: Eppur si muove,''
  Nucl.\ Phys.\  B {\bf 732} (2006) 118
  [arXiv:hep-th/0502077].

\bibitem{Kodama:2005fz}
  H.~Kodama and K.~Uzawa,
  ``Moduli instability in warped compactifications of the type IIB
  supergravity,''
  JHEP {\bf 0507} (2005) 061
  [arXiv:hep-th/0504193].

\bibitem{Kodama:2005cz}
  H.~Kodama and K.~Uzawa,
  ``Comments on the four-dimensional effective theory for warped
  compactification,''
  JHEP {\bf 0603} (2006) 053
  [arXiv:hep-th/0512104].

\bibitem{Kodama:2006ay}
  H.~Kodama and K.~Uzawa,
  ``Moduli instability in warped compactification,''
  arXiv:hep-th/0601100.

\bibitem{Binetruy:2007tu}
  P.~Binetruy, M.~Sasaki and K.~Uzawa,
  ``Dynamical D4-D8 and D3-D7 branes in supergravity,''
  Phys.\ Rev.\  D {\bf 80} (2009) 026001
  [arXiv:0712.3615 [hep-th]].

\bibitem{Binetruy:2008ev}
  P.~Binetruy, M.~Sasaki and K.~Uzawa,
  ``Dynamical solution of supergravity,''
  arXiv:0801.3507 [hep-th].

\bibitem{Maeda:2009tq}
  K.~i.~Maeda, N.~Ohta, M.~Tanabe and R.~Wakebe,
  ``Supersymmetric Intersecting Branes in Time-dependent Backgrounds,''
  JHEP {\bf 0906} (2009) 036
  [arXiv:0903.3298 [hep-th]].

\bibitem{Maeda:2009zi}
  K.~i.~Maeda, N.~Ohta and K.~Uzawa,
  ``Dynamics of intersecting brane systems -- Classification and their
  applications --,''
  JHEP {\bf 0906} (2009) 051
  [arXiv:0903.5483 [hep-th]].

\bibitem{Gibbons:2009dr}
  G.~W.~Gibbons and K.~i.~Maeda,
  ``Black Holes in an Expanding Universe,''
  Phys.\ Rev.\ Lett.\  {\bf 104} (2010) 131101
  [arXiv:0912.2809 [gr-qc]].

\bibitem{Maeda:2009ds}
  K.~i.~Maeda and M.~Nozawa,
  ``Black Hole in the Expanding Universe from Intersecting Branes,''
  Phys.\ Rev.\  D {\bf 81} (2010) 044017
  [arXiv:0912.2811 [hep-th]].

\bibitem{Maeda:2010yk}
  K.~i.~Maeda, N.~Ohta, M.~Tanabe and R.~Wakebe,
  ``Supersymmetric Intersecting Branes on the Waves,''
  JHEP {\bf 1004} (2010) 013
  [arXiv:1001.2640 [hep-th]].

\bibitem{Maeda:2010ja}
  K.~i.~Maeda and M.~Nozawa,
  ``Black Hole in the Expanding Universe with Arbitrary Power-Law Expansion,''
  Phys.\ Rev.\  D {\bf 81} (2010) 124038
  [arXiv:1003.2849 [gr-qc]].

\bibitem{Minamitsuji:2010fp}
  M.~Minamitsuji, N.~Ohta and K.~Uzawa,
  ``Dynamical solutions in the 3-Form Field Background in the
  Nishino-Salam-Sezgin Model,''
  Phys.\ Rev.\  D {\bf 81} (2010) 126005
  [arXiv:1003.5967 [hep-th]].

\bibitem{Maeda:2010aj}
  K.~i.~Maeda, M.~Minamitsuji, N.~Ohta and K.~Uzawa,
  ``Dynamical $p$-branes with a cosmological constant,''
  Phys.\ Rev.\  D {\bf 82} (2010) 046007
  [arXiv:1006.2306 [hep-th]].

\bibitem{Minamitsuji:2010kb}
  M.~Minamitsuji, N.~Ohta and K.~Uzawa,
  ``Cosmological intersecting brane solutions,''
  Phys.\ Rev.\  D {\bf 82} (2010) 086002
  [arXiv:1007.1762 [hep-th]].

\bibitem{Minamitsuji:2010uz}
  M.~Minamitsuji and K.~Uzawa,
  ``Cosmology in $p$-brane systems,''
  Phys.\ Rev.\  D {\bf 83} (2011) 086002
  [arXiv:1011.2376 [hep-th]].

\bibitem{Maeda:2011sh}
  K.~i.~Maeda and M.~Nozawa,
  ``Black hole solutions in string theory,''
  Prog.\ Theor.\ Phys.\ Suppl.\  {\bf 189} (2011) 310
  [arXiv:1104.1849 [hep-th]].

\bibitem{Minamitsuji:2011jt}
  M.~Minamitsuji and K.~Uzawa,
  ``Dynamics of partially localized brane systems,''
  Phys.\ Rev.\  D {\bf 84} (2011) 126006
  [arXiv:1109.1415 [hep-th]].

\bibitem{Maeda:2012xb}
  K.~i.~Maeda and K.~Uzawa,
  ``Dynamical brane with angles : Collision of the universes,''
  Phys.\ Rev.\  D {\bf 85} (2012) 086004
  [arXiv:1201.3213 [hep-th]].

\bibitem{Blaback:2012mu}
  J.~Blaback, B.~Janssen, T.~Van Riet and B.~Vercnocke,
  ``Fractional branes, warped compactifications and backreacted 
  orientifold planes,''
  JHEP {\bf 1210} (2012) 139
  [arXiv:1207.0814 [hep-th]].
  
\bibitem{Cvetic:2000cj}
  M.~Cvetic, H.~Lu, C.~N.~Pope and J.~F.~Vazquez-Poritz,
  ``AdS in warped spacetimes,''
  Phys.\ Rev.\  D {\bf 62} (2000) 122003
  [arXiv:hep-th/0005246].
  
\bibitem{Youm:1999ti}
  D.~Youm,
  ``Partially localized intersecting BPS branes,''
  Nucl.\ Phys.\ B {\bf 556} (1999) 222.

\bibitem{Bergshoeff:1994cb}
  E.~Bergshoeff, R.~Kallosh and T.~Ortin,
  ``Duality versus supersymmetry and compactification,''  
  Phys.\ Rev.\ D {\bf 51} (1995) 3009  [hep-th/9410230].
  
\bibitem{Bergshoeff:1995as}
  E.~Bergshoeff, C.~M.~Hull and T.~Ortin,
  ``Duality in the type II superstring effective action,''
  Nucl.\ Phys.\  B {\bf 451} (1995) 547
  [arXiv:hep-th/9504081].

\bibitem{Breckenridge:1996tt}
  J.~C.~Breckenridge, G.~Michaud and R.~C.~Myers,
  ``More D-brane bound states,''
  Phys.\ Rev.\  D {\bf 55} (1997) 6438
  [arXiv:hep-th/9611174].

\bibitem{Costa:1996zd}
  M.~S.~Costa and G.~Papadopoulos,
  ``Superstring dualities and $p$-brane bound states,''
  Nucl.\ Phys.\  B {\bf 510} (1998) 217
  [arXiv:hep-th/9612204].
  
\bibitem{Cvetic:1999xx}
  M.~Cvetic, S.~S.~Gubser, H.~Lu and C.~N.~Pope,
  ``Symmetric potentials of gauged supergravities in diverse dimensions and
  Coulomb branch of gauge theories,''
  Phys.\ Rev.\  D {\bf 62} (2000) 086003
  [arXiv:hep-th/9909121].

\bibitem{Cvetic:1999un}
  M.~Cvetic, H.~Lu and C.~N.~Pope,
  ``Gauged six-dimensional supergravity from massive type IIA,''
  Phys.\ Rev.\ Lett.\  {\bf 83} (1999) 5226
  [arXiv:hep-th/9906221].  
  
\bibitem{Nastase:2003dd}
  H.~Nastase,
  ``On D$p$-D$(p+4)$ systems, QCD dual and phenomenology,''
  arXiv:hep-th/0305069.

\bibitem{Brandhuber:1999np}
  A.~Brandhuber and Y.~Oz,
  ``The D4-D8 brane system and five dimensional fixed points,''
  Phys.\ Lett.\  B {\bf 460} (1999) 307
  [arXiv:hep-th/9905148].

\bibitem{Behrndt:1999mk}
  K.~Behrndt, E.~Bergshoeff, R.~Halbersma and J.~P.~van der Schaar,
  ``On domain-wall/QFT dualities in various dimensions,''
  Class.\ Quant.\ Grav.\  {\bf 16} (1999) 3517
  [arXiv:hep-th/9907006].

\bibitem{Nunez:2001pt}
  C.~Nunez, I.~Y.~Park, M.~Schvellinger and T.~A.~Tran,
  ``Supergravity duals of gauge theories from F(4) gauged supergravity in  six
  dimensions,''
  JHEP {\bf 0104} (2001) 025
  [arXiv:hep-th/0103080].

\bibitem{Itsios:2012dc}
  G.~Itsios, Y.~Lozano, E.~.O Colgain and K.~Sfetsos,
  ``Non-Abelian T-duality and consistent truncations in type-II supergravity,''
  JHEP {\bf 1208} (2012) 132
  [arXiv:1205.2274 [hep-th]]. 

\bibitem{Bergman:2012kr}
  O.~Bergman and D.~Rodriguez-Gomez,
  ``5d quivers and their AdS${}_6$ duals,''
  JHEP {\bf 1207} (2012) 171
  [arXiv:1206.3503 [hep-th]].

\bibitem{Itzhaki:1998uz}
  N.~Itzhaki, A.~A.~Tseytlin and S.~Yankielowicz,
  ``Supergravity solutions for branes localized within branes,''
  Phys.\ Lett.\  B {\bf 432} (1998) 298
  [arXiv:hep-th/9803103].

\bibitem{Uzawa:2012}
  K.~Uzawa, in preparation.

\end{thebibliography}
\end{document}